\documentclass[reprint]{revtex4-2}
\usepackage[utf8]{inputenc}

\usepackage{natbib}
\usepackage{graphicx}
\usepackage{amsmath}
\usepackage{amsthm}
\usepackage{color}
\usepackage{amssymb}

\usepackage{hyperref}

\newcommand{\bes}{\begin{subequations}}
\newcommand{\ees}{\end{subequations}}

\newtheorem{lemma}{Lemma}

\begin{document}

\title{Precision of quantum simulation of all-to-all coupling in a local architecture}
\author{Evgeny Mozgunov }
\address{Information Sciences Institute, University of Southern California, Marina del Rey, CA, 90292, USA}
\begin{abstract}
We present a simple 2d local circuit that implements all-to-all interactions via perturbative gadgets. We find an analytic relation between the values $J_{ij}$ of the desired interaction and the parameters of the 2d circuit, as well as the expression for the error in the quantum spectrum. For the relative error to be a constant $\epsilon$, one requires an energy scale growing as $n^6$ in the number of qubits, or equivalently a control precision up to $ n^{-6}$. Our proof is based on the Schrieffer-Wolff transformation and generalizes to any hardware. In the architectures available today, $5$ digits of control precision are sufficient for $n=40,~ \epsilon =0.1$. Comparing our construction, known as paramagnetic trees, to ferromagnetic chains used in minor embedding, we find that at chain length $>3$ the performance of minor embedding degrades exponentially with the length of the chain, while our construction experiences only a polynomial decrease.
\end{abstract}

\maketitle

\section{Introduction}

Quantum simulation can be performed on a future fault-tolerant gate-based computer with minimal overhead \cite{haah}. Yet, we believe there will always be use cases for analog devices that, instead of quantum gates, implement the simulated Hamiltonian directly. In the NISQ era, they are the only ones available at the system sizes of interest \cite{king2022quantum,ebadi2022quantum}, and in the future competition with the fault-tolerant gate-based approaches, they may still prove to be more economical. There are direct applications of such analog quantum simulators to the study of many-body physics in search for insights for material science \cite{king2022quantum}, as well as the alternative computing approach where a physical system is driven to solve an abstract computational problem, best exemplified by quantum annealing and its application to binary optimization \cite{ebadi2022quantum}. An obstacle on the path to those two applications is the inevitable difference between the hardware interaction graph of the quantum simulator and the desired interaction graph of the target system of interest. This obstacle can be circumvented by embedding the logical qubits of the target Hamiltonian into a repetition code in the hardware Hamiltonian \cite{choi2008minor, cai2014practical}. The performance of the quantum simulators after such an embedding suffers: the scaling of the time-to-solution of the embedded optimization problems becomes far worse \cite{Kowalsky_2022} than that of the native ones \cite{Mandr__2018}, and the accessible range of transverse fields flipping the value stored in the repetition code becomes exponentially reduced \cite{king2022quantum} with the length of the repetition code. This has been a major obstacle to demonstrating a clear advantage of the analog quantum simulators on a problem of practical interest, despite large qubit numbers and a promising performance on the native problems \cite{ebadi2022quantum}.

We present a solution to the exponential decrease in performance with the length of the repetition code: instead of a ferromagnetic repetition code, one needs to use a paramagnetic chain in its ground state as the interaction mediator, together with a single well-isolated hardware qubit serving as a logical qubit. This idea has already appeared under the name of {\emph{paramagnetic trees}} \cite{Kerm, article}, and here we provide a theoretical justification for this approach. We observe that such a mediator is a type of perturbative gadget \cite{kempe:1070}, and analyze it via an exact version of perturbation theory: a Schrieffer-Wolff transformation \cite{B1}. Perturbative gadgets were previously used to implement a many-body interaction using only two body terms \cite{Cao:17}. Here we use them instead to implement a long-range two-body interaction using only nearest-neighbor two-body terms \cite{article}. The mediator can be any physical system. We investigate several cases, focusing our attention on the transmission line with bosonic degrees of freedom as all the relevant quantities can be found analytically. A fermionic or spin chain near its critical point would have worked just as well. We note that other methods \cite{kempe:1070,Cao:15,Bausch:20} for the study of perturbative gadgets can provide better performance guarantees than the Schrieffer-Wolff, but we choose to use it as its application is straightforward and it maintains the information about the basis change induced by the presence of the gadget.

Our result did not appear in the literature to the best of our knowledge. The works \cite{Lechner:2015,Puri:2016aa} constructed a classical all-to-all system that would generally exhibit different quantum properties from the target system when the quantum terms are turned on. Schrieffer-Wolff has been applied to the circuit model of interactions between a pair of qubits\cite{Consani_2020}, but not for long-range interactions or a large interaction graph.
In Sec. \ref{sec:tar} we define the problem of quantum simulation of an all-to-all coupling, and in Sec. \ref{sec:main} we present our solution to it: a physically realistic 2d layout of circuit elements on a chip. Our other results for variations of this problem are summarized in Sec. \ref{sec:oth}. Our method is a version of a perturbation theory introduced in Sec. \ref{sec:pert} and proven in App. \ref{app:Bravo}. We illustrate its use in an example of the effect of non-qubit levels on a qubit quantum simulator in App. \ref{warmup}, before stating in  Sec. \ref{general} and proving in App. \ref{proofG} the all-to-all gadget theorem at the center of this work. 
The calculations for applications of our theorem to various architectures of an all-to-all gadget can be found in App. \ref{sec:app}. The Sec. \ref{sec:opt} and App. \ref{40g} are the most practically relevant to the applications of our gadget on current quantum annealers such as D-Wave \cite{boothby2021architectural}. We discuss how to quantify the accuracy of a quantum simulator from the application perspective in App. \ref{PCP}, and present an in-depth study of the circuits of our gadget in App. \ref{tlSec}.

\section{Problem setting}
\label{sec:tar}
The target qubit Hamiltonian of $n$ qubits we wish to implement is the transverse field Ising model on arbitrary graphs $G$ of degree $2s$ that can be as big as the number of qubits $n-1$ (such that the number of edges is $ns$):
\begin{equation}
    H_{\text{target}} =\sum_{ij \in G} J_{ij}\sigma_i^z \sigma^z_j + \sum_i h_i \sigma^z_i + t_i \sigma^x_i \ .\label{targ1}
\end{equation}
The fields and interactions $h_i,t_i, J_{ij}$ can take any values in $[-1,1]$. Note that some of the $J_{ij}$ can be $0$, which means models that do not have a regular interaction graph can be cast into the form above. For the purposes of this work, the smallest nontrivial graph is $n=4$ complete graph with $s=1.5$. We do not see the need for interaction gadgets for $s=1$ graphs consisting of rings. Thus the range of $n,s$ for this work is $n\geq 4, ~s\geq 1.5$.

By implementing Eq. (\ref{targ1}) we mean that another quantum system will have all $2^n$ levels of the quantum spectrum of Eq. (\ref{targ1}) to some set precision. This faithful reproduction of the quantum spectrum of the desired Hamiltonian is the key difference from embedding methods \cite{Lechner:2015,Puri:2016aa} that would only reproduce the $\sigma^z$ part of $H_{\text{targ}}$. The implementation may also have additional levels above the $2^n$ levels we use. We will propose several architectures that are possible on a chip, that is, in a 2d plane with elements that are qubits and circuit elements such as inductors and capacitors. It is not easy to formally define what the set of allowed architectures in this lumped element description is. To justify proposed architectures, we will draw parallels between them and the devices available today. The reason why all-to-all coupling in Eq. (\ref{targ1}) is impossible on a chip is that to implement it, the qubits on which $H_{\text{target}}$ is defined will have to be extended objects, which will lead to the failure of the qubit approximation, as well as uncontrollable levels of noise. Instead, we take inspiration from the idea of {\emph{paramagnetic trees}} \cite{article,Kerm} where the qubits are well isolated and highly coherent, and the extended objects connecting different qubits are the mediators of interactions. 

\section{Analytic expressions for the gadget}
\label{sec:main}
\begin{figure}
\centering
\includegraphics[width=\columnwidth]{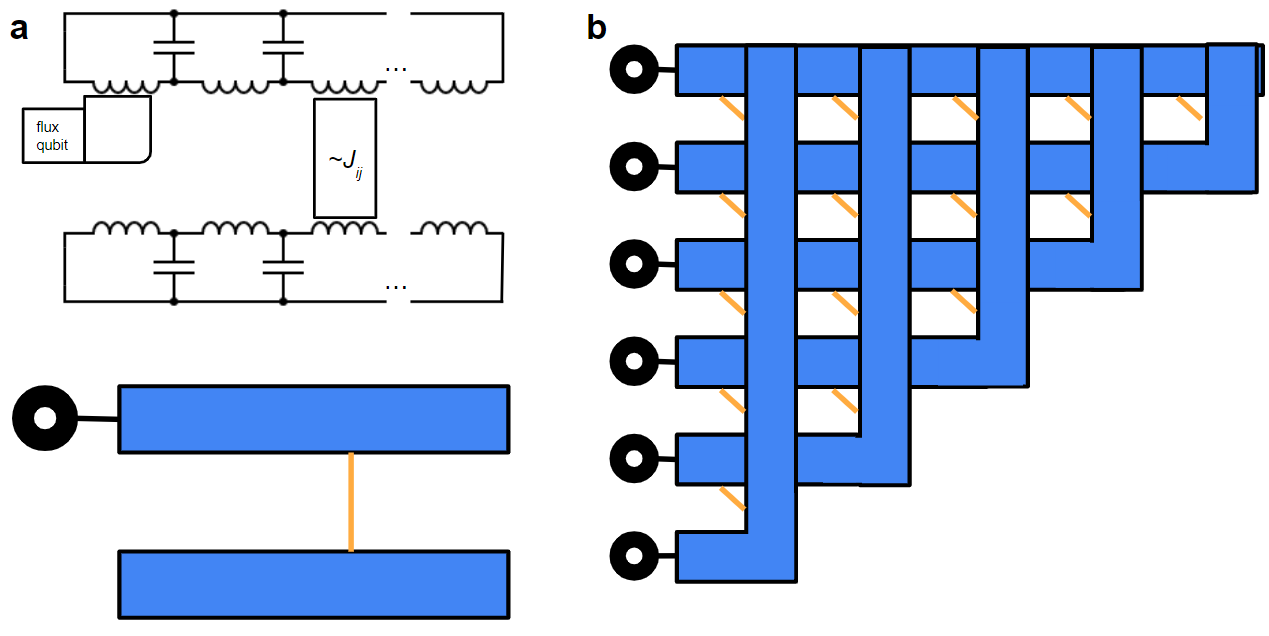}
\caption{ (a) An explanation of the graphical notation for the circuit elements. Transmission lines are shown in blue, the black circles are flux qubits, and the black line indicates the inductive coupling between the two. The yellow lines are tunable inductive couplings between the segments of the transmission lines. (b) The
layout for the analytic construction of the all-to-all gadget. Yellow couplings encode the desired values of the all-to-all coupling $J_{ij}$ with a coefficient $\alpha \chi^{-2}$.}
\label{circ}
\end{figure}

We present a construction of a scalable paramagnetic tree \cite{article,Kerm}. Specifically, we show that a circuit on a chip with a constant density of elements shown in Fig.~\ref{circ} implements the desired all-to-all interaction (Eq. (\ref{targ1})) to a controlled precision. 

This circuit illustrated in Fig. \ref{circ} contains $n$ flux qubits and $n$ transmission lines, each modeled as $n$ segments of an inductance and a capacitance, with an extra inductance before closing the loop at the end. A single harmonic oscillator can be associated with an LC circuit, and similarly, it can be shown (see App. \ref{tlSec}) that a transmission line is equivalent to the Hamiltonian of a chain of coupled harmonic oscillators:
\begin{equation}
    H_{ci} = \sum_{l=1}^n p_{ci,l}^2 + 2x_{ci,l}^2- 2\sum_{l=1}^{n-1}x_{ci,l}x_{ci,l+1}+  Z_i x_{ci,1}\ .
\end{equation}
Here the index $i$ denotes which of the $n$ transmission lines is considered. In our notation, $Z_i,~X_i$ are Pauli operators on $i$'th flux qubit, and $x_{ci,l}, p_{ci,l}$ are the canonically conjugate coordinate and momentum of the $l$'th harmonic oscillator of the $i$'th chain. The last term is responsible for the coupling of the $i$'th chain to its designated qubit. The degrees of freedom $x_{ci,l}$ are such that the flux through $l$'th inductor is $x_{ci,l}-x_{ci,l-1}$, except for $l=1$ where the flux coincides with $x_{ci,1}$ so that we can couple the qubit to $x$ directly. The coupling between such transmission lines is due to the mutual inductances, and the location in the $j$'th transmission line where it couples to the $i$'th transmission line is given by
\begin{equation}
    r_j(i) =
    \begin{cases}
	 i,~ i>j; &\\         i+1,~ i<j\ . & \end{cases} \label{rDef}
\end{equation}
Define the interaction term:
\begin{equation}
    I_{i,j} =x_{ci,r_i(j)} -x_{ci,r_i(j)-1} \ .
\end{equation}
Note that the number of harmonic oscillators in our model of the transmission line was set to $n$ for this convenient definition of $I_{i.j}$. More generally, since the length of the transmission line is $\sim n$, the number of harmonic oscillators will be $\sim n$ with some coefficient. That coefficient, together with the characteristic energy scales of $x^2, p^2$, and $Zx$ terms needs to be informed by the hardware constraints outlined in App. \ref{tlSec}. Optimizing these parameters will give a prefactor improvement to the performance of our gadget, but we do not expect it to change the scaling we obtain below.

We propose the following  Hamiltonian $H_0 +V$ for our gadget:
\begin{align}
    V =    \alpha\sum_i (h_i  Z_i  + F^{-1}t_i  X_i+ \chi^{-2}\sum_{j>i}J_{ij} I_{i,j}I_{j,i} ),\\
     H_0 =  \sum_i H_{ci}  \ . 
    \end{align}
The coefficients $\alpha, F,\chi$ are the parameters of the gadget. The reduction factor $\alpha$ corresponds to the reduction in the energy scale between $\sim 1$ terms in $H_{\text{targ}}$ and $\sim \alpha$ terms of the effective Hamiltonian of our gadget.  We assume that our implementation of $H+V$ is imperfect, that is, we implement the Hamiltonian $H+V +V_n$, and our control noise $V_n$ is:
\begin{align}
    V_n =\sum_i (\delta h^c_{i}  Z_i  + \delta t^c_{i}  X_i+ \delta_{zx,i} Zx_{ci,1}+\\+ \sum_{j>i}\delta f_{ij} I_{i,j}I_{j,i}  + \sum_{l}\delta_{x,il} I_{i,l}  ) \ .
\end{align}
The individual errors are unknown, but their strength is characterized by  $ \delta_{H,\text{loc}}, \delta_1, \delta$. Here $\delta_1\geq|\delta_{zx,i}|$, $\delta\geq |\delta h^c_{i}|, |\delta t^c_{i}|,|\delta f^c_{ij}|$, and $ \delta_{H,\text{loc}} \geq |\delta_{x,il}|$ is the local error of implementation of the transmission line $H_{ci}$. The following statement specifies the values of these parameters that guarantee that the gadget fulfills the task:

\emph{Main result:}
The gadget effective Hamiltonian satisfies:
    \begin{align}
   \| H_{\text{eff}} - \alpha H_{\text{target}}\|\leq \alpha \epsilon n s\ ,
\end{align}
if the error $\epsilon$ and the control errors satisfy the inequality:
  \begin{align}
   \sqrt{2} n \delta_{H,\text{loc}}  +\left(1+\sqrt{\text{ln} n }\right)\delta_1  +  3\delta \leq \frac{0.01 \epsilon^2}{n(n+1)^5}
    \ .
\end{align} 
We are free to choose any such $\epsilon$, and we used the following values of the remaining parameters in our construction:
\begin{equation}
    \alpha_o =\frac{0.035\epsilon  }{ns(n+1)^5} \ .
\end{equation}
The factor $F$ is given via a sum:
\begin{equation}
    F^{-1} = \text{exp} \frac{1}{4(n+1)}\sum_{k=1}^{n} \frac{\cos^2 \frac{k\pi}{2(n+1)} }{\sin\frac{k\pi}{2(n+1)}} \leq e^{1/8}n^{1/4}\ .
\end{equation}
The extra factor for the interactions is:
   \begin{align}
     \chi = 1/(n+1) \ .
\end{align}
The gap of $H_0$ is
\begin{equation}
    \Delta = 2 \sin \frac{\pi}{2(n+1)} \ . 
\end{equation}

This establishes theoretically that an all-to-all interaction of an arbitrary number $n$ of qubits can be realized in 2D hardware at the cost of a polynomial ($1/n^6$) reduction in the interaction strength compared to the physical energy scale. Equivalently, to get unit interaction strength, the energy scale of the hardware should scale as $n^6$.

Moreover, the lowest $2^n$ eigenvalues of the quantum spectrum match between the circuit Hamiltonian and the target, and the gap $\sim \Delta$ separates them from the other eigenvalues. The rigorous meaning of the effective Hamiltonian is discussed in Sec. \ref{sec:pert} and App. \ref{app:Bravo}. The control errors $ \delta_{H,\text{loc}}, \delta_1, \delta$ are required to be polynomially small as well ($1/n^7$ for the elements of the transmission line, $1/n^6$ up to logarithmic factors for everything else). The specific power of the scaling takes into account the chosen allowance for error  $\epsilon n s$, treating $\epsilon$ as a constant. The motivation for allowing this extensive error and the initial comparison with the gate-based approach to quantum simulation are presented in App. \ref{PCP}. If instead, we require a constant global error $\| H_{\text{eff}} - \alpha H_{\text{target}}\|\leq \alpha \epsilon_G$, we can use $\epsilon = \epsilon_G/ns$ in the inequalities of this paper to obtain the corresponding control precision requirements. For this gadget, one obtains $n^{-9}$ and $n^{-8}$ for respective $\delta$'s. 

The powers of $n$ in our rigorous result can also be obtained by the following back-of-the-envelope calculation. Each mediator is a distributed circuit element with $n$ effective degrees of freedom. The linear response $\chi$ of the ground state to a qubit attached to its end will be distributed evenly as $1/n$ at each of the degrees of freedom. Since each interaction between qubits involves four elements: qubit-mediator-mediator-qubit, the interaction strength between two mediators needs to be $\chi^{-2}$ times higher than its target value for qubits. We use the reduction factor $\alpha$ to get into the range of applicability of the perturbation theory, s.t. the magnitude of the perturbation $V$ can be estimated as $\alpha \chi^{-2}sn$. Even without control errors, the second order of the perturbation theory $\sim V^2/\Delta$ needs to be within our error budget $\alpha \epsilon n s$. Plugging in $V \to\alpha \chi^{-2}sn$, we obtain:
\begin{equation}
    \alpha \sim \frac{\Delta \chi^4 \epsilon }{ns}\ .
\end{equation}
For a constant $\Delta $ the response of most 1d mediators decays exponentially, so its optimal to take $\Delta\sim 1/n$ to get the response $\chi\sim 1/n$, which leads to $\alpha \sim 1/n^6$. With that, the error budget becomes $\sim\epsilon^2 /n^5 $, and the control errors $n^2 \delta_{H,\text{loc}} + n(\delta_1 +\delta)$ (estimated by counting the number of terms) need to be at least less than the error budget, resulting in $\delta_{H,\text{loc}} \sim1/n^7$ and $\delta_1, \delta \sim 1/n^6$. The main result of our work is making this back-of-the-envelope calculation rigorous and obtaining an analytic expression for the required controls.  

Note that while the expression for $F$ is not analytically computable, there is a sequence of approximate analytic expressions for it that correspond to progressively smaller errors in $t^*$. This error becomes smaller than $\epsilon$ for some order of the analytic expression, or we can numerically compute $F$ and get the exact value of $t^*$ for that $n$.

Numerical investigation \cite{urlCode} shows that: 
\begin{align}
   c_l(n+1)^{1/2\pi}\leq F^{-1} \leq c_u(n+1)^{1/2\pi}\ ,\\
    c_l =e^{\frac{(\gamma - 1 -\text{ln}(\pi/4))}{2\pi}} \approx 0.9716\ ,\\
     c_u =\frac{e^{\frac{1}{8\sqrt{2}}}}{2^{\frac{1}{2\pi}}}\approx 0.9783\ .
\end{align}
Using either the left or the right bound instead of the exact expression for $F^{-1}$ introduces only $<1\%$ relative error in $t^*$. So if $\epsilon> 0.01$, using the approximate analytic expression won't significantly change the overall error.

\section{List of other results}
\label{sec:oth}
\begin{itemize}
    \item First, as a warm-up exercise, we use our machinery to estimate the effect of the non-qubit levels present in every implementation of a qubit quantum simulator. We seek to reproduce the quantum spectrum of a problem native to the hardware graph for this example. Unlike the other problems studied in this work, no interaction mediators are involved.
For a hardware implementation with $n$ qubits and a graph of degree $2s$, let $\delta$ be the usual control errors, $r$ the norm of the term in the Hamiltonian connecting to the third level, and $\omega_p$ is the gap to the non-qubit levels. For the precise definitions, see App. \ref{warmup}. As long as
$ \omega_p \leq 32n(s+2)$,
the best solution we found requires $\delta =O(1/n)$ for any $r\in[0,1]$. The dependence on $\epsilon$ for $ r\geq \frac{16\epsilon s }{7 (2+s)}$ is $r\delta = O(\epsilon^2/n)$. For the complete expressions and the solutions found for other values of $\omega_p, r$ see App. \ref{warmup}.

For the realistic values of parameters, we find that a rigorous reproduction of the quantum spectrum with $\epsilon =0.1$ accuracy requires three digits of control precision $\delta\leq 0.8\cdot 10^{-3}$ for $n=4$ qubits and four digits of control precision $\delta\leq 0.8\cdot 10^{-4}$ for $n=40$. 

\item We also prove a general theorem (see Sec. \ref{general}) applicable for any mediators defined by their Hamiltonians $H_{m,i}$ and their coupling to the qubits $Z_iI_{m,i}$, as well as to other mediators $I_{i,j}$. 
It is also applicable to any control errors as long as $\|P\delta H_{m,i} \| \leq \delta_H ,~ \|P \delta I_{m,i}\| \leq \delta_I$, where $P$ is the projector onto a $2^n$-fold degenerate ground state subspace of $\sum_i H_{m,i} + X_i I_{m,i}$. We sometimes omit the index $i$ when working with an individual mediator. The direct consequences of the theorem are, besides the above result for a transmission line, two simpler results for a qubit mediator and an LC circuit mediator presented in App. \ref{sec:app}:
\item 
The qubit case is the simplest possible case, where each  qubit of our quantum simulator is coupled to a qubit coupler as follows:
 \begin{equation}
    H_m=\sqrt{1-J^2} X_{qc}, \quad  I_m =    J Z_{qc} \ .
 \end{equation}
 The qubit couplers are extended objects that have small mutual inductances where they overlap: 
 \begin{equation}
    V_c= \sum_{i>j}f_{ij}I_{i,j} I_{j,i}, \quad I_{i,j } =Z_{qc,i}\ .
 \end{equation}
 This is inspired by the Chimera and Pegasus architectures of D-Wave \cite{boothby2020next}, with the only difference that here the qubits are only connected to one coupler each, while each coupler is coupled to $2s$ other couplers. We obtain the following relationship between the control precision and the target precision:
 \begin{equation}
    \delta  \leq \text{max}_J\frac{(s\epsilon)^2 0.95 (3 +\sqrt{1-J^2} +sJ^2)^{-1}}{12\cdot7n(1 + \sqrt{1-J^2}^{-1} + s J^{-2} )^2} \ ,
\end{equation}
where $\delta$ is the control precision of all the qubit and qubit coupler parameters.

\item We also consider a harmonic oscillator (LC-circuit) mediator. Define the Hamiltonian of each mediator: 
\begin{equation}
    H_m = a^\dag a , \quad I_m = J(a + a^\dag)\ ,
\end{equation}
and $I_{i,j} =a_i+a^\dag_i$ independent of $j$.
The errors $\delta_H, \delta_I$ defined in the theorem and control errors $\delta$ limiting the terms in $V$ are related to the target precision as follows:
\begin{align}
    \delta_H +\delta_I +  \delta (1 +e^{-2J^2} +s(2J)^2) \leq \\ \leq \frac{0.9936(s\epsilon)^2}{12\cdot 7n(1  +e^{2J^2} +s  (\frac{1}{2J} +1 )^2)^2}\ .
\end{align}
We can vary $J\in [0,1]$ to find the best values of $\delta$'s.
We see that $\delta$'s are still $\sim 1/n$, which means the massive increase in the power of $n$ is due to the distributed nature of the transmission line, not due to the difference between linear (LC) and nonlinear (qubit) elements.

\item Our general theorem favored simplicity of expression as opposed to the optimality of the bound. We also try to push the bound to the limit for a specific example of an $n=40$, degree $2s=4$ random graph implemented via qubits and qubit couplers arranged as in the Pegasus architecture. For $\epsilon=0.1$ we find the required qubit control precision to be $10^{-5}$, which is 3 orders of magnitude away from the experimental values \cite{boothby2021architectural}, and of roughly the same order as what is projected for the future fault-tolerant architectures (though one uses flux qubits and another - transmons, so a direct comparison of control precision is not available). The details of this calculation can be found in Appendix \ref{q40}. 
\item Finally, in Sec. \ref{sec:opt}
we discuss the application of our gadget to quantum annealing. We present the schedules required to operate our all-to-all gadget, concluding that the minimal required adjustment to the current capabilities of the D-Wave \cite{boothby2021architectural} is to allow for a third, constant anneal schedule on some of the terms. We also demonstrate how the minimal gap along the anneal of a commonly used minor embedding \cite{cai2014practical,choi2008minor} method decreases exponentially with the length of the chains $k$ used in the embedding. In contrast, our method sees only a polynomial decrease in $k$. The prefactors are such that our method is advantageous already for $k=4$. We believe this approach will bridge the gap between the D-Wave performance on the native graph problems \cite{Mandr__2018} and the highly-connected application-relevant problems \cite{Kowalsky_2022}.
\end{itemize}

\section{Perturbation theory used}
\label{sec:pert}
Let $H_0\geq 0$ be a Hamiltonian over possibly infinite-dimensional Hilbert space, and choose the energy offset such that its (possibly degenerate) ground state has energy $0$. Let $0$ be an isolated eigenvalue of the spectrum of $H_0$, separated by a gap $\Delta$ from the rest of the spectrum. Denote the projector onto the finite-dimensional ground state subspace as $P$, s.t. $PH_0=0$.

We will formulate a version of degenerate perturbation theory with explicit constants in the bounds on its applicability and accuracy. Allow the perturbation $V$ to have unbounded operator norm ($\|V\|=\infty$ is allowed). We will need another constraint to separate physical $V$'s from unphysical ones. We define a custom norm $\|V\|_c$ for all operators $V$ to be the smallest number s.t.:
\begin{equation}
    -\|V\|_c (1 +H_0) \leq V \leq \|V\|_c (1 +H_0) \label{vBound}
\ .\end{equation}
Here $1$ is the identity operator. Instead of the exact value $\|V\|_c$, we will use its upper bound: some value $v$ s.t. we can prove $v \geq \|V\|_c$. More details on this norm can be found in App. \ref{app:Bravo}.

Define the adjusted gap $\Delta_V =\Delta - v(1+\Delta)$ and the projector $Q=1-P$.
We will use the following perturbation theory result:
\begin{lemma}\label{SimpLem}{\emph{ (properties of SW, simplified)}} For any $H_0 + V$ as above, such that $\Delta_V>0$ and $\|PV\|/\Delta_V  < 1/32$, the following holds.
There exists a rotation $U_{\text{SW}}$  that makes the Hamiltonian block-diagonal
\begin{equation}
    U_{\text{SW}}(H_0 + V)U_{\text{SW}}^\dag = H_{SW} = PH_{SW}P +QH_{SW} Q
\ .\end{equation}

The low-energy block is approximately $PVP$:
\begin{equation}
    \|P(H_{SW} -V)P\| \leq  7 \|PV\|^2/\Delta_V
\ .\end{equation}
 \end{lemma}
While many rotations satisfy the above,  $U_{SW}$ possesses an additional property of being close to an identity (a bound $\|U_{SW} -1\| =O(\|PV\|/\Delta_V)$ is given in App. \ref{app:Bravo}), which means the physical measurements are close to the measurements done in the basis defined by $U_{SW}$. We will interpret $PH_{SW}P$ as the effective Hamiltonian in the subspace corresponding to $P$.
For a special case of a finite-dimensional  Hamiltonian $H_0 + V$, one can use a simpler statement without requiring Eq. (\ref{vBound}): 
\begin{lemma}{\label{lem2}\emph{ (finite-dimensional case, simplified)}} For any $H_0 $ and $ V$, let $P$ be the projector onto the ground state subspace of $H_0$. Let the ground state of $H_0$ be separated by a gap $\Delta$ from the rest of the spectrum, and shift the energy s.t. $PH_0=0$. If $\|V\|/\Delta  < 1/16$, the first order degenerate perturbation theory for states in $P$ has the following error:
\begin{equation}
    \|P(H_{SW} - V)P\| \leq  3.5 \|PVQ\|\|V\|/\Delta \leq 3.5\|V\|^2/\Delta
\ .\end{equation}
 \end{lemma}

The statements of the finite-dimensional Lemma closely follow the results of \cite{B1}. We present a more detailed statement and proof of both in App. \ref{app:Bravo}. Though we formulated the perturbation theory for the case of $PH_0=0$, these lemmas can be straightforwardly generalized to non-degenerate eigenvalues. Following \cite{B1}, it is also possible to extend it to higher orders in $V$ for finite-dimensional systems. We are unaware of a simple way to obtain higher orders in $V$ for infinite-dimensional systems.

\section{Statement of the general theorem}
\label{general}
Consider the Hamiltonian $H_0+V$, where:
\begin{equation}
    H_0  =\sum_i H_{m,i} + Z_i I_{m,i}\ ,
\end{equation}
with the ground state subspace of states $|g_b,b\rangle$ labeled by a string $b$ of $\pm 1$ describing the corresponding qubit computational basis state. The projector onto the ground state subspace is $P = \sum_b P_b P_{g_b}$.
The perturbation is: 
\begin{equation}
    V= \sum_i h_i^c Z_i +t_i^c X_i+\delta H_{m,i} + Z_i \delta I_{m,i}  +\sum_{i>j} f_{ij} I_{i,j} I_{j,i}  
\ .\end{equation}
Here an operator $I_{i,j}$ acts on mediator $i$ and is responsible for interaction with the mediator $j$. In the simple case of a qubit coupler or an LC circuit, $I_{i,j}\sim I_{m,i}$ is independent of $j$. Generally, we assume that for every mediator, the operators $H_{m,i},I_{i,j}, I_{m,i}$ have a symmetry $S_i$ such that $S_iH_{m,i}S_i^\dag =H, ~ SIS^\dag =-I$ for all $I$ in the $i$'th mediator. We will use the gap of $H_0$ denoted as $\Delta$ (each $H_{m,i} \pm I_{m,i}$ has the same gap) and its adjusted version $\Delta_V = \Delta -v(1+\Delta)$ that depends on the chosen $V$.

We define the errors $\delta$ : 
\begin{align}
\forall i:\quad \|P\delta H_{m,i} \| \leq \delta_H ,\quad \|P \delta I_{m,i}\| \leq \delta_I
\ .\end{align}
Note that $\delta_H$ and $\delta_I$ are potentially nontrivial functions of $n$. Determination of the quantity $v$ in $\Delta_V =\Delta -v(1+\Delta)$ will also require $\|\delta H_{m,i}\|_c,~ \|\delta I_{m,i}\|_c$ defined in Eq. (\ref{vBound}) to be finite, but these norms will only appear in the following theorem through $\Delta_V$. The parameters $h_i^c,t_i^c,f_{ij}$ of the perturbation are considered to be implemented imprecisely, with the error $ \delta h = \delta t = \delta f  =\delta $. For simplicity, we assume that their error never increases their magnitude beyond the maximum possible exact value within the context of our construction so that we can use the exact expression for $V$ in the second order of the error bound in App. \ref{proofG}. Moreover, we consider the scenario where the graph is fabricated to match the degree $2s$ graph of the specific problem, and it is possible to have other couplings exactly $0$ with no control error. This is the most optimistic expectation of the hardware since our architecture has every pair of mediators crossing each other, and realistically there would be some cross-talk. We will comment on the behavior in the realistic case at the end of App. \ref{proofG}.

The intermediate functions we use are as follows:
\begin{align} 
 \chi_{i,j}=\langle g_{b_i}| I_{i,j}|g_{b_i}\rangle|_{b_i =1}\ ,\quad
\|PI_{i,j} P\|= |\chi_{i,j}|\ , \\ \|PI_{i,j}\| \leq i_{i,j}  \ , \quad 
      F = \langle g_{b_i=1}|g_{b_i=-1}\rangle
\ .\end{align}
Here $i_{i,j}$ is any upper bound on $\|P I_{i,j}\|$. One such bound can be derived as $i_{i,j} = |\chi_{i,j}| +i_m$:
\begin{equation*}
    \|PI_{i,j}\| \leq|\chi_{i,j}| + \|PI_{i,j}Q\|~, \quad \|PI_{i,j}Q\|\leq i_m\ ,
\end{equation*}
where $i_m$ is any upper bound on $\|PI_{i,j}Q\|$.

\emph{Theorem:} For any $ \epsilon\leq 7/16$ choosing the parameters of the gadget as $h_i^c = \alpha_o h_i, ~ t_i^c =\alpha_o F^{-1} t_i, ~ f_{ij} = \alpha_o J_{ij}^*/\chi_{i,j}\chi_{j,i}$ with the reduction factor $\alpha_o$:
\begin{equation}
    \alpha_o = \frac{s\epsilon \Delta_V}{3\cdot 7n(1+F^{-1} + s  ~ \text{max} ~\frac{i_{i,j}i_{j,i}}{|\chi_{i,j}\chi_{j,i}|})^2} \ ,
\end{equation}
ensures that the error is rigorously bounded as $\|H_{\text{targ}}-H_{\text{eff}}\| \leq \epsilon ns$ (for $H_{\text{eff}}$ in the logical basis defined via SW transformation, and the bound on how close it is to the qubit computational basis can be obtained using the Lemma in App. \ref{app:Bravo}) as long as the following inequalities are satisfied by some choice of $v$:
\begin{align}
    \delta_H +\delta_I +  \delta (2 +s~  \text{max} ~|\chi_{i,j}\chi_{j,i}|) \leq \\ \leq  \frac{\Delta_V(s\epsilon)^2}{12\cdot 7n(1  +F^{-1} +s  ~ \text{max} ~\frac{i_{i,j}i_{j,i}}{|\chi_{i,j}\chi_{j,i}|})^2}\ , \\
    \pm V \leq v(1+H_0)
\ .\end{align}
In practice, we will always be able to prove that the correction to $\Delta$ is subleading, i.e., for the purposes of scaling, one may think of $\Delta_V$ as $\Delta/2$. We prove the theorem in App. \ref{proofG}, and present a version of the theorem with an explicit choice of $v$ in App. \ref{plan}.

\section{Comparison with minor embedding}
\label{sec:opt}
For theory applications, it is sufficient that the control errors scale polynomially with the system size $n$.
For practical applications, the $n^{-6}$ scaling of control errors required for the transmission line construction is unrealistic. We note that this scaling results from building a complete graph of extended mediators. For intermediate $n =40\dots 100$, there are more economical hardware graphs that effectively host a wide range of fixed degree $2s$ random problem graphs. Chimera and Pegasus architectures implemented in D-Wave \cite{boothby2020next} are prime examples of such graphs. Our construction applies to the following practical cases: (i) quantum simulation of a $n =40\dots 100$ system that requires faithful reproduction of the quantum spectrum.  We will derive the bound on the control errors for a specific example of an $n=40$, degree $2s=4$ random graph in the App. \ref{40g} (ii) optimization of a classical $n=100\dots 1000$ problem that D-Wave was originally intended for, which is the focus of this section. In both cases, our construction enables a boost in performance compared to the existing method, colloquially referred to as {\emph{minor embedding}}. For both, we need an embedding: an association between groups of qubits of the hardware graph and individual qubits of the problem graph, such that for interacting problem qubits, there is at least one interaction between the two corresponding groups in the hardware. In the case of minor embedding, hardware qubits within a group are used as a classical repetition code for the corresponding problem qubit, which we discuss in more detail later in this section. In our construction, there is an extra step where one qubit of a group is selected as a problem qubit, while the other qubits within that group are used as the mediator for that problem qubit. This, in principle, allows designing an architecture where the selected problem qubits have better coherence properties than qubits used as mediators, at the cost of less flexibility during the embedding stage. For our estimates here, we will assume that all qubits are the same, as is the case in the current hardware.

Let us first describe how to apply the result of our general theorem in practice. There is one straightforward generalization that our theorem needs: not all qubits will need a mediator, as some can be connected directly to all of their problem graph neighbors. Thus the couplings have three types: qubit-qubit, qubit-mediator, and mediator-mediator. The values of susceptibility $\chi_{i,j}$ can be computed just by considering one connected component of the mediator (there may be noninteracting parts of one mediator), while the value of the overlap $F$ requires considering all connected components of the mediator of the qubit in question. These computations are still feasible classically for the problem sizes we consider since the individual chain length (size of the group associated with one logical qubit) of the embedding stays within the exact diagonalization range. Even beyond that range, a method such as DMRG \cite{schollwock2005density}  can provide the values of $\chi_{i,j}$ and $F$. The hardware couplings for qubit-qubit, qubit-mediator, and mediator-mediator cases are set respectively to:
\begin{equation}
    f_{ij} =\{\alpha J_{ij},\quad  \alpha J_{ij}/\chi_{i,j}, \quad  \alpha J_{ij}/(\chi_{i,j}\chi_{j,i}) \}
\ .\end{equation}
Only the last term previously appeared in our theorem. The other terms, such as the transverse field, are unchanged and still contain the appropriately defined overlap $F$. According to our theorem, such a construction will work for sufficiently small control noise, giving a precision $\epsilon$ as a function of control noise, assuming an appropriate choice of $\alpha$. Knowing the control noise, one can estimate the range of possible $\epsilon$ and the $\alpha$ required for them by our theorem. In practice, it is expected that the inequalities in our theorem are not satisfied for the hardware control noise for any $\epsilon<1$, and we have no guarantees on the gadget's performance. As our bounds are not tight, and $\alpha$ is a free parameter, we argue that choosing it according to the formula with some arbitrary $\epsilon'>1$ may still demonstrate the physical effects of interest for the case of quantum simulation, or boost the success of optimization. To push the gadget to the limits of its performance, we note that the expression for the allowed control errors as the functions of $\epsilon$ depends on the internal parameters of the gadget and can be maximized with respect to them. The optimal values obtained can be used for all $\alpha$, including those outside the guaranteed performance region. We note that this parameter optimization only requires simulating a single mediator, not the whole gadget, which means it can be performed classically. 

For applications to optimization problems via quantum annealing, our method suggests a new schedule for controlling the device parameters. Let us use our method to implement the quantum spectrum of the traditional anneal schedule faithfully:
\begin{equation}
    H(s) = A(s) \sum_i X_i  + B(s) (\sum_i h_i Z_i + \sum_{ij}J_{ij}Z_iZ_j)
\ .\end{equation}
We note that this doesn't mean the effective Hamiltonian of the dynamics is as above since the geometric terms due to rotation of the effective basis need to be included, for which we refer to Sec. VI of our recent work on adiabatic theorem \cite{mozgunov2023quantum} and leave further developments to future work. We, however, have a guarantee on the spectrum at every point, thus on the minimal gap along the anneal. According to our method, the hardware Hamiltonian is $H_0 +V$, where:
\begin{equation}
    H_0  ={\sum_i}' H_{m,i} + JZ_i Z_{m,q(i)}
\ .\end{equation}
Here the sum is over the qubits that have mediators, $q(i)$ is the point of attachment of the qubit to the mediator, and $H_{m,i}$ is some Hamiltonian on the coupler qubits that can in principle be optimized, but for simplicity, we can take $H_{m,i} = J^*\sum_{i,j\in m} Z_{m,i}Z_{m,j} + \sum_{i_\in m} X_{m,i}$, where $J^*$ corresponds to the approximate location of the critical point for this finite-size transverse field Ising model. In particular, if the mediator is a chain or a collection of chains, then $J^* =1$.
The perturbation is:
\begin{equation}
    V= \sum_i \alpha(s) (B(s)h_i Z_i +F_i^{-1}A(s) X_i)+\sum_{i>j} (f ZZ)_{i,j}\ .\end{equation}
Here $f_{i,j}$ is given by
\begin{equation}
    f_{ij} =B(s)\alpha(s)\{ J_{ij},\quad  J_{ij}/\chi_{i,j}, \quad   J_{ij}/(\chi_{i,j}\chi_{j,i}) \}\ ,
\end{equation}
depending on the coupling type. The $(fZZ)_{i,j}$ is a shorthand notation for a weighted sum of the various couplings between qubits $i$ and $j$ or the coupler qubits in their respective mediators. The weights in the sum weakly affect the bound on $\|PV\|$ that is used for our theorem and can thus be optimized. Intuitively, we always prefer to use direct couplings instead of mediators whenever possible. We observe that there are the following separate schedules that are required for $X$ and $ZZ$ terms:
\begin{center}
\begin{tabular}{ |c|c|c| } 
 \hline
 ~ & problem & mediator \\ 
 X & $\alpha(s)F_i^{-1}A(s)$ & 1 \\ 
 ZZ & $\alpha(s)B(s) \{J_{i,j},J_{i,j}/\chi_{i,j} \dots\}$ & $J,J^*$ \\ 
 Z & $\alpha(s)B(s)h_i$ & $0$ \\ 
 \hline
\end{tabular}
\end{center}
We see that the mediator qubit controls must be kept constant while the problem experiences an anneal schedule. The transverse field controls generally have different overlap factors in front of them, but if the hardware constraints them to be the same, the change in the anneal schedule of the effective Hamiltonian is not substantial:
\begin{equation}
    H(s) = A(s) \sum_i F_i X_i  + B(s) (\sum_i h_i Z_i + \sum_{ij}J_{ij}Z_iZ_j)
\ .\end{equation}
That reduces the number of independent schedules to 3: $\alpha(s) A(s), ~ \alpha(s) B(s)~, 1 $. As $\alpha(s)$ is a free parameter in our construction that determines which error $\epsilon$ can we guarantee, we can set $\alpha(s) =$const for simplicity. This highlights that the only missing capability from the current D-Wave devices is holding some of the $X$ and $ZZ$ terms constant throughout the anneal. For some polynomially small $\alpha$ and control errors that satisfy our theorem, we guarantee that our construction preserves the polynomially small features of the spectrum. In particular, a polynomially small minimal gap above the ground state along the anneal is preserved by this construction, albeit polynomially reduced. As we will see below, the traditional minor embedding, in general, makes that gap exponentially small in the size of the mediator.

We note that using our scheme for optimization also has a disadvantage: the final classical effective Hamiltonian at the end of the anneal has its energy scale reduced by a polynomially small factor of $\alpha$. It only has an extensive error $\epsilon$ for a polynomially small control noise. We lose all guarantees on the error past a certain system size for a constant control noise. In contrast, minor embedding retains extensive error of the ground state of the classical Hamiltonian at the end of the anneal, even for a constant control noise. We expect a tradeoff between the errors due to non-adiabatic effects and the errors of the implementation of the effective Hamiltonian to result in an optimal schedule that uses some combination of the two schemes.


In minor embedding, a repetition code is used for each qubit, and the field $X$ is applied with the same schedule $A(s)$ everywhere. The repetition code is enforced by $B(s)ZZ$ terms (the largest allowed scale in the problem), while the problem interactions and longitudinal fields are all reduced as $B(s)J_{i,j}/M$ and $B(s)h_{i}/M$, where $M$ is a free parameter. The longitudinal fields and, when possible, the problem interactions are distributed between the hardware qubits representing one problem qubit. We note that minor embedding does not adjust the coupling depending on the location; thus, there are no factors of $\chi$ in the hardware Hamiltonian, in contrast with our construction. For a special case where the factors of $\chi$ are always the same in our construction, minor embedding becomes a special case of our construction at each $s$, with an $s$-dependent factor $M$. Our construction corresponds to $M\leq 1$ since the hardware qubits for each individual problem qubit are in a paramagnetic state. We believe that when extended to $k$-qubit chains, the advantage of our paramagnetic gadget vs. the ferromagnetic repetition code is exponential in $k$. For instance, the minimal gap of the logical problem will experience only polynomial in $k$ reduction for our method, while the reduction will be exponential in $k$ for minor embedding. 

\begin{figure}
\centering
\includegraphics[width=\columnwidth]{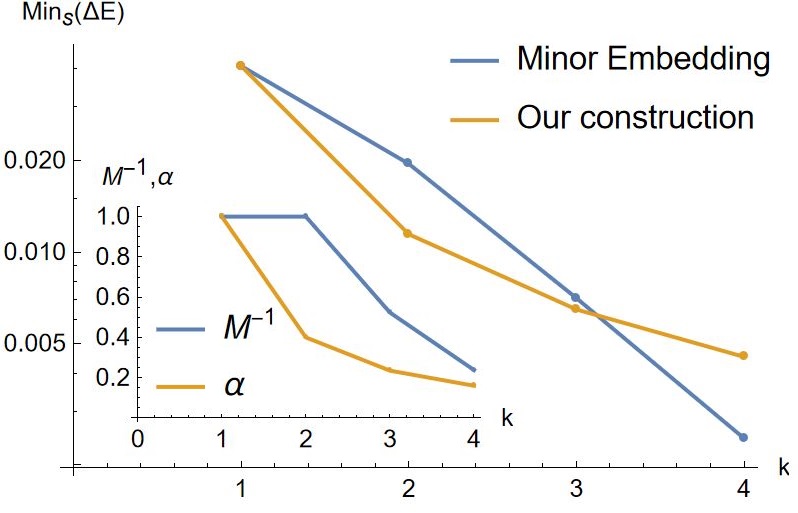}
\caption{Minimal gap along the anneal for a 3-qubit problem embedded in a ring of $3k$ hardware qubits, with chains of length $k$ for minor embedding and mediators of length $k-1$ for our construction. We see that the minimal gap of minor embedding decreases exponentially with $k$, and our construction is advantageous for $k>3$. Inset: the overall problem energy scale also decreases for both constructions. Here we plot the optimal values of the problem energy scale used for the minimal gap plotted in the main plot.}
\label{Mig}
\end{figure}

While a naive extension of our perturbative results into the non-perturbative regime can be done by just increasing $\alpha$ as described above, it is essential to push the gadgets to the limit of their performance. We investigate this for $n=3,~k=1,2,3,4$ when both the gadget and the system are only allowed to have terms limited in magnitude ($|h|,|t|, |J|\leq 1$), and the geometry is fixed as a ring of $3k$ hardware qubits: which schedule on the gadget and the system leads to the best minimal gap? We use the minor embedding schedules to compare our results. For the method outlined above, an improvement over minor embedding is seen in Fig. \ref{Mig} for $k>3$. The code producing these results can be found in \cite{urlCode}. We note that interpolation between the two methods will likely produce even better improvement. For this example, we only optimized $\alpha$ and kept $J,J^*=1$. A full optimization will also likely improve these results. Here the optimization involved full system simulation, but we believe the mediator optimized for a collection of small examples like this will still perform well when used as a building block in a large $n$ system. Such a generalization must, however, be wary that a high enough system scale $\alpha$ (or $M^{-1}$ for minor embedding) can change the ground state at the end of the anneal. In our example, the ground state was preserved well above the optimal values of $\alpha$ and $M^{-1}$.

\section{conclusions}
We have proven that a physical system can be an accurate quantum simulator. Specifically, we first made sure that the proposed architecture is realistic: it is a 2d layout with a fixed density of elements, and the elements we use are the standard building blocks of superconducting circuits today. We then presented rigorous proof that an all-to-all system is accurately simulated for all system sizes $n$. The geometry of its interaction graph can be infinitely more complicated than 2d or 3d space, yet the low energy physics of our quantum simulator on a chip will reproduce it accurately. While the scaling of the required control errors $n^{-6}$ is very costly, and there are likely practical limits to a control precision of a physical system, there are no immediate fundamental limits on it. Future theory work may rely on our construction whenever a low-energy model with complicated geometry is needed to exist in a 3d world.

We studied our gadgets and perturbation theory in the context of superconducting qubits. However, the theorem we prove is more general: any type of qubit used in quantum simulators can be connected to a faraway qubit perturbatively using mediators, and our theorem will describe the highly connected limit of that system.
In the current D-Wave architecture, relatively short chains can already embed large all-to-all graphs that are intractable classically. Other types of hardware for quantum simulation may be even more efficient than D-Wave for this task. Coupling via the transmission line has yet to be scaled to a large number of qubits, but we already have a promising demonstration of using qubits as couplers. We propose a minimal schedule adjustment needed for that: some of the terms are to be kept constant during the anneal. Our method is expected to close the performance gap between native and application problems for quantum optimization, opening the way for quantum advantage on the latter. Another fruitful direction is to benchmark a variant of the Chimera and Pegasus graphs where the distinction between qubits and qubit couplers is fixed at fabrication and to propose better graphs with more economical embeddings in this setting.  

A surprising result of this work is that there is no apparent difference in performance between linear (bosons with a quadratic Hamiltonian) and nonlinear mediators (qubit couplers). Investigating it further is a promising direction for future work, along with improving the scaling and the value of the required control precision. The next step in developing all-to-all gadgets is to investigate qubit chain mediators, which are most likely the simplest to implement experimentally. It is an important future theoretical milestone to obtain a specification on circuit parameters required for qubit couplers.

This material is based upon work supported by the Defense Advanced Research Projects Agency (DARPA) under Agreement No. HR00112190071. Approved for public release; distribution is unlimited.

 \bibliography{refs}
\bibliographystyle{apsrev4-2}
\onecolumngrid
\appendix

\section{Perturbation theory lemma following Bravyi et al.}
\label{app:Bravo}
Consider a possibly infinite-dimensional Hilbert space, and a Hamiltonian $H_0$ where $H_0 \geq 0$, and we set the ground state energy to $0$ and denote the ground state subspace projector as $P$, s.t. $PH_0=0$ (though our proof can be straightforwardly generalized for projectors onto subspaces corresponding to a collection of eigenvalues). Let $P$ have finite rank, and let the spectral gap $\Delta$ separate the ground state from the rest of the spectrum. 

We introduce the perturbation $V$ that can be unbounded s.t. a usual norm $\|V\| = \infty$, meaning that for a normalized $|\phi\rangle$ the expectation value $\langle \phi| V|\phi\rangle$ can be arbitrarily large. In mathematical literature, it's a convention to assume that $V$ is bounded in a different sense: choosing a reference Hamiltonian to be $H_0$, we require
\begin{equation}
    \|V\|_{H_0 } = \text{sup}_{|\phi\rangle}
    \frac{\|V\phi\|}{\sqrt{\|\phi\|^2 +\|H_0\phi\|^2}}= \text{sup}_{|\phi\rangle} \frac{\sqrt{\langle \phi| V^2|\phi\rangle }}{\sqrt{\langle \phi| \phi\rangle +\langle \phi| H_0^2|\phi\rangle}}  \label{mathNorm}
\end{equation}
to be bounded. Most physical perturbations of infinite-dimensional Hilbert spaces obey this, one notable exception being a shift in a position of a hydrogen atom potential. It will be convenient to use a slightly different "custom" norm $\|V\|_c$ as our starting point:
\begin{equation}
     \|V\|_c=\left\|\sqrt{|V|}\right\|_{\sqrt{H_0}}^2 = \text{sup}_{|\phi\rangle} \frac{|\langle \phi| V|\phi\rangle|}{\langle \phi| \phi\rangle +\langle \phi| H_0|\phi\rangle} 
\ .\end{equation}
This means that for any  $\phi$:
\begin{equation}
   - \|V\|_c (\langle \phi| \phi\rangle + \langle \phi| H_0|\phi\rangle)\leq  \langle \phi| V|\phi\rangle \leq \|V\|_c (\langle \phi| \phi\rangle + \langle \phi| H_0|\phi\rangle) \label{productForm}
\ .\end{equation}
In a shorthand notation of matrix inequalities where $A \leq B \Leftrightarrow  B-A$ has a nonnegative spectrum, we get:
\begin{equation}
    -\|V\|_c (I +H_0) \leq V \leq \|V\|_c (I +H_0)
\ .\end{equation}
Showing this relation for some $v$:
\begin{equation}
    -v (I +H_0) \leq V \leq v (I +H_0)\ ,
\end{equation}
establishes an upper bound $\|V\|_c \leq v$. We will use $v$ from now on, and present a sketch of its computation for specific cases of the LC circuit and the transmission line appearing in our paper. For a special case of the bounded $\|V\|$, $v =\|V\|$ can be used in the following. 

We will now prove a result about a shift in an isolated eigenvalue due to such $V$ (finite degeneracy can be resolved by adding a small perturbation and then sending it to zero):
\begin{lemma} {\emph{ (Eigenvalue shift under perturbation)}}
Assuming $v< 1$, an isolated eigenvalue $\lambda$ of $H_0$ shifts to $\lambda(1)$ as $\epsilon$ goes from $0$ to $1$ in $H_0 +\epsilon V$, and
\begin{equation}
   | \lambda(1) -\lambda| \leq v(1+\lambda) 
\ .\end{equation}
\end{lemma}
{\bf{Proof.}} An eigenvalue satisfies $\lambda(\epsilon) = \langle \psi(\epsilon)| H +\epsilon V |\psi(\epsilon)\rangle$, where $|\psi(\epsilon)\rangle$ is the corresponding normalized eigensate: $\langle \psi(\epsilon)|\psi(\epsilon)\rangle = 1$. Taking a derivative of the latter, we obtain $\langle \psi'|\psi\rangle +\langle \psi|\psi'\rangle   =0$. Taking a derivative of the former, we obtain a differential equation on $\lambda$:
\begin{equation}
    \lambda'(\epsilon) = \lambda(\epsilon)(\langle \psi'|\psi\rangle +\langle \psi|\psi'\rangle ) + \langle \psi| V|\psi\rangle   = \langle\psi(\epsilon)| V|\psi(\epsilon)\rangle \ .
\end{equation}
$\lambda_{\uparrow}(\epsilon)$ and $\lambda_{\downarrow}(\epsilon)$ upper and lower bound $\lambda(\epsilon)$ if $\lambda_{\uparrow}(0)=\lambda_{\downarrow}(0) =\lambda(0)$ and their derivatives satisfy $\lambda_{\uparrow}'\geq \lambda',\quad \lambda_{\downarrow}' \leq\lambda'$ for all $\epsilon$. Using Eq. (\ref{productForm}) with $v$, we know that:
\begin{equation}
    -v(1 + \langle \psi(\epsilon)| H_0| \psi(\epsilon)\rangle ) \leq \langle \psi(\epsilon)| V| \psi(\epsilon)\rangle  \leq v(1 + \langle \psi(\epsilon)| H_0| \psi(\epsilon)\rangle )
\ .\end{equation}
Adding or subtracting $v\epsilon \langle \psi(\epsilon)| V| \psi(\epsilon)\rangle$, we get:
\begin{equation}
     -v(1 + \lambda(\epsilon) ) \leq (1-v\epsilon) \lambda', \quad (1+v\epsilon) \lambda'  \leq v(1 + \lambda(\epsilon) )\ .
\end{equation}
We can show that solving
\begin{equation}
      (1-v\epsilon) \lambda_{\downarrow}'=-v(1 + \lambda_{\downarrow}(\epsilon) ) , \quad (1+v\epsilon) \lambda_{\uparrow}' = v(1 + \lambda_{\uparrow}(\epsilon) )
\end{equation}
satisfies the properties of $\lambda_{\uparrow, \downarrow}$ and ensures a bound $ \lambda_{\downarrow}(1)\leq \lambda(1) \leq\lambda_{\uparrow}(1) $. Indeed if we assume that there are points where it is not satisfied, then for inf of those points we reach a contradiction with the bounds above. The solutions are:
\begin{equation}
   \frac{1 + \lambda_{\downarrow}(\epsilon) }{1 + \lambda(0)} = 1-v\epsilon, \quad  \frac{1 + \lambda_{\uparrow}(\epsilon) }{1 + \lambda(0)} = 1+v\epsilon \ .
\end{equation}
For $\epsilon =1$ the first solution exists as long as $v<1$, and we get:
\begin{equation}
\lambda_{\downarrow} -\lambda(0) = (1-v)(1+\lambda(0)) - 1 -\lambda(0), \quad   \lambda_{\uparrow} -\lambda(0) = (1+v)(1+\lambda(0)) - 1 -\lambda(0)
\ .\end{equation}
Plugging that into the bound gives the magnitude of the eigenvalue shift in the lemma.
$\square$

Define $Q =1-P$. The operator $H_0$ is block-diagonal: $H_0 = QH_0Q$, while $V$ can have nontrivial matrix elements in all blocks. An operator $V_P = V -QVQ$. satisfies $\|V_P\| \leq \|PVP\| + \|PVQ\|\leq 2\|PV\|$, and we require these quantities to be finite (they are if $\|V\|_{H_0}$ in Eq. (\ref{mathNorm}) is finite).  We are ready to prove the rigorous version of Lemma \ref{SimpLem} from the main text:

\begin{lemma}{\emph{ (properties of SW)}} For any $H_0 + V$ as above, such that $\|V_P\|/(\Delta - v(1+\Delta)) =x < 1/16$, the following holds.
There exists a rotation $U_{\text{SW}}$  that makes the Hamiltonian block-diagonal
\begin{equation}
    U_{\text{SW}}(H_0 + V)U_{\text{SW}}^\dag = H_{SW} = PH_{SW}P +QH_{SW} Q
\ .\end{equation}

The low-energy block is approximately $PVP$:
\begin{equation}
    \|P(H_{SW}  -V)P\| \leq  c(x) \|PVQ\|\|V_P\|/(\Delta - v(1+\Delta))
\ .\end{equation}
  Here $ 2\leq c(x) =  \frac{1}{x}\text{tan}\frac{-1}{4}\text{ln}(1 -\frac{8 x}{1- 2x})<16 \text{tan}(\frac{1}{4} \text{ln}(7/3))\leq 3.441$ is a known universal function of $x$.
 The rotation itself is formally given by
\begin{equation}
     U_{\text{SW}} = \sqrt{(2P-1)(2P_V-1)} \ ,
\end{equation}
where $P_V$ is the exact projection onto the low-energy eigenvalues of the perturbed system $H_0 +V$, s.t. $P_V = U_{SW}^\dag P U_{SW}$. The rotation $U_{SW}$ can be bounded as follows:
\begin{equation}
    \| U_{SW}- 1 \| \leq c_S (x) \|V_P\| /(\Delta - v(1+\Delta))\ ,
\end{equation}
where $4\leq c_S(x)=\frac{1}{x}((1 -\frac{8 x}{1- 2x}  )^{-\frac{1}{2}} -1) < 16(\sqrt{7/3}-1) < 8.441 $

\end{lemma}

In the text of the paper, we use the constant upper bound $c=3.5$, but we keep the functional form for a tighter bound that can in principle give better gadget guarantees. In the main text, we only use a simplified bound $\|PVQ\| \|V_P\| \leq 2\|PV\|^2$. Note that while both $\|PV\|$ and $\|V_P\|$ are $ \leq \|PVP\| + \|PVQ\|$ (for $\|V_P\|$ it can be seen by singular value decompositon of $PVQ$), it may be that $\|V_P\| > \|PV\|$. For a finite-dimensional Hilbert space,
a perturbative series for log $U_{\text{SW}}$ can be found in \cite{B1} along with further terms of a series expansion for $H_{\text{SW}}$. Our proof closely follows that of \cite{B1}, and at the end of this Section, we present a comparison with their notation.

\subsection{Proof}

In the perturbation theory that follows, we split the perturbation $V$ into $V_P =V- QVQ$ and $QVQ$, and include the latter into the new bare Hamiltonian $H_{0,P} = H_0 + QVQ$. We now define a reduced gap $\Delta_V =\Delta -v(1+\Delta)$ that accounts for possible eigenvalue shift of the first excited state (not counting the degenerate ground state), or a third state coming down from the spectrum in $H_{0,P}$. Note that $\|QVQ\|_x=$sup $\langle \phi| Q VQ| \phi \rangle/ (\langle \phi|  \phi \rangle+ \langle \phi| H| \phi \rangle)$. Let $\phi_Q$  be the $|\phi\rangle$ achieving the supremum. It is one of the allowed $|\phi\rangle$ in the expression $\|V\|_x = $sup $\langle \phi| V| \phi \rangle/ (\langle \phi|  \phi \rangle/+ \langle \phi| H| \phi \rangle)$, which means $\|V\|_x \geq \|QVQ\|_x$ and we can use $v \geq \|V\|_{x}$ for $\Delta_V$.

Consider a contour $\gamma$ in the complex plain around $0$ (the ground state of $H_0$, which is still the ground state of $H_{0,P}$), and recall that $P$ is the projector onto the ground state subspace of $H_0$ and $H_{0,P}$ associated with the eigenvalue $0$. Let the contour pass right in the middle of the gap $\Delta_V$, s.t. the contour's length $|\gamma| = \pi \Delta_V$. A  standard perturbative expansion of the resolvent expression  $P_V = \frac{1}{2\pi i} \oint dz (zI -H_{0,P} -V_P)^{-1}$ (assuming the eigenvalues don't shift by more than $\Delta_V/2$) for the projector $P_V$ onto the low-energy subspace of $H_{0,P} +V_P = H_0 +V$ is:
\begin{equation}
 P_V =  P + \sum_{j=1}^\infty P_j, \quad P_j =  \frac{1}{2\pi i} \oint dz (zI -H_{0,P} )^{-1} (V_P(zI -H_{0,P} )^{-1} )^j
\ .\end{equation}
We can bound 
\begin{equation}
    \|P (1-P_V)\| = \|P - P P_V\| \leq \sum_{j=1}^\infty \|P_j\| \leq 
    \frac{|\gamma|}{\pi\Delta_V}\sum_{j=1}^\infty \left(\frac{\| V_P\|}{\Delta_V /2}\right)^j =  \frac{\frac{2 \|V_P\|}{\Delta_V}}{1-\frac{\|V_P\|}{\Delta_V /2} } \label{Zstart}
\ .\end{equation}
 The term $\| QP_V\| =\|\sum_{j=1}^\infty QP_j \| $ is bounded in the same way. Define the finite-dimensional subspace of the Hilbert space that contains vectors corresponding to $P_V$ and $P$, and the corresponding projectors $\overline{P},~ \overline{Q} =1-\overline{P}$. 
By Definition 2.2 of \cite{B1} and the following arguments applied to this finite-dimensional subspace, we know that as long as $\|P-P_0\| <1$ (which we can now check using the above), we can
define
\begin{equation}
    Z = -2(P(1-P_V) +QP_V), \quad U_{SW }  =\sqrt{I + Z} \label{Zdef}\ ,
\end{equation}
and
{\bf{Fact 1}}(Lemma 2.2 of \cite{B1}) holds:
\begin{equation}
   U_{SW}P_V U_{SW}^\dag = P
\ .\end{equation}
In the full space, we add $1$'s s.t. $U_{SW} = \overline{P} U_{SW} + \overline{Q}$ only acts nontrivially within the first term. The total Hamiltonian $H= H_{0,P} +V_P$  conjugated by $U_{SW}$ contains the following blocks w.r.t. $\overline{P}$:
\begin{equation}
    U_{SW} H U_{SW}^\dag = U_{SW}\overline{P}  H \overline{P}U_{SW}^\dag  + R,\quad R = \overline{P} U_{SW} H  \overline{Q} + \overline{Q} H U_{SW}^\dag \overline{P} + \overline{Q}H\overline{Q}
\ .\end{equation}
Using Fact 1 and $[P,\overline{P}] =0$, we can show that $PRQ =QRP =0 $, so we only need to investigate the effect of $U_{SW}$ on $H_{\square} =\overline{P}H \overline{P}$. That finite dimensional Hamiltonian is a sum of $H_{0,\square} =\overline{P}H_{0,P} \overline{P}$ and the perturbation $\overline{P}V_{P} \overline{P}$. By commutation, we confirm the block-diagonal structure of both:
\begin{equation}
    H_{0,\square} = PH_{0,\square}P + QH_{0,\square}Q, \quad H_{\square} = P_VH_{\square}P_V + (1-P_V) H_{\square}(1-P_V)
\ .\end{equation}
We can now apply the arguments of \cite{B1} directly to this finite-dimensional Hamiltonian. Let $U_{SW} =e^S$, where $S$ is augmented by 0's outside the subspace of $\overline{P}$.
Define a notation for a commutator superoperator: $\hat{Y}(X) = [Y,X]$. Functions of $\hat{Y}$ are defined via Taylor series. In particular, $e^S He^{-S} = e^{\hat{S}}(H)$. Splitting $V$ into its diagonal and off-diagonal parts with respect to $P_0$, we can use  Eq. 3.5 of \cite{B1}  to arrive at {\bf{Fact 2}}:
\begin{align}
V_d = PVP  \ ,\quad V_{od} =PVQ + QVP \ ,\quad 
U_{SW}H_{\square}U_{SW}^\dag =H_{0,\square}+ V_d +\left(\text{tanh}\hat{S}/2\right) (  \overline{P}V_{od} \overline{P})
\ .\end{align}
This shows that we have successfully block-diagonalized $H_\square$, and $H$ itself.
Multiplying everything by $P$, we obtain:
\begin{equation}
    P(H_{SW} - H_0 - V) P = P\left(\left(\text{tanh}\hat{S}/2\right) (  V_{od} ) \right)P
\ .\end{equation}
Taking the norm of that establishes the bound we're proving. We will first bound $Z$ and $S$.

Denote $x = \|V_P\|/\Delta_V$. Using Eq. (\ref{Zstart}) and the definition of $Z$ in Eq. (\ref{Zdef}), we get:
\begin{equation}
   \|Z\| \leq  \frac{8 x}{1- 2x} \quad \text{for} \quad x=\frac{\|V_P\|}{\Delta_V} \label{sixtyfour}
\ .\end{equation}
For the following functions of $Z$ to be well-defined, we need $\|Z\| <1$, which translates into $x< 1/16$.

First we compute
\begin{equation}
    S = \frac12 \text{ln}(I + Z) = -\sum_{k=1} c_k (-Z)^k\ ,
\end{equation}
where  $c_k = \frac{1}{2k}> 0$ are half of the Taylor series for the logarithm $-$ln$(1-x)=2\sum_k c_k  x^k$. From this we find the norm bound:
\begin{equation*}
    \|S\| \leq \sum_{k=1} c_k \|Z\|^k = -\frac12 \text{ln}(1-\|Z\|) \leq  -\frac12\text{ln}(1- \frac{8 x}{1- 2x})
\ .\end{equation*}

We can also bound $\|e^S-1\|$ using the Taylor series for $e^x-1$ that has nonnegative coefficients $e_n$ :
\begin{equation}
    \|e^S-1\| \leq \sum_n |e_n|\|S\| \leq  \text{exp}(-\frac{1}{2}\text{ln}(1 -\frac{8 x}{1- 2x}  )) -1 = c_S(x) \|V_P\|/\Delta_V \ ,
\end{equation}
where $c_S(x) =\frac{1}{x}((1 -\frac{8 x}{1- 2x}  )^{-\frac{1}{2}} -1)$, and in the range of $x$ we use it is $4\leq c_S(x) \leq 16(\sqrt{7/3}-1) \leq 8.441 $.

For $\|\text{tanh} \frac{\hat{S}}{2} V_{od}\|$, we use the fact that tanh has a Taylor series tanh$(z) = \sum_n t_n z^n$ that after taking absolute values becomes tan$(z) = \sum_n |t_n| z^n$. Using these two facts we can establish convergence with explicit constants.  
\begin{align}
    \|\text{tanh} \frac{\hat{S}}{2} V_{od}\| \leq \| V_{od}\| \sum_n |t_n| \|S\| ^n \leq \| V_{od}\|\text{tan}\frac{-1}{4}\text{ln}(1 -\frac{8 x}{1- 2x}) \leq c(x) \| V_{od}\| \|V_P\|/\Delta_V\\c(x) =\frac{1}{x}\text{tan}\frac{-1}{4}\text{ln}(1 -\frac{8 x}{1- 2x}), \quad 2 < c(x) < c(1/16) =16 \text{tan}(\frac{1}{4} \text{ln}(7/3))\leq 3.441\ ,
\end{align}
which together with the observation that $\|V_{od}\| = \|PVQ\|$ completes the proof.

\subsection{Finite-dimensional case}
Repeating the proof above for a finite-dimensional system, where we do not split $V$ into $V_P$ and $QVQ$, and the bare Hamiltonian and the gap $\Delta$ is unchanged, we get:

\begin{lemma}{\emph{ (finite-dimensional case)}} For any finite-dimensional $H_0 $ and $ V$, let $P$ be the projector onto the ground state subspace of $H_0$. Let the ground state of $H_0$ be separated by a gap $\Delta$ from the rest of the spectrum, and shift the energy s.t. $PH_0=0$. If $\|V\|/\Delta =x < 1/16$, the following holds.
There exists a rotation $U_{\text{SW}}$  that makes the Hamiltonian block-diagonal
\begin{equation}
    U_{\text{SW}}(H_0 + V)U_{\text{SW}}^\dag = H_{SW} = PH_{SW}P +QH_{SW} Q
\ .\end{equation}

The low-energy block is approximately $PVP$:
\begin{equation}
    \|P(H_{SW}  -V)P\| \leq  c(x) \|PVQ\|\|V\|/\Delta 
\ .\end{equation}
  Here $ 2\leq c(x) =  \frac{1}{x}\text{tan}\frac{-1}{4}\text{ln}(1 -\frac{8 x}{1- 2x})<16 \text{tan}(\frac{1}{4} \text{ln}(7/3))\leq 3.441$ is a known universal function of $x$.
 The rotation itself is formally given by
\begin{equation}
     U_{\text{SW}} = \sqrt{(2P-1)(2P_V-1)} \ ,
\end{equation}
where $P_V$ is the exact projection onto the low-energy eigenvalues of the perturbed system $H_0 +V$, s.t. $P_V = U_{SW}^\dag P U_{SW}$. The rotation $U_{SW}$ can be bounded as follows:
\begin{equation}
    \| U_{SW}- 1 \| \leq c_S (x) \|V\| /\Delta\ ,
\end{equation}
where $4\leq c_S(x)=\frac{1}{x}((1 -\frac{8 x}{1- 2x}  )^{-\frac{1}{2}} -1) < 16(\sqrt{7/3}-1) < 8.441 $

\end{lemma}

\subsection{Notation comparison with Bravyi et al.}

The work \cite{B1} considers a Hamiltonian $H_0$ over a finite-dimensional Hilbert space $\mathcal{H}$, that has a gap $\Delta$ separating its eigenvalues of eigenvectors from the subspace $P_0\mathcal{H}$ from others. (p. 14.)

The full Hamiltonian is $H = H_0 + \epsilon V$, and the subspace of interest $P\mathcal{H}$ is still separated by a nonzero gap from other eigenvalues and has the same dimension as $P_0\mathcal{H}$. The rotation $U$ between the two is well-defined if the two subspaces have nonzero overlap. In what follows on p. 16 onwards a perturbative expansion for an operator $S$ is constructed that allows one to compute $U = e^S$. The details of this construction are not formulated as separate lemmas, so we will quote specific equations. Earlier lemma 2.3 by \cite{B1} establishes that the operator $S$ is block off-diagonal, and the transformed Hamiltonian $e^S He^{-S}$ is block-diagonal with respect to $P_0$.

Splitting $V$ into its diagonal and off-diagonal parts with respect to $P_0$, Bravyi et al. arrive at the following expression (Eq. 3.5 of \cite{B1}):
\begin{align}
V_d = P_0VP_0 +Q_0 VQ_0 \ ,\quad V_{od} =P_0VQ_0 + Q_0VP_0 \ ,\quad 
e^{\hat{S}}(H) =H_0+\epsilon V_d +\left(\text{tanh}\hat{S}/2\right) (\epsilon V_{od})
\ .\end{align}

Here it was assumed that the perturbation series converges. Bravyi also investigates when this convergence happens in Lemma 3.4:

The series for $P_0 e^{\hat{S}}(H)$ and $S$ converge absolutely for \begin{equation}
    |\epsilon| <\frac{\Delta}{16\|V\| ( 1 + \frac{2|I_0|}{\pi\Delta})}\ ,
\end{equation}
where $|I_0| = \lambda_{P, \text{max}} - \lambda_{P, \text{min}}$, the energy difference between the largest and the smallest eigenvalue of $H_0$ in $P_0 \mathcal{H}$. In our case, it is zero.

\subsection{From $\epsilon$ to $\|V\|/\Delta$}
The construction by Bravyi et al presents three operator series for $P_0 e^{\hat{S}}(H)$ , $S = \frac{1}{2} \text{ln} (I+Z)$ and $Z$ in terms of parameter $\epsilon$, and the only dependence on $\|V\|$ and $\Delta$ is in the radius of convergence of those series. To illustrate how this approach is complementary to ours, we will explicitly show the following dependence of the series for $Z$ on the small parameter $\|V\|/\Delta$ for $\epsilon =1$ and $|I_0| =0$ as follows:

\begin{align}
    Z =  O(\|V\|/\Delta)\ , \label{sBigO}
\end{align}
Strictly speaking, it does not follow from Lemma 3.4 as stated in \cite{B1}. It only proves that $Z =e^{2S}-1$ converges absolutely in the defined disk of $\epsilon$. A series $Z = \sum_{k=1}^\infty \epsilon^k Z_k$ converging absolutely means that $\sum_{k=1}^\infty |\epsilon|^k \|Z_k\|$ is bounded by some constant $C(\epsilon_x,\|V\|/\Delta)$ for $\epsilon \leq \epsilon_x <\frac{\Delta}{16\|V\| ( 1 + \frac{2|I_0|}{\pi\Delta})}$. That extra dependence on $\|V\|/\Delta$ can translate to arbitrary other terms in the big-O notation of Eq. (\ref{sBigO}).  However, in the proof of that lemma, the construction is strong enough to prove the statement that we are making. Bravyi et al show that for a perturbative series $Z = \sum_{k=1}^{\infty}\epsilon^k Z_k$, the following holds:
\begin{equation}
   \sum_{k=1}^{\infty}|\epsilon|^k \|Z_k\| < 1 \quad \text{for} \quad \epsilon  <\frac{\Delta}{16\|V\| ( 1 + \frac{2|I_0|}{\pi\Delta})} 
\ .\end{equation}
The authors then claim that if this is true, then absolute convergence holds for $S =\frac{1}{2} \text{ln} (I+Z)$ and $P_0 e^{\hat{S}}(H)$ as well in the same open disk. Indeed, the cut of the logarithm starts for $-1$ eigenvalue of $Z$, and the inequality $\|Z\| \leq  \sum_{k=1}^{\infty}|\epsilon|^k \|Z_k\| < 1$ just barely keeps the function $S$ within its analytic regime. $C(\epsilon_x,\|V\|/\Delta)$ defined as above would diverge for $\epsilon_x$ approaching the radius of convergence. We can use a slightly smaller radius in the inequality for $\epsilon$ and plug it into the last line of equations before the end of the proof of Lemma 3.4 in \cite{B1}, obtaining:
\begin{equation}
      \sum_{k=1}^{\infty}|\epsilon|^k \|Z_k\| \leq 4x_0/(8-x_0)   ~ ~ ~\text{for} ~ ~ ~  \epsilon \leq  x_0 \frac{\Delta}{16\|V\| ( 1 + \frac{2|I_0|}{\pi\Delta})} 
\ .\end{equation}
Unlike \cite{B1}, we only concern ourselves with $|I_0|=0$ and $\epsilon =1$ case. Moreover, translating the assumption of our lemma into this notation we get $\|V\|/\Delta =x < 1/16$, which means $\epsilon \leq x_0/16x$. To have $\epsilon =1$ included, it's enough to take $x_0= 16x$. We get the bound \begin{equation}
   \|Z\|\leq \sum_{k=1}^{\infty}|\epsilon|^k \|Z_k\| \leq \frac{8x}{1-2x} 
\ .\end{equation}
This coincides with Eq. (\ref{sixtyfour}) in our proof. We did not use the absolute convergence for the first-order error bound, it is only needed for a good bound on higher orders of the perturbative expansion. Formally the higher orders in our setting coincide with the expressions 3.11, 3.23 in \cite{B1}, but they are not practically useful since we used $H_{0,P} = H_0 + QVQ$ as the bare Hamiltonian and its excited states in the subspace corresponding to $Q$ are generally unknown. 

Finally, we note that \cite{B1}  also proves a result for systems on infinite lattices. Though also infinite dimensional, they require quite different formalism from our approach. The resulting perturbative expansion only provides information about the ground state, not the entire spectrum of the effective Hamiltonian.

\section{On desired precision}
\label{PCP}
\subsection{Tasks that require full spectrum simulation}

Approximating ground states may be easier than the entire spectrum: as \cite{B1} shows, the g.s. energy of a lattice system of size $n$ is given by a perturbative expansion with finite local precision, requiring only a constant gap even though the spectrum is $O(n)$ wide. We, however, require precision for approximating the whole spectrum.
There might be a result that makes our bounds less stringent under the assumption that instead of the full $O(n)$ width of the spectrum of $H$, only $\sim T$ fraction of it (still $O(n)$) needs to be accurate. Varying the effective temperature (energy density) $T$ will be an extra handle on the precision required. Unfortunately, we don't know of a readily available perturbative method that will allow such flexibility, so we always require the {\emph{full}} spectrum of $H$ to be faithfully reproduced.

One of the possible applications for a faithful simulator of the full spectrum is the task of quantum simulation, where the experiment we perform may be approximated by $\mathcal{E}_m (\mathcal{E}_t(\mathcal{E}_p(|b\rangle \langle b|))) $, where $b$ is any bitstring, $\mathcal{E}_{m,p}$ are quantum channels corresponding to ramps required before measurement and after state preparation, respectively. The measurement is in the computational basis. The $\mathcal{E}_t$ is the evolution for a time $t$ with the fixed Hamiltonian, such as the effective Hamiltonian computed in this paper.  Though the real dynamics is dissipative, there's a range of time $t$. ($\sim[0,50ns]$ for D-Wave \cite{king}) where the unitary approximation $\mathcal{E}_t = U_t$ is valid. Currently, this range is likely incompatible with the specific reverse annealing protocol we suggest and is not publicly available, as the minimum time interval for the schedule is $0.5\mu s$. Ideally, the preparation and measurement ramps between the problem Hamiltonian and the $s=1$ (zero transverse field) one are instantaneous, so the maps $\mathcal{E}_{m,p}$ are identity. In reality, they are far from identity, taking $0.5 \mu s$. However, they are still nontrivial - different bitstrings result in different states at the start of the unitary evolution, and there are differences in the bitstring distributions obtained from varying $b,t$. One important property of any nonadiabatic process like this is the constant energy density. That means that generically there's $O(n)$ portion of the spectrum of $H$ that needs to be correct so that the device still reproduces our theoretical expectation $\mathcal{E}_m (\mathcal{E}_t(\mathcal{E}_p(|b\rangle \langle b|)))$ for some tractable noise model $\mathcal{E}$. Only then it deserves to be called a quantum simulator. One may attempt to relax this definition and instead call a quantum simulator any device with an output distribution close to the ideal $U_t(|b\rangle \langle b|)$, but for most current quantum simulators the output distribution is maximally far away from the ideal one in any reasonable metric, and only retains some of the qualitative features of the physical phenomena being simulated.



Here we propose the first new idea: let the task for quantum simulation with a potential speedup shift from solving the ideal problem to solving any problem in a family of different noise models with the same ideal part. The family is defined so that there are no cheating noise models (such as dividing the system into small pieces), and a good model of the quantum device is one of them, and all the models have the same or smaller level of noise in some sense. Thus the quantum device solves this problem, and the classical effort can be directed at the noise model that makes the computation simplest instead of the real one. The question is what measure of noise and family of noise models to use here and how to exclude cheating rigorously. We leave these questions for future work. Exploring this further will guide the future experimental claims of quantum advantage of a particular quantum simulator for the task of simulating the dynamics itself.

A different task of obtaining a specific dynamical property of the system might be more feasible than showing the quantum advantage of the simulation itself as described above. Note that quantities like the critical exponents are known to be universal, i.e. independent of small variations in the system. Noise can break that universality, but sufficiently small noise can still lead to some universal behavior. Suppose that the noise is both unknown and, though small, produces large deviations in apparent evolution from the ideal case. Thanks to the universality, polynomial postprocessing on the experimental outcomes may still return the right answer, while the efficient classical algorithms for simulating the same system with a different and tractable noise return a different value of the universal quantity. 
There's also the question of heuristics here: can we train some model to guess the answer better than our trust in our quantum computer with unknown noise?
Surprisingly, universal critical exponents may not require faithful spectrum simulation. Indeed, even though a simple estimate points out that among the states involved in the density of defects experiment on D-Wave \cite{king} there should be some non-qubit states, it does not invalidate their claim of quantum simulation since the specific critical exponent they focus on turned out to be insensitive to those non-qubit states. 

The starting point for measuring a dynamical quantity is often the thermal state. Having the approximately correct spectrum does not necessarily translate into the approximately correct thermal expectation values even for static quantities. 
We're looking at the exponentiation $e^{iHt}$ and $e^{-\beta H}$, which may blow up the originally small errors. 
Using Duhamel's formula, we get $e^{i(H+\delta H)t} = e^{iH t}  + i t\int_0^1\int_0^1 ds d \tau  e^{i(H+s\delta H)  t\tau} \delta H e^{i(H+s\delta H)  t(1-\tau)}$ and derive an error bound:
\begin{align}
    \|e^{i(H+\delta H)t} - e^{iHt}\| = t \|\int_0^1\int_0^1 ds d \tau  e^{i(H+s\delta H)  t\tau} \delta H e^{i(H+s\delta H)  t(1-\tau)} \|  \leq t\|\delta H\| \label{timebou}
\ .\end{align}
From Eq. (\ref{timebou}) we obtain the local error in unitary evolution generated by evolving with $\alpha H_{\text{targ}} +\delta H$ where $\|\delta H\| \leq \alpha \epsilon n s$ for a time $t/\alpha$:
\begin{equation}
 \|\psi_{\text{true}} - \psi_{\delta H} \| \leq t \epsilon n s   
\ .\end{equation}
We will attempt to naively compare this with the gate-based model implementation of the quantum evolution that works on all initial states. The two steps involved here are first to use the result by \cite{haah} to note that the circuit depth required on the logical graph is only logarithmically longer than the quantum simulator implementation where we can turn on the Hamiltonian directly, which means it does not affect the power of $n$. The dependence on target error is logarithmic as well. For a constant local precision at a time $t$, it suffices to use a depth $t$ circuit. Second is the observation that translating any circuit of depth $D$ on $n$ qubits to a circuit of nearest neighbor gates on a ring of $n$ qubits can be done with SWAP's in $O(Dn)$ gates. Indeed, at each step of the logical circuit, the applied 2-qubit gates define two sets of qubits: the first qubit of each gate and the second qubit of each gate. Swapping the first qubits of each gate with depth-$O(n)$ circuit of swaps allows every qubit to have its pair as a neighbor. This shows that $O(tn^2)$ ideal gates implement the desired local precision. The gate errors $\delta_g$ add a total error of $O(tn^2\delta_g)$, which suggests $\delta_g =O( 1/n)$ is required to match our local precision with this simple construction. This scaling is substantially better than $1/n^6$ that we found for the transmission line. We note that the control errors of a quantum simulator and gate errors of a quantum computer are not directly comparable. We also expect large overheads from the construction of \cite{haah}, which suggests the quantum simulator approach is superior at intermediate $n$.

The bound is a bit more complicated for the thermal one. Define $\rho_s =Z_s^{-1}e^{-\beta (H+s\delta H) } $. Introduce a bath $H_B$ and a system bath interaction $V_{SB}$ of variable strength $v$, so that the total Hamiltonian is:
\begin{equation}
    h_{s,v} = H + s\delta H +  vV_{SB} +H_{B}
\ .\end{equation}
Define a thermalization scale $T_{s,\epsilon,v}$ as the smallest number such that for any $t >T_{s,\epsilon,v}$:
\begin{equation}
    \|\rho_s -\text{tr}_B e^{ih_{s,v}t}(\rho_{i}\otimes \rho_B) e^{-ih_{s,v}t} \|_1 \leq \epsilon
\ .\end{equation}
and we assume that the bath is chosen big and generic enough for the above condition to have a solution as long as $v \leq v_\epsilon$. Here $\rho_i$ is any initial state of our choice. The problem then reduces to the closed system case for the evolution of the total system:
\begin{equation}
    \|\rho_0 - \rho_1 \|_1 \leq \text{min}_{\epsilon,v\leq v_\epsilon} (2 \epsilon + \|\delta H\| \text{max}(T_{0,\epsilon,v},T_{1,\epsilon,v}))
\ .\end{equation}

In particular, one can relax the bound by defining $\epsilon^*$ such that:
\begin{equation}
     \epsilon^* =\|\delta H\| \text{max}(T_{0,\epsilon^*,v^*},T_{1,\epsilon^*,v^*}) \ ,
\end{equation}
for some $v^*$. Let the thermalization timescale defined implicitly in this way be $\tau = \text{max}(T_{0,\epsilon^*,v^*},T_{1,\epsilon^*,v^*})$. The bound then assumes a simple form:
\begin{equation}
    \|\rho_0 - \rho_1\|_1 \leq 3\|\delta H\| \tau
\ .\end{equation}
The exact value of $\tau$ depends on the specific bath and initial state chosen. The value of this bound is in providing intuition for the behavior of the thermal state (and the expectation values of the observables). The thermalization timescale can be exponential in the system size for glassy systems, and even for non-glassy systems, it is still nontrivial to derive the fast thermalization (i.e. that $\tau$ is polynomial in system size) defined above. We can also derive a weaker bound that does not rely on thermalization:
\begin{align}
    \|\rho_1 - \rho_0\|_1 = \beta \left\|\int_0^1\int_0^1 ds d \tau  \rho_s^\tau \delta H \rho_s^{1-\tau}  + \rho_s \text{tr} \rho_s \delta H\right\|_1 \leq \\ \leq \beta \int_0^1\int_0^1 ds d \tau  \|\rho_s^\tau \delta H \rho_s^{1-\tau}\|_1  +  \text{tr} \rho_s \| \delta H\| \leq  \beta \|\delta H\|  +  \beta\text{max}_s{\pi_s} \|\delta H\|_1\ ,
\end{align}
where $\pi_{s} = \|\rho_s\|$ is the ground state probability, and we have used $\|\rho_s^\tau \delta H \rho_s^{1-\tau} \rho_s \|_1 \leq \|\rho_s^\tau\|\|  \delta H \rho_s^{1-\tau}\|_1 \leq \|\rho_s\|\|  \delta H \|_1 $. The trace $\|\delta H\|_1$ norm would generally contain the Hilbert space dimension rendering this bound useless for most temperatures.

We will also derive a linear response in $\epsilon$ for $\delta H = \epsilon V$. We first establish that $\rho_s - \rho_0 = O(\epsilon)$ for $s\in[0,1]$. Using the above bound:
\begin{align}
    \|\rho_1 - \rho_0\|_1  \leq  \beta\epsilon \|V\|  +  \beta \epsilon\text{max}_s{\pi_s} \|V\|_1
\ .\end{align}
Since $\pi_s \leq 1$, we get:
\begin{equation}
    \|\rho_1 - \rho_0\|_1  \leq \beta \epsilon (\|V\| + \|V\|_1) = O(\epsilon)
\ .\end{equation}
Same argument shows that $\rho_s - \rho_0 = O(\epsilon)$ and $\rho_s^\tau - \rho_0^\tau = O(\epsilon)$  for $s,\tau\in[0,1]$. Now we examine the exact expression again, taking out the $\epsilon$-dependence:
\begin{align}
    \rho_1 - \rho_0 = -\beta \epsilon \int_0^1\int_0^1 ds d \tau  (\rho_0^\tau +O(\epsilon)) V (\rho_0^{1-\tau}+O(\epsilon))  + (\rho_0+O(\epsilon)) \text{tr} (\rho_0 + O(\epsilon)) V = \\
    = O(\epsilon^2) -\beta \epsilon \int_0^1 d \tau  \rho_0^\tau  V \rho_0^{1-\tau}  + \rho_0 \text{tr} \rho_0  V
\ .\end{align}
For the traceless $V$ we obtain the following linear response:
\begin{equation}
    \rho_1  = \rho_0 -\beta \epsilon \int_0^1 d \tau  \rho_0^\tau  \overline{F} \rho_0^{1-\tau}  +O(\epsilon^2)
\ .\end{equation}

Note that for the Gibbs state bound above we didn't consider the effect of the thermal population of the non-qubit levels inside the mediators of our constructions. Even if a more sophisticated proof technique will provide a guarantee of a faithful simulation of the Gibbs state for all sources of error, such states are either effectively simulated by Quantum Monte-Carlo \cite{landau2021guide} for the topology of the current annealing devices or are intractable both classically and quantumly due to spin glass behavior. A hardware adjustment needed to separate quantum from classical is the non-stoquasticity \cite{hen2021determining}, but even that may be within reach of classical tensor network states. In view of this, the dynamics has fewer obstacles for a quantum advantage of a quantum simulator, as the bound $t\|\delta H\|$ can be sufficiently small, the non-qubit levels don't contribute for static $H$, the non-stoquasticity is not required and the classical tensor network methods are not equally well developed.


\subsection{Norm of the target Hamiltonian}
\begin{figure}
\centering
\includegraphics[width=1.\columnwidth]{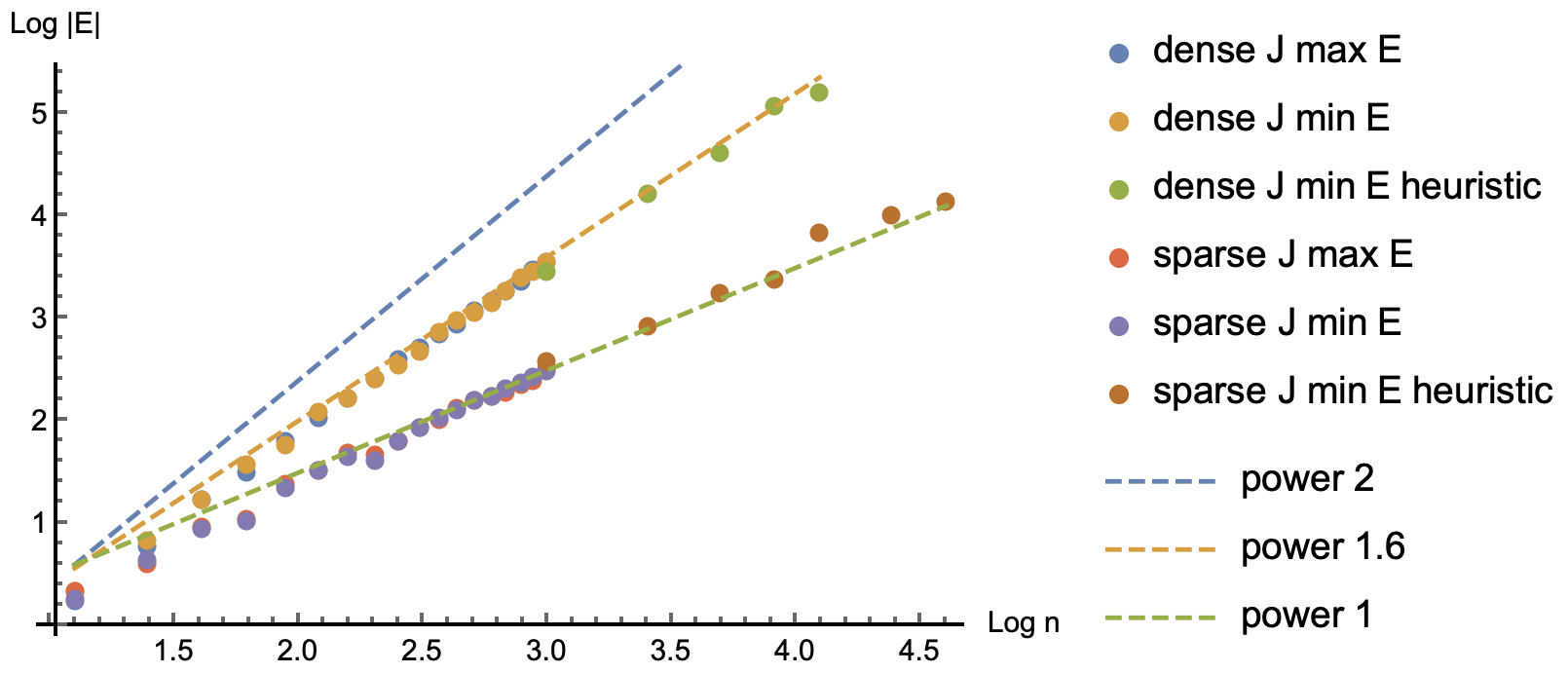}
\label{powers}
\caption{Norm of the classical random all-to-all and sparse (degree $2s=3$) Ising Hamiltonian, dots: PT-ICM results for $n >20$, exact diagonalization for $n\leq20$. Dashed lines: observed scaling compared to $O(n^2)$ scaling given by the bound on the norm we use for $s=(n-1)/2$.}
\end{figure}

The desired precision was chosen as a fraction of the norm of the target Hamiltonian. The main contribution to that comes from the classical part  $\sum_{ij}J_{ij} Z_iZ_j$, which we bounded as 
\begin{equation}
  \|\sum_{ij}J_{ij} Z_iZ_j \| \leq ns 
\ ,\end{equation}
and in the special case of a complete graph, we used $s=(n-1)/2$. Here we note that this norm can be an overestimate, as local contributions may cancel each other, resulting in a factor lesser than $s$. This consideration can improve the bounds derived in this paper, which we leave for future work. To estimate the scaling of the norm, we need to specify the distribution that $J_{ij}$ is drawn from.  One class is where $J_{ij}\in [-1,1]$ is an i.i.d. random variable for each pair $i,j$, drawn from a uniform distribution on $[-1,1]$. Setting all $Z_i \to 1$ we get the energy of that bitstring to be a random variable with zero mean and spread $O(n)$. The norm requires maximum and minimum energy over all bitstrings, so its scaling may differ. While this problem may have been studied in the past, for our purposes it is sufficient to calculate $\|\sum_{ij}J_{ij} Z_iZ_j\| \sim n^a$  numerically to obtain an estimate of the power $a$. In Fig. \ref{powers} we present an exact numerical calculation of the minimum and maximum energy of $\sum_{ij}J_{ij} Z_iZ_j$ built as described above for up to $n=20$ spins, for 40 disorder realizations at each size. We also use our heuristic PT-ICM solver \cite{Mandr__2018} up to $n=60$ using default (suboptimal) solver parameters to obtain an approximation of the ground state. For these larger sizes, we only use 1 disorder realization for each size, relying on self-averaging of the ground-state energy. The runtime is chosen so the entire data collection takes 1 minute on a single CPU. We observe that at that number of PT steps the heuristic is not to be trusted to find the ground state for $n>60$, and an increase in the computation time is needed. The check is done by repeating the entire optimization 10 times and comparing the outcomes. If half of the outcomes disagree with the minimum, the algorithm is considered to miss the ground state. Of course, even if the outcomes agree it comes with no guarantees that the true minimum is found. 
Fits show that the power is $a \approx 1.6$ and does not appear to have significant finite-size effects.

Another problem class is where the degree of the interaction graph is restricted. The easiest way to achieve that is to multiply each $J_{ij}$ by an i.i.d. random variable that takes values $\{0,1\}$ with $1$ having the probability $s/n$, where $s=O(1)$ is a desired degree. For this case, the energy of all $Z_i \to 1$ is a random variable with zero mean and spread $O(\sqrt{sn})$. Our bound on the norm is $\leq sn$, and we expect it to be tight in $n$ since the energy should be extensive, at least for all the lattice interaction graphs that fall into this category. We also check that numerically for $2s=3$ and plot it in Fig. \ref{powers} The PT-ICM for the set runtime and parameters becomes unreliable in the sense explained above after $n=100$. We do not find significant deviations from the extensive scaling of the ground state energy.

Note that even though the graph's degree is 3 on average, there's no general way to draw it on a plane with all the links local (of finite length) and finite density of vertices. Indeed, a tree is a 3 local graph that takes up at least $O(\sqrt{n})$ linear space, but any two elements are connected by a log$(n)$ sequence of edges, which means there has to be an edge of length $O(\sqrt{n}/$log$n)$ which is not local. Embedding is required, and highly non-planar architectures and long-range connectivity are desirable properties of the hardware graph.


\subsection{Relation to PCP theorem}
Our construction focuses on the approximate reproduction of all $2^n$ eigenvalues of the quantum spectrum, however, it can also be applied to the special case of the ground state. The precision parameter $\epsilon$ we define guarantees that the ground state energy has an extensive error at most  $\epsilon s n$ for a problem on a graph of degree $2s$. Even if we had a way to prepare that approximate ground state, it would have $O(n)$ energy difference with the true ground state of the target problem.
One may be concerned that returning a $O(n)$ energy excited state of an $O(s)$-sparse Hamiltonian is not computationally interesting, but there is a surprising result in the computational complexity that in fact for some $\Theta(1)$ constants $\epsilon$ and $s$ it's $NP$-hard to return a state within $\epsilon n $ of the ground state of a sparse Hamiltonian on a graph of degree $2s$. This result is based on the certain hardness of approximation results, which are equivalent to the PCP theorem of computational complexity. Below we will sketch its proof based on \cite{dinur2007pcp}.

We first define the complexity class in question. NP is a class of decision problems, where the length of the question is (power of) $n$, the answer is just one bit and the proof is also poly$(n)$-long. Here proof is a bitstring needed to verify the answer $1$, which can be done efficiently. Answer $0$ is returned by the verifier algorithm by default if no working proof is provided, but there's no way to have a poly$(n)$-long "proof" of $0$ (unless NP = coNP). Efficiently here means that the runtime of the verifier is poly$(n)$. Being in  NP means that there is such a verifier. A family of problems is NP-complete if any problem in the NP family can be reduced to this problem family. That means, in particular, that a hypothetical black box that solves (returns 0 or 1) any problem within an NP-complete family can also be used to solve any other problem in NP. Finally, NP-hard refers to more general problems where the answer is not necessarily binary, and the verifier need not be possible, as long as a black box solving that can also be used to solve any problem in NP.

An example of an NP-complete problem is a constraint satisfaction problem. A constraint is a truth table on $\leq K = O(1)$ variables. Given $n$ binary variables and $n$ constraints (some of the variables may be unused and some of the constraints may always be true), the binary decision question is if there's a way to satisfy all the constraints. The proof is a bitstring that supposedly satisfies all the constraints and the verifier is an algorithm that checks one by one if the constraints are satisfied. The PCP theorem states that in fact, a weaker version of this is NP-hard: for some constant $\epsilon =\Theta(1)$
we are given a promise that in all problems allowed in the family either all constraints are satisfied, or there are $>\epsilon n$ unsatisfied constraints. The decision is, given a problem, to return which of the two possibilities is true. 

Let's elaborate on various aspects of this statement. Note that the PCP theorem says only that this promise decision problem is NP-hard. Saying that it is in NP is trivial since the question is the same (are all constraints satisfied), but it's usually not mentioned that it is in NP to avoid confusion with the non-promise version of the problem. The nontrivial fact is that this promise decision problem is in fact NP-complete. Compare: the original statement of NP-completeness of CSP says that among the family of all constraint satisfaction problems, deciding if all constraints can be satisfied is NP-complete. The PCP version says that in fact even if we restrict the family only to the decision problems that satisfy the promise (the minimum number of unsatisfied constraints is either 0 or $\geq \epsilon n$), the NP-completeness is still preserved. In other words, one can encode every CSP with an arbitrary minimum number of unsatisfied constraints into a CSP$'$ that satisfies the promise, possibly at the cost of some overhead in $K$ and $n$.

An alternative formulation in terms of a black box may be insightful: consider a black box that returns an integer within $\pm \epsilon n/2$ of the minimum number of unsatisfied constraints of any CSP problem. This black box can be used on a CSP satisfying the promise to decide the question of the PCP formulation (0 or $\geq \epsilon n$ minimum number of unsatisfied constraints). Since the PCP theorem states that this question is NP-hard, the black box is also NP-hard. It is easy to reduce it to a problem in NP if necessary, thus making it NP-complete. We will now prove that this result can also be applied to Ising models. That is, a black box is NP-hard that takes in an Ising model $\sum_{ij} J_{ij}Z_iZ_j + h_i Z_i $ on $n$ spins, where only $O(n)$ matrix elements of $J_{ij}$ are nonzero, and returns its energy to within $\epsilon n$. This is easy to prove since a CSP can be translated into an Ising model, with each constraint incurring O(1) overhead in ancillary spins and the number of $J_{ij}$ (but possibly exponential in K). The specific construction is a matter of taste, but we need to first write a Hamiltonian $H_C = \sum_i C_i$ where $C_i=0,1$ if the constraint is satisfied or unsatisfied, and then replace each $C_i$ by some Ising gadget $H_{C,i}$ that is $0$ if the constraint is satisfied, and $1\leq H_{C,i} \leq f(K)$ if it is not. This gadget acts on $K$ spins involved in the constraint and possibly $O(2^K)$ ancillae used for this particular constraint. See an example of the construction of such a gadget below. 

List the bitstrings that satisfy the constraint as $b_j$, $j=1..j_{\text{max}}$. Place an ancilla $a_j$ per $j$. Place FM or AFM links between the spins $s_i$ involved in the constraints and this ancilla, depending on bits of $b_j$, and offsets such that the Hamiltonian contains terms $\hat{s}_i(\hat{a}_j+1)(b_j-1/2)$. This makes sure that only $a_j=-1$ enforces the constraint. Finally, add a penalty Hamiltonian on ancillae that enforces a "one-hot" encoding:
\begin{equation}
    H_p = (\sum a_j)^2 -1.5 j_{\text{max}}\sum a_j \ ,
\end{equation}
after an appropriate constant shift of energy, this construction satisfies the requirements on $H_{C,i}$, which concludes the proof. Note that the magnetic field can be equivalently replaced by an extra ancilla coupled to every spin subjected to the magnetic field, which doubles the degeneracy of the ground state but doesn't affect the decision problem.

Now to make it more relevant to the construction in this paper, we prove the following:
A black box that, given $J$ and $h$, returns the ground state energy of $\sum_{ij} J_{ij}Z_iZ_j + h_i Z_i  + r$ for some unknown $r$ s.t. $\|r\|\leq \epsilon n$ ( $J_{ij}$ is still sparse with only $O(n)$ nonzero elements) is NP-hard. Indeed, that ground state is guaranteed to be within $\epsilon n$ of the true ground state (of the $\sum_{ij} J_{ij}Z_iZ_j + h_i Z_i$), thus reducing the problem to the previous one. The black box that returns the excited (by at most $\epsilon n/2$ in energy ) state of noisy hamiltonian with noise $\leq \epsilon n/2$ is also NP-hard. This inspires us to approximate $\sum_{ij} J_{ij}Z_iZ_j + h_i Z_i$ to an error of norm $\leq \epsilon n$ since the corresponding ground state energy is NP-hard to know. We use this extensive target precision in our construction.

We note that a construction obtaining a target nonlocal Hamiltonian ground state bitstring of a classical Hamiltonian (no transverse field) $\sum_{ij} J_{ij}Z_iZ_j + h_i Z_i$ on a 2D local graph $\sum_{<ij>} J_{ij}Z_iZ_j + h_i Z_i$ is well known as "minor embedding" \cite{cai2014practical,choi2008minor}. Each qubit is represented as a chain with ferromagnetic penalties enforcing consistency, and a non-planar lattice is needed. This construction only copies the classical energies below the penalty scale but is in principle exact for those states. 
From the point of view of complexity, 2D planar lattices have ground states that are easy (decisions about energy can be done in P)\cite{fisher1966dimer}, nonplanar 2D lattice ground states have NP-hardness to approximate to within $\epsilon \sqrt{n}$ of the ground state energy while approximating within $ \geq
\Theta(n/\text{log} n)$ is easy (decisions can be done in P) by dividing into squares of appropriate size and finding the ground state within each square. The scaling $\epsilon \sqrt{n}$ can be obtained by minor-embedding the NP-hard family from the all-to-all result with $\epsilon n'$ approximation being hard. Embedding $n'$ logical qubits on a complete graph generally requires $n = O(n'^2)$ physical qubits on a lattice, which is where the scaling comes from.

In this paper, instead of minor embedding, we advocate for using paramagnetic chains. Minor embedding reproduces the classical spectrum exactly under the penalty gap, but the quantum spectrum suffers from an avoided crossing gap reduction that scales exponentially with the chain length, or as exp($-cn^b$) in system size. Paramagnetic chains only reproduce the spectrum approximately and will require the overall scale of the system energy, thus also the minimal gap scale to go as $1/n^a$, but after that is accounted for, the full system spectrum is reproduced faithfully. Thus we lose some accuracy of the classical problem but gain in the potential of quantum tunneling. One may wonder if there's a middle ground that is both more accurate on the classical spectrum and doesn't reduce the minimal gap of the avoided crossings of the quantum one, while possibly not reproducing the quantum spectrum of the logical problem since it doesn't matter how we get there as long as the gaps are not too small. We leave these considerations for future work. 

 \section{Warm-up: non-qubit levels on a native graph}
\label{warmup}

A polynomial in $n$ reduction in the energy scale in the results above may seem expensive, but it is unavoidable for the task of reproducing the full quantum spectrum of the target Hamiltonian. To illustrate that, we consider a simpler system without gadgets, where qubits themselves have extra high energy levels, neglected in the qubit description. We use this case as an illustration of the general method that follows.

Let $H_0$ be a Hamiltonian of non-interacting $k$-level systems, where each of them is implementing a qubit in its two lowest levels $|0\rangle, |1\rangle$, degenerate in $H_0$, and has the gap $\omega_p$ to the remaining levels $|2\rangle\dots |k-1\rangle$:
\begin{equation}
    H_0 = \sum_{i=1}^n\sum_{e=2}^{k-1} \omega_p|e_i\rangle \langle e_i|
\ .\end{equation}
Our goal is to use this system as a quantum simulator, and our control capabilities allow us to switch on $H_{\text{target}}$ of Eq.(\ref{targ1})  on the qubit subspace $P =\prod P_i, ~ P_i =|0_i\rangle \langle 0_i| + |1_i\rangle \langle 1_i|$. While the scale of terms $H_{\text{target}}$ goes $\leq 1$, we are free to introduce a reduction factor $\alpha$, such that the physical Hamiltonian is $\alpha H_{\text{target}}$. We aim to study the effect of control imprecision as well as the presence of non-qubit levels on the deviation of the spectrum of the system from that of $\alpha  H_{\text{target}}$. Both $\alpha H_{\text{target}}$ and the control errors will be included in perturbation:
\begin{equation}
    V = \alpha H_{\text{target}} + V_c
\ .\end{equation}
Both $H_{\text{target}}$ and $V_c$ are made of one and two-qubit terms on the interaction graph. The norm of $H_{\text{target}}$ is just $\|H_{\text{target}}\| \leq (2+s)n$ counting the number of terms.
We now split the noise into the qubit control noise $PVP$ of strength $\|PV_cP\| \leq \delta n(2+s)$ where $\delta$ can be associated with the control precision of individual $ZZ$ couplings and $X$ and $Z$ local fields. Those terms lie within the block of qubit states, for the one or two qubits involved, and are zero outside the block. The rest of the $k$-level systems enter these interaction terms as $\otimes 1$. We let the remaining terms $PV_cQ +QV_cP$ and $QV_cQ$ (where $Q= 1-P$) be bounded as:
\begin{equation}
    \|PV_cQ\| \leq r\alpha\|H_{\text{target}}\| \leq r\alpha n(2+s), \quad \|QV_cQ\| \leq \delta n(2+s) \label{generalEr}
\ .\end{equation}
Note that the term mixing the blocks only appears as we turn on the target Hamiltonian. We will need to assume that there is some smallness in $r$, but to see its physical meaning we need to look at the details of specific hardware, unlike the more general definition of $\delta$ characterizing control errors $\|PV_cP\|,~\|QV_cQ\| \leq \delta n(s+2)$. Below we define $V$ for the flux qubit hardware \cite{harris2010experimental,Krantz:19}.  

The starting point is the flux description where each system has an infinite number of levels, and the dynamics of the flux $\Phi$ is that of a 1d particle in a potential $V(\Phi)$, which has two potential wells. We will now simplify it to the lowest $k=4$ levels in that potential. The kinetic energy is characterized by $E_C$ and the potential energy in each potential well by $E_J$. The plasma frequency is the energy of the first excited state in each of the wells and is $\sim \sqrt{E_J E_C}$. We are operating in the units where the flux quantum $\Phi_0 =1$, which means the potential term is $\sim E_J (\Phi +0.5)^2 +E_J (\Phi -0.5)^2 $. The characteristic length scale $\Delta \Phi$ of the wavefunction in one of the wells is given by
\begin{equation}
 \Delta\Phi \sim ( E_C/E_J)^{1/4}
\ .\end{equation}
The operator for $k=4$ lowest levels of the individual flux qubit is $H_0 =0.5\tau_z \omega_p $, where we introduced Pauli matrices $Z$ to indicate which well the system is in, and $\tau_z$ to indicate whether it is in a qubit state or non-qubit state. The flux operator $\Phi$ used for $hZ +JZZ$ terms in $H_{\text{target}}$ has extra matrix elements between the qubit and non-qubit states in the same well $\Phi \pm 0.5 \sim  \Delta \Phi (a+a^\dag)$ in terms of the creation and annihilation operators of the harmonic oscillator corresponding to each well. In terms of the Pauli matrices, we get:
\begin{equation}
    \Phi = 0.5 (Z + rZ \tau_x)\ ,
\end{equation}
where $r \sim \Delta \Phi$. We ignored $\sim r Z \tau_z$ terms as they will turn out to be subleading. The full Hamiltonian of $n$ flux qubits takes the form:
\begin{align}
   H_0 +V= \sum_i\frac{\omega_p}{2} \tau_{zi}  + (\alpha h_i + \delta h_i) Z_i (1 + r \tau_{xi}) + (\alpha t_i + \delta t_i) X_i (1+c(\tau_{zi}+1)) +\\+\sum_{ij}(\alpha J_{ij} + \delta J_{ij}) Z_iZ_j (1 + r \tau_{xi})(1 + r \tau_{xj})
\ .\end{align}
Here we also used the tunneling rates of the excited and ground states of the well $(\alpha t_i + \delta t_i), ~ (\alpha t_i + \delta t_i) (1 +2c)$ respectively. Their computation depends on the specific shape of the barrier between wells, and the constant $c=\Theta(1)$ for most cases. We will not discuss that computation in this work, referring the reader to Sec. V of our previous work \cite{mozgunov2023quantum}.

The errors $\delta h_i, \delta t_i, \delta J_i$ are the control errors satisfying  $|\delta h_i|, |\delta t_i|, |\delta J_i| \leq \delta$. We see that after separating the terms into $H_0,~ \alpha H_{\text{target}}$ and $V_c$, the latter is found to be:
\begin{align}
   V_c = \sum_i \alpha r h_i Z_i  \tau_{xi} + \delta h_i Z_i (1 + r \tau_{xi}) + \alpha c(\tau_{zi}+1) t_i X_i + \delta t_i X_i (1+c(\tau_{zi}+1))+ \\+\sum_{ij}\alpha J_{ij}  Z_iZ_j ( r^2 \tau_{xi}  \tau_{xj} + r \tau_{xi} + r \tau_{xj})  + \delta J_{ij} Z_iZ_j (1 + r \tau_{xi})(1 + r \tau_{xj})
\ .\end{align}
The leading terms (in powers of $r$, $\delta$) of the bounds on various blocks of $V_c$ are as follows:
\begin{align*}
    \|PV_cP \| \leq \delta (2+s)n + o(\delta,r)\ ,\\ \|QV_cQ\| \leq \alpha|2c|n + \delta (1 +|1+2c|+s)n +o(\delta,r)\ , \\
    \|PV_cQ\| \leq \alpha r(1+s)n + o(\delta,r)\ .
\end{align*}
We have checked that even for a choice of the bare Hamiltonian where the effects of $t_i$ and $h_i$ are included, the error bounds due to interaction retain the same order in $r$ \cite{urlCode}. Note that $\alpha \|H_{\text{target}}\| \leq \alpha (2+s) n$. For $c=0$ the leading terms are bounded exactly as we assumed in Eq. (\ref{generalEr}) for the general case, and nonzero $c$ will only introduce a $\Theta(1)$ constant but will not affect the scaling of our results. Because of that, we proceed using Eq. (\ref{generalEr}).
We would like to take the realistic values $r \sim \Delta \Phi \sim  ( E_C/E_J)^{1/4}$, which is $10^{-1}$ for the latest reported CJJ flux qubits (\cite{harris2010experimental} describes the state of the art in 2010, and \cite{khezri2021anneal} summarizes the recent progress), and $\omega_p\sim \sqrt{E_J E_C} \sim 10$ times the characteristic qubit energy. The best reported $\delta =10^{-2}$ \cite{boothby2021architectural}. 

The many-body result by Bravyi et al \cite{B1} applied to this system establishes that independent of $n$ the ground state energy of $H_{\text{target}}$ is reproduced to a local precision which is $O((\alpha +\delta +r)/\omega_p)$ (note that as the perturbation is turned on the ground state may switch with an excited state, but the energy of both doesn't go too far from the non-perturbed values). The dependence on the degree of the graph is not explicitly stated. In this work, we attempt to reproduce the whole spectrum for the task of quantum simulation, which means we will need to have a $1/n$ reduction factor at the very least so that the spectrum fits under the gap $\omega_p$ of $H_0$. The specifics will be computed below. 

We apply the finite-dimensional lemma \ref{lem2}, assuming that $\|V\|/\omega_p \leq 1/16$ (we will check it later).  The norm of $V$ is:
\begin{equation}
    \|V\| =(\alpha(1+r)  +\delta )(2+s) n 
\ .\end{equation}
We now require that both the control errors as well as the perturbative errors due to Lemma \ref{lem2} are extensive with a bound $\epsilon$, which is the desired precision:
\begin{equation}
    \delta(2+s)n  +3.5 \|PVQ\|\|V\|/\omega_p \leq \alpha n s \epsilon
\ .\end{equation}

This leads to:
\begin{equation}
    \delta(2+s)n  +3.5 r\alpha n^2(2+s)^2(\alpha(1+r)  +\delta ) /\omega_p \leq \alpha n s \epsilon
\ .\end{equation}

We obtain a quadratic inequality in $\alpha$:
\begin{align}
   3.5r (1+r)n(2+s)^2\alpha^2  + (3.5rn(2+s)^2\delta- \epsilon s \omega_p)\alpha  +\delta (2+s)\omega_p \leq 0\ ,
\end{align}
or
\begin{align}
    &(1+r)A_1 \alpha^2 + (A_2 -\epsilon s \omega_p) \alpha + A_3 \leq 0 \ ,\label{orig0} \\
    &A_1 = 3.5r n(2+s)^2\ ,\quad 
    A_2 =3.5rn(2+s)^2\delta\ ,\quad 
    A_3 = \delta (2+s)\omega_p
\ .\end{align}
As $r\leq 1$, we get rid of a factor $1+r$ by requiring a stronger inequality to hold (using $1/(1+r) \leq 1, ~ -1/(1+r) \leq 2$):
\begin{align}
    A_1 \alpha^2 + (A_2 -\mathcal{E}) \alpha + A_3 \leq 0 \label{orig} \ ,\quad  
    \mathcal{E} = \frac{1}{2}\epsilon s \omega_p
    \ .\end{align}
If the roots $\alpha_{1,2}$ of the quadratic polynomial are real, the solutions lie in the interval  $[\alpha_1, \alpha_2]$. The center of the interval is $\alpha_c =( \mathcal{E}-A_2)/2A_1$. The range of applicability $\frac{\|V\|}{\omega_p} \leq \frac{1}{16}$ becomes:
\begin{align}
    \frac{1}{\omega_p} ((1+r)\alpha +\delta) (2+s)n \leq 
    \frac{1}{16}
\ .\end{align}
We will use a stronger constraint obtained by replacing $1+r \to 2$. This defines the allowed range of $\alpha$:
\begin{equation}
    \alpha \leq \frac{1}{2}\left(\frac{\omega_p}{16n(s+2)} -\delta \right) \label{alphaR}
\ .\end{equation}
We seek to find some of the solutions to the inequalities
\begin{align}
\begin{cases}
     \alpha \leq \frac{1}{2}\left(\frac{\omega_p}{16n(s+2)} -\delta \right)\ , \\ A_1 \alpha^2 + (A_2 -\mathcal{E}) \alpha + A_3 \leq 0 \ ,
\end{cases}
\end{align}
in the region $\alpha,\delta,r,\epsilon\in[0,1]$. Given $\epsilon$, we aim to present a region of $\delta,r$ where a solution exists, together with an example of $\alpha$ expressed via $\delta,r,\epsilon$ that belongs to a solution. We allow the region of $\delta,r$ to be missing some of the solutions, as long as its shape has the right scaling with $n$ and is expressed concisely. There are two forms of $\alpha$ that we will use. One is the boundary value saturating Eq. (\ref{alphaR}), while another is $\alpha_c=( \mathcal{E}-A_2)/2A_1$, the center of the solutions of Eq. (\ref{orig}). The latter depends on $\epsilon$.  Eq. (\ref{orig}) is satisfied if:
\begin{align}
   ( \mathcal{E}- A_2)^2 - 4A_1 A_3 \geq 0
\ .\end{align}
Out of the two signs for $ \sqrt{( \mathcal{E}- A_2)^2}$, we choose the positive side $ \mathcal{E} \geq A_2 +2\sqrt{A_1A_3}$ that gives positive $\alpha$.  Repeating the same arguments with Eq. (\ref{orig0}), we obtain a tighter inequality:
\begin{align}
    \epsilon s \omega_p  =3.5rn(2+s)^2\delta+2 \sqrt{  3.5r (1+r)n(2+s)^3 \delta \omega_p} \label{exprWsqrt}
\ .\end{align}
This is the form we will use for the numerical estimates. 

We will now show how for the choice of $\alpha =\alpha_c$ the inequality $ \mathcal{E} \geq A_2 +2\sqrt{A_1A_3}$ is simplified to a stronger inequality $4A_1 A_3 \leq (8 \mathcal{E}/9)^2$. Substituting the values of $A_1,A_3,\mathcal{E}$ into the latter, we obtain:
\begin{align}
   4 \cdot 3.5r n(2+s)^3 \delta \omega_p \leq \frac{16}{81}(\epsilon s \omega_p)^2 
\ .\end{align}
We can now obtain a bound on $A_2$:
\begin{equation}
    A_2 =3.5r n(2+s)^2 \delta \omega_p \leq \frac{8}{81}\frac{\epsilon s \omega_p}{2} \frac{\epsilon s }{(2+s)} \leq 
    \frac{1}{9}\mathcal{E} \ ,
\end{equation}
since $\epsilon\leq 1, r\leq 1, s\geq 2$. This shows that $ \mathcal{E} \geq A_2 +2\sqrt{A_1A_3}$ follows from $2\sqrt{A_1 A_3} \leq 8 \mathcal{E}/9$.

The reduction factor $\alpha = \alpha_c = (\mathcal{E} - A_2)/2A_1$ is lower bounded as $\alpha_c \geq \alpha_o =\sqrt{A_3/A_1} =2\sqrt{A_1A_3}/2A_1$:
\begin{equation}
    \alpha_o = \frac{8 \mathcal{E}_o/9}{2A_1}= \frac{ 4\epsilon s \omega_p/9}{7rn(2+s)^2} \label{alpo}
\ .\end{equation}
This ensures that $\alpha_c > 0$ in the region of solutions  $ \mathcal{E} \geq A_2 +2\sqrt{A_1A_3}$, and in its subregion $2\sqrt{A_1 A_3} \leq 8 \mathcal{E}/9$.
We conclude that the solutions of Eq. (\ref{orig}) for the choice of $\alpha =\alpha_c$ exist if  $4A_1 A_3 \leq (8 \mathcal{E}/9)^2$ (though they may exist outside that range as well). However, $\alpha_c$ does not always satisfy Eq. (\ref{alphaR}). If that happens, we use the boundary value $\alpha_b = \frac{1}{2}\left(\frac{\omega_p}{16n(s+2)} -\delta \right) $ instead. Define $r_b,\delta_b$ as the value of $r, \delta$ at the intersection point of $\alpha_c = \alpha_b$ and $4A_1 A_3 = (8 \mathcal{E}/9)^2$. We find that:
\begin{equation}
    r_b = \frac{16\epsilon s }{7 (2+s)} ,\quad \delta_b  = \frac{\mathcal{E}}{ 81 n(2+s)^2 }
\ .\end{equation}
We will now show that in the region $\alpha_c \geq \alpha_b$, and $\delta\leq \delta_b$, the Eq. (\ref{orig}) is satisfied for $\alpha_b$. First, note that in the $\delta \leq \delta_b$ region of solutions we have positive
$\alpha_b(\delta) \geq \alpha_b(\delta_b) = \alpha_c(\delta_b,r_b) >0$ as shown above. Dividing Eq. (\ref{orig}) by $\alpha_b$, we get:
\begin{equation}
    A_1 \alpha_b + A_2 -\mathcal{E}  + A_3 \alpha_b^{-1} \leq 0 
\ .\end{equation}
Note that $\mathcal{E}  = \frac{\epsilon s \omega_p}{2}\leq \frac{s\omega_p}{2}$, and:
\begin{align}
    \alpha_b \leq \frac{\omega_p}{32n(s+2)}, \quad \alpha_b^{-1} \leq \frac{2}{ \frac{\omega_p}{16n(s+2)} - \delta_b}   \leq   \frac{32n(s+2)}{\omega_p \left(1 - \frac{8s}{81 (2+s)}\right)}  \leq  \frac{81 \cdot 32n(s+2)}{73\omega_p}
\ .\end{align}
Using the above and noting that $\alpha_c \geq \alpha_b$ translates into $r\leq r_b$, we see that it suffices to show the following inequality:
\begin{equation}
    A_1(r_b) \frac{\omega_p}{32n(s+2)} + A_2(r_b,\delta_b)   + A_3(\delta_b) \frac{81 \cdot 32n(s+2)}{73\omega_p} \leq \mathcal{E}
\ .\end{equation}
Plugging in the expressions for $A_{1,2,3}$ into the r.h.s., we get:
\begin{equation}
   r_b \frac{3.5(s+2)\omega_p}{32} + 3.5n(2+s)^2r_b\delta_b   +  \delta_b \frac{81 \cdot 32n(s+2)^2}{73}
\ .\end{equation}
Finally, we use the expressions for $r_b, \delta_b$:
\begin{equation}
  \frac{\mathcal{E}}{4} + \left(\frac{8\epsilon s }{ 81 (2+s)}   +  \frac{ 32}{73}\right)\mathcal{E} \leq \mathcal{E}
\ .\end{equation}
We conclude by putting together the condition $4A_1 A_3 \leq (8 \mathcal{E}/9)2$ for $r_b \leq r\leq 1$, or explicitly:
\begin{equation}
    \delta r  \leq \frac{8(\epsilon s)^2 \omega_p}{7\cdot 81 n(2+s)^3 }\ , 
\end{equation}
and $\delta \leq \delta_b  \sim 1/n$ for $0\leq r \leq r_b$, that the best solutions we find require $\delta=O(1/n)$ for all $r$. Indeed, even for a qubit system with $r=0$ the control precision $\delta\sim 1/n$ is required.

We have yet to check that $\alpha_c$, $\alpha_b$ we have used are $\leq 1$. This adjusts the region of solutions. 
The condition $\alpha_b \leq 1$ requires:
\begin{equation}
\frac{\omega_p}{16n(s+2)}  \leq 2 +\delta\ ,
\end{equation}
while $\alpha_c \leq 1$ becomes:
\begin{equation}
    \frac{\epsilon s \omega_p}{7 rn(2+s)^2}  \leq 2 + \delta
\ .\end{equation}
We note that the first condition always implies the second since $r\geq r_b$ can be used in the second. Strengthening it, we arrive at the result that as long as:
\begin{equation}
    \omega_p \leq 32n(s+2) \ ,
\end{equation}
the best solution we found requires $\delta =O(1/n)$.

Now suppose that 
one of the two conditions is violated for some $\delta,r,\epsilon$: either $\alpha_c\geq 1$ or $\alpha_b\geq 1$.  We will determine the new allowed region of $\delta,r$. 

We choose $\alpha=1$ in this situation. Let's see if we can show the inequality
\begin{align}
  A_1 +A_2 +A_3 \leq \mathcal{E}\\
  3.5r n(2+s)^2
    (1 +\delta)
+ \delta (2+s)\omega_p \leq 
\frac12 \epsilon s \omega_p
\ .\end{align}
For $r \geq r_b$ we can use $\alpha_c \geq 1$ to show: 
\begin{equation}
    \text{r.h.s.} \geq 3.5 rn(2+s)^2(1+\delta) + \frac{ \epsilon s \omega_p}{2(2+\delta)}
\ .\end{equation}
 The remaining terms that we need to show the inequality for are:
\begin{equation}
   \frac{ \epsilon s \omega_p}{2} \geq \delta(2+\delta) (2+s) \omega_p
\ .\end{equation}
For $r\leq r_b$ we use that and $\alpha_b\geq 1$ to show:
\begin{align}
    \text{l.h.s}\leq r_bn(2+s)^2(1+\delta) +  \delta (2+s)\omega_p, \\ \text{r.h.s} \geq 8n\epsilon s(2+s)(1+\delta)+\frac{ \epsilon s \omega_p}{2(2+\delta)}
\ .\end{align}
After substituting $r_b$ and the resulting cancellations, the remaining terms are the same as in the $r \geq r_b$ case:
\begin{equation}
   \frac{ \epsilon s \omega_p}{2} \geq \delta(2+\delta) (2+s) \omega_p \ ,
\end{equation}
or, requiring a stronger condition:
\begin{equation}
  \delta \leq  \frac{ \epsilon s }{6(2+s)} \label{bigop}
\ .\end{equation}
For $r\leq r_b$, it is straightforward to check that for $\alpha_b \geq 1$ the r.h.s is $< \delta_b$, so there will be an interval of $\delta$ between the interval where $\alpha_b\leq 1$ and solutions exist, and the above. The region of solutions we specify may become disconnected into two regions. This is a small artifact of our approximations and does not substantially affect the scaling.

The final result is as follows. Given $\epsilon \leq 1$, we have found the following conditions on $r,\delta$, such that for the choice of $\alpha$ described below the inequalities are satisfied and the gadget works. For each point in $0\leq \delta,r \leq 1$ we first compare $r$ with $r_b = \frac{16\epsilon s }{7 (2+s)}$.
If $r\leq r_b$, we compare $\alpha_b$ and $1$. If $\alpha_b\leq 1$, the point is the solution with $\alpha =\alpha_b$ if $ \delta \leq \delta_b$.  For $r\geq r_b$ we compare $\alpha_c$ and $1$. If $\alpha_c\leq 1$, the point is a solution with $\alpha =\alpha_c$ if $\delta r \leq \frac{8(\epsilon s)^2 \omega_p}{7\cdot 81 n(2+s)^3 } $. In case the $\alpha\geq 1$ on either side of $r_b$, the point is a solution with $\alpha =1$ if $\delta \leq \frac{ \epsilon s }{6(2+s)} $.

 We note that the $\omega_p \to \infty$, $\alpha=1$ limit requires just $\|PVP\| \leq ns \epsilon$, which translates into $ \delta (s+2) \leq s \epsilon $, which coincides with Eq. (\ref{bigop}) up to a factor. This is the case for a qubit system with only control errors.

The simplified region we found illustrates the $1/n$ scaling of $\delta$, but to investigate the required $\delta,r$ numerically, we will use the full expression Eq. (\ref{exprWsqrt}) instead. We will also need some realistic parameters. Consider that we're simulating degree $2s=3$ graph, and $\omega_p =10,~ \delta =10^{-2}, ~r= 0.1$. Plugging in the numbers, we find that for $\epsilon >0.1$ we are in the regime $r \leq r_b$. Using $\alpha_b$ in Eq. (\ref{orig0}, we get $\epsilon \geq 1.4$ for $n \geq4$ ( $n=4$ is the smallest system size that allows degree $2s=3$). Now using $\alpha_c$ for $\epsilon=0.1$ in Eq. (\ref{exprWsqrt}) with $\omega_p =10,~ r=0.1,~n=4$, we find solutions for $\delta \leq 8\cdot 10^{-4}$. For $n=40$, keeping the rest of the parameters the same, the solutions are found for $\delta \leq 8\cdot 10^{-5}$.


Our result can also be used as a back-of-the-envelope estimate for the result \cite{B1} about the perturbative ground state of a lattice. He showed that a lattice of perturbative gadgets can simulate the ground state energy to a constant local precision. In other words, if gadgets work locally on a patch of a system, they would also globally reproduce the ground state up to a constant factor of error. For a rough estimate, we will ignore this extra constant factor, and use the formula derived above with a patch $n=2s+1$. For not too small $\epsilon$ such that $r\leq r_b$, which is expected to be the relevant regime for realistic parameters, our $\delta \leq \delta_b$ result suggests the maximum allowed $\delta$ scales as $\epsilon \omega_p/(2+s)^2$. We note that such a quantum simulator of the ground state is quite far from chemical accuracy, and requiring a rigorous proof of the chemical accuracy $\epsilon<10^{-3}$ with the method \cite{B1} will demand unrealistically small $\delta$.

\section{Proof of the general theorem}
\label{proofG}

This is the derivation of the main theorem with a general mediator (with possibly infinite-dimensional Hilbert space) and a relatively general connectivity. 

The bound on the error $r =H_{\text{targ}}-H_{\text{eff}}$ is given by Lemma \ref{SimpLem}:
\begin{align}
    \|r\| \leq  \|P(\sum_i \delta h_i^c Z_i +\delta t_i^c X_i + \delta H_{m,i} + Z_i \delta I_{m,i} +\sum_{i>j}\delta f_{ij} I_{i,j} I_{j,i})P\|+   7\frac{\|PV\|^2}{\Delta_V}\label{genErr}
\ .\end{align}
It will be insightful to keep track of another version of error $\delta_H^P, ~ \delta_I^P$:
\begin{align}
\|P\delta H_{m,i} P\| \leq \delta_H^P \leq \delta_H \ ,\quad  \|P \delta I_{m,i}P\| \leq \delta_I^P \leq \delta_I 
\ .\end{align}
In the final result, we will use $\delta_H^P = \delta_H, ~ \delta_I^P = \delta_P$, while in Appendix \ref{q40} we will use the intermediate version.
The $\|PV\|$ contains:
\begin{align}
    \|PV\| \leq \|P\sum_i\delta H_{m,i} + Z_i \delta I_{m,i}\| +  \sum_i |h_i^c|  +|t_i^c|  +\sum_{i>j} |f_{ij}| \| PI_{i,j}\| \|P I_{j,i}\|  
\ .\end{align}
Putting everything together: 
\begin{align}
 &\|P(\sum_i\delta H_{m,i} + Z_i \delta I_{m,i})P\| \leq n(\delta_H^P +\delta_I^P)\ , \\ &\|P(\sum_i \delta h_i^c Z_i +\delta t_i^c X_i  +\sum_{i>j}\delta f I_{i,j} I_{j,i})P \| \leq  n \delta (1+ F +s ~ \text{max} ~|\chi_{i,j}\chi_{j,i}|)\ ,  \\
     &\|PV\| \leq n(\delta_H +\delta_I +  \alpha (1  +F^{-1} +s  \text{max} ~i_{i,j}i_{j,i}/|\chi_{i,j}\chi_{j,i}|))
\ .\end{align}

Plugging the expressions above into Eq. (\ref{genErr}), we get a quadratic inequality on $\alpha$ (not strictly quadratic due to the dependence of $\Delta_V$ on $\alpha$, but we would draw the intuition from quadratic inequalities):
\begin{align}
   \delta_H^P +\delta_I^P +  \delta (1+F +s ~\text{max} ~|\chi_{i,j}\chi_{j,i}|) + \frac{7n}{\Delta_V}(\delta_H +\delta_I + \alpha (1  +F^{-1} +s ~ \text{max} ~\frac{i_{i,j}i_{j,i}}{|\chi_{i,j}\chi_{j,i}|}))^2 \leq \alpha \epsilon s
\ .\end{align}
We rewrite it as:
\begin{equation}
    nG_1(\alpha + G_2)^2 +G_4 -\alpha s \epsilon \leq 0 \label{pseudo} \ ,
\end{equation}
where:
\begin{align}
    G_1 = 7(1  +F^{-1} +s  ~\text{max} ~\frac{i_{i,j}i_{j,i}}{|\chi_{i,j}\chi_{j,i}|} )^2/\Delta_V\ , \\
    G_2 = \frac{\delta_H +\delta_I}{1  +F^{-1} +s~  \text{max} ~\frac{i_{i,j}i_{j,i}}{|\chi_{i,j}\chi_{j,i}|}}\ , \\
    G_4 =\delta_H^P +\delta_I^P +  \delta (1+F +s~\text{max} ~|\chi_{i,j}\chi_{j,i}|)
\ .\end{align}
The second inequality we need to satisfy is the range of applicability of the perturbation theory $\|PV\|/\Delta_V \leq 1/32$. We aim to solve the system of inequalities:
\begin{equation}
     \begin{cases}n(\delta_H +\delta_I +  \alpha (1  +F^{-1} +s ~ \text{max} ~\frac{i_{i,j}i_{j,i}}{|\chi_{i,j}\chi_{j,i}|} )) \leq \frac{\Delta_V(\alpha)}{32} \ ,\\
 nG_1(\alpha)\alpha^2  +(2nG_1(\alpha) G_2-s\epsilon)\alpha +nG_1(\alpha)G_2^2 +G_4  \leq 0
     \end{cases}\ ,
\end{equation}
where we indicated that $G_1$ weakly depends on $\alpha$ via $\Delta_V$.  Define $\alpha_c$ as we would the center of the interval of solutions for a quadratic inequality:
\begin{equation}
    \alpha_c = \frac{s\epsilon -2nG_1(\alpha_c) G_2}{2n G_1(\alpha_c)}
\ .\end{equation}
Assuming $\Delta_V$ and thus $G_1$ only weakly depend on $\alpha_c$, there is always a solution. Plugging that in Eq. (\ref{pseudo}), we obtain:
\begin{equation}
    \frac{(s \epsilon -2n G_1(\alpha_c) G_2)^2}{4n G_1(\alpha_c)} \geq G_4 + \frac{ 4nG_1^2(\alpha_c) G_2^2}{4n G_1(\alpha_c)} 
\ .\end{equation}
We obtain the same range of solutions for $\alpha =\alpha_c>0$ as we would for a quadratic inequality in $\alpha$ (we will omit the argument in $G_1(\alpha_c)$ and $\Delta_V(\alpha_c)$ from now on):
\begin{equation}
    s \epsilon \geq 2(nG_1 G_2 +\sqrt{nG_1 G_4 + n^2G_1^2 G_2^2}) \label{bigForm}
\ .\end{equation}
For scaling analysis, we will use a stronger condition $G_1G_4 \leq (s\epsilon)^2/12n$ and show that $G_1 G_2 \leq s\epsilon/6n$ and the above follows. 
The definitions we have so far are:
\begin{align}
    &G_1G_4 \leq (s\epsilon)^2/12n\ , \\ &G_1 G_4 =7(1  +F^{-1} +s ~ \text{max} ~\frac{i_{i,j}i_{j,i}}{|\chi_{i,j}\chi_{j,i}|})^2  (\delta_H^P +\delta_I^P +  \delta (1 +F +s~\text{max} ~|\chi_{i,j}\chi_{j,i}|))/\Delta_V \ , \\
    &G_1 G_2 = 
    7(1  +F^{-1} +s ~ \text{max} ~\frac{i_{i,j}i_{j,i}}{|\chi_{i,j}\chi_{j,i}|})(\delta_H +\delta_I)/\Delta_V
\ .\end{align}
After simplifying $\delta_H^P = \delta_H, ~ \delta_I^P = \delta_P$, 
 we note that 
 \begin{align}
     G_1 G_2 \leq 7(1  +F^{-1} +s  ~ \text{max} ~\frac{i_{i,j}i_{j,i}}{|\chi_{i,j}\chi_{j,i}|})(\delta_H +\delta_I +\delta (1 +F +s~\text{max} ~|\chi_{i,j}\chi_{j,i}|))/\Delta_V=\\ =  \frac{G_1G_4}{(1  +F^{-1} +s  ~ \text{max} ~\frac{i_{i,j}i_{j,i}}{|\chi_{i,j}\chi_{j,i}|})} \leq \frac{G_1 G_4}{s}\ ,
 \end{align}
 where we have used $i_{i,j} \leq |\chi_{i,j}| + i_m$. Using the simplified condition $G_1G_4 \leq (s\epsilon)^2/12n$, we get:
 \begin{equation}
      G_1 G_2 \leq \frac{\epsilon}{2} \cdot\frac{s\epsilon}{6n} \leq \frac{s\epsilon}{6n}\ ,
 \end{equation}
since $\epsilon<1$.  Rewriting $G_1G_4 \leq (s\epsilon)^2/12n$ explicitly:
\begin{align}
    \delta_H +\delta_I +  \delta (1 +F +s~ \text{max} ~|\chi_{i,j}\chi_{j,i}|) \leq  \frac{\Delta_V(s\epsilon)^2}{12\cdot 7n(1  +F^{-1} +s~  \text{max}~\frac{i_{i,j}i_{j,i}}{|\chi_{i,j}\chi_{j,i}|})^2}
\ .\end{align}
This gives the main result of our theorem: the noise should be small as above for the error in the quantum simulator to be bounded as $\epsilon n s$. Note that if all the parameters of the mediator are $n$-independent, that would suggest $1/n$ scaling required from the control errors. We now check the range of applicability of the perturbation theory $\|PV\| \leq \Delta_V/32$: 
\begin{equation}
    n(G_2 +  \alpha_c ) G_1(1  +F^{-1} +s ~ \text{max} ~\frac{i_{i,j}i_{j,i}}{|\chi_{i,j}\chi_{j,i}|} )^{-1}\leq \frac{7}{32} 
\ .\end{equation}
Plugging in the expression for $\alpha_c$, we get:
\begin{equation}
   s \epsilon (1  +F^{-1} +s ~ \text{max} ~\frac{i_{i,j}i_{j,i}}{|\chi_{i,j}\chi_{j,i}|} )^{-1}\leq \frac{7}{16} 
\ .\end{equation}
Using $i_{i,j} \leq |\chi_{i,j}| + i_m$ again, we arrive at $\epsilon \leq 7/16$ which is a small reduction in the range of allowed $\epsilon$.
For the optimal $\epsilon$ given by the equality in Eq. (\ref{bigForm}), the reduction factor $\alpha_x$ is:
\begin{equation}
    \alpha_x = \sqrt{\frac{ G_4}{nG_1} + G_2^2} \sim 1/n
\ .\end{equation}
The $\alpha_c$ we use is always $\geq \alpha_x$, which ensures its positivity. 

From  $G_1G_2 \leq (s\epsilon)/6n$ we know that $\alpha_c \geq s \epsilon /3nG_1$. Denote $s \epsilon /3nG_1 =\alpha_o$ which has a simpler form we will use
for the theorem:
\begin{equation}
    \alpha_o = \frac{s\epsilon \Delta_V}{3cn(1+F^{-1} + s   ~ \text{max} ~\frac{i_{i,j}i_{j,i}}{|\chi_{i,j}\chi_{j,i}|} )^2}
\ .\end{equation}
We will now show that it also satisfies the original inequality.
We will use the conditions $G_1G_4 \leq (s\epsilon)^2/12n$ and $G_1G_2 \leq (s\epsilon)/6n$. Plugging that in, we get:
\begin{equation}
   (s\epsilon)^2\left( \frac{1}{9nG_1} + \frac{1/3 -1}{3nG_1} + \frac{1}{36nG_1} + \frac{1}{12nG_1}\right)=0 \leq0
\ .\end{equation}
Finally, $\alpha_c \geq \alpha_o$ ensures it also satisfies the range of applicability. 

Suppose we want to implement a truly all-to-all Hamiltonian, which would correspond to setting $s=(n-1)/2$, with the worst-case norm of the interaction part of $H_{\text{target}}$ being $n(n-1)/2$. We still use a constant $\epsilon$ corresponding to an extensive error. Our requirement on control errors becomes:
\begin{align}
    \delta_H +\delta_I +  \delta (1 +F +(n-1)~ \text{max} ~|\chi_{i,j}\chi_{j,i}|/2) \leq  \frac{\Delta_V((n-1)\epsilon)^2}{48\cdot 7n(1  +F^{-1} +\frac{n-1}{2}~  \text{max}~\frac{i_{i,j}i_{j,i}}{|\chi_{i,j}\chi_{j,i}|})^2}
\ .\end{align}
The most notable change is the extra power of $n$ in the required scaling of $\delta$. In particular, for $n$-independent mediator parameters, $\delta$ is required to scale as $1/n^2$, which is expected since there are $\sim n^2$ terms in the interaction. The expression for $\alpha_o$ will also increase its scaling by $\times \frac{1}{n}$.

We now consider a slight variation in our derivation, that did not appear in the theorem. Let the target interaction graph have a small $s$, but the gadget is fully programmable, that is all $f_{ij}$ can be controlled. Then setting most of them to $0$ incurs an error in the same way as if $s= (n-1)/2$. The inequality on $\alpha$ will take the form:
\begin{align}
   &\delta_H^P +\delta_I^P +  \delta (1+F +(n-1) ~\text{max} ~|\chi_{i,j}\chi_{j,i}|/2) +\\&+ \frac{7n}{\Delta_V}(\delta_H +\delta_I + \delta \frac{n-1}{2}\text{max}~ i_{i,j}i_{j,i} +\alpha (1  +F^{-1} +s ~ \text{max} ~\frac{i_{i,j}i_{j,i}}{|\chi_{i,j}\chi_{j,i}|}))^2 \leq \alpha \epsilon s
\ .\end{align}

The range of applicability of the perturbation theory $2\|PV\|/\Delta_V \leq 1/16$ becomes: 
\begin{align}
    n(\delta_H +\delta_I +\delta \frac{n-1}{2}\text{max}~ i_{i,j}i_{j,i}+  \alpha (1  +F^{-1} +s ~ \text{max} ~\frac{i_{i,j}i_{j,i}}{|\chi_{i,j}\chi_{j,i}|} )) \leq \frac{\Delta_V(\alpha)}{32}
\ .\end{align}
We see that by constraining $\delta$ to be $1/n$ times the value obtained in the bound from the theorem, we can hope to satisfy these inequalities as long as the original system was satisfied. We leave the rigorous derivation and the study of examples for this case of fully tunable architecture to future work.

In our bounds on the control precision, there may be free parameters of the mediator in the values of $i_{i,j},\chi_{i,j},F,\Delta$. Ideally, an optimization of $\epsilon$ for a given set of control errors, or an optimization of control errors for a given $\epsilon$ would reveal their optimal value.

\subsection{Value of $v$ for the general theorem}
\label{plan}
We will use the inequalities of the theorem from Sec. (\ref{general}) to present an explicit choice of $v$:
\begin{align}
    \delta_H +\delta_I +  \delta (1+F +s~  \text{max} ~|\chi_{i,j}\chi_{j,i}|) \leq  \frac{\Delta_V(s\epsilon)^2}{12 \cdot 7n(1  +F^{-1} +s  ~ \text{max} ~\frac{i_{i,j}i_{j,i}}{|\chi_{i,j}\chi_{j,i}|})^2}\ , \\
    \pm V \leq v(1+H_0)
\ .\end{align}
Note that $\Delta_V \leq \Delta$, which means a solution to the above inequalities will also satisfy:
\begin{align}
    \delta_H +\delta_I +  \delta (1+F +s~  \text{max} ~|\chi_{i,j}\chi_{j,i}|) \label{delDupe}  \leq  \frac{\Delta(s\epsilon)^2}{12\cdot 7n(1  +F^{-1} +s  ~ \text{max} ~\frac{i_{i,j}i_{j,i}}{|\chi_{i,j}\chi_{j,i}|})^2}\ , \\
    \alpha_o \leq \frac{s\epsilon \Delta}{3\cdot 7n(1+F^{-1} + s  ~ \text{max} ~\frac{i_{i,j}i_{j,i}}{|\chi_{i,j}\chi_{j,i}|})^2}
\ .\end{align}
 We will now bound $v$. We need to assume that $\|\delta H_m\|_c,~ \|\delta I_m\|_c$ and $\delta_H, \delta_I$ are linearly related:
 \begin{equation}
     \|\sum_i\delta H_{m,i}\|_c \leq C_H n\delta_H,~ \|\sum_i\delta I_{m,i}\|_c \leq C_In \delta_I
 \ .\end{equation}
The interaction term will be assumed to be bounded as:
\begin{equation}
\|\sum_{i>j}|J_{ij}^*I_{i,j} I_{j,i}|\|_c \leq ns C_J^2
\ .\end{equation}
We obtain the following inequality of $V$:
\begin{align}
\| V \|_c\leq n(C_H  \delta_H +C_I  \delta_I + \alpha_o(2 +C_J^2s\text{max}\frac{1}{|\chi_{i,j} \chi_{j,i}|} ))  \leq 
 \frac{\Delta s \epsilon(\frac{1}{4}\text{max}(C_H,C_I)s\epsilon  +(2 +C_J^2s\text{max}\frac{1}{|\chi_{i,j} \chi_{j,i}|} ))}{3\cdot 7(1+F^{-1} + s  ~ \text{max} ~\frac{i_{i,j}i_{j,i}}{|\chi_{i,j}\chi_{j,i}|})^2} \leq \\ \leq
\frac{\Delta  \epsilon(\frac{1}{4}\text{max}(C_H,C_I)\epsilon  + C_J^2\text{max}\frac{1}{|\chi_{i,j} \chi_{j,i}|} )}{3\cdot 7   (\text{max} ~\frac{i_{i,j}i_{j,i}}{|\chi_{i,j}\chi_{j,i}|})^2} =v\leq\frac{\Delta}{\Delta+1}  \label{vExpl}\ ,
\end{align}
which is a valid choice of $v$ for sufficiently small $\epsilon$:
\begin{equation}
    \epsilon \leq \text{min}(\frac{7}{16}, \frac{3\cdot 7   (\text{max} ~\frac{i_{i,j}i_{j,i}}{|\chi_{i,j}\chi_{j,i}|})^2}{(\Delta+1)  (\frac{7}{64}\text{max}(C_H,C_I) + C_J^2\text{max}\frac{1}{|\chi_{i,j} \chi_{j,i}|} )})
\ .\end{equation}
The theorem will now say that for any epsilon in the range above and the choice of $v$ in Eq. (\ref{vExpl}), the gadget works as long as the inequality (\ref{delDupe}) on errors $\delta_H,\delta_I, \delta$ is satisfied.

\section{Application of the theorem} \label{sec:app}

\subsection{Definitions and results for a qubit coupler}

 This is the simplest possible case, where each  qubit of our quantum simulator is coupled to a qubit coupler: 
 \begin{equation}
    H_m=\Omega X_{qc}, \quad I_m = J Z_{qc} 
 \ .\end{equation}
 The qubit couplers are extended objects that have small mutual inductances where they overlap:
 \begin{equation}
    V_c= \sum_{i>j}f_{ij}Z_{qc,i}Z_{qc,j}
 \ .\end{equation}
 Finally, each qubit of the simulator has a local field:
 \begin{equation}
    V_q = h_i^cZ_{q,i} +t_i^c X_{q,i} 
 \ .\end{equation}
 It will turn out that $t, f, h $ are all $\sim 1/n$, while $J,\Omega $ is $O(1)$. The smallness of $f$ ensures that if the extended coupler qubit is a flux qubit, its inductance is more than enough to account for all those $f$ mutuals: imagine a long coil split into $n$ segments, and each brought in proximity with the corresponding coil of another coupler qubit. Two coils can wind together on a chip. The bare Hamiltonian is just the qubit couplers and their couplings to the qubit:
  \begin{equation}
     H_0 = \sum_i JZ_{q,i} Z_{qc,i} + \Omega X_{qc,i}
 \ .\end{equation}

 The perturbation includes the noise in the bare Hamiltonian, as well as the remaining terms. As discussed in the general proof, they can also have noise, but we'll just use upper bounds on them except for the first-order error. The full perturbation is:
 \begin{equation}
     V =V_q + V_c + \sum_i \delta J_i Z_{q,i} Z_{qc,i} + \delta\Omega_i X_{qc,i}
 \ .\end{equation}
To apply our result, we define the logical subspace $P$ in the bare Hamiltonian. Out of the 4 states of each qubit-coupler pair, we choose 2 as follows. Note that $Z_q$ is the integral of motion. For each of the two eigenvalues $\pm1$ of $Z_q$, choose the corresponding ground state. We will not use the state itself, only the projector. Let $P_0, P_1$ be the projectors onto the qubit states in the computational basis (with $\pm1$ eigenvalues of $Z_q$). The projector onto (in other words, pure state density matrix of) the corresponding coupler state is given by:
\begin{equation}
    \rho_{0,1} = \frac{1}{2}(1 - \frac{\Omega X_{qc} \pm JZ_{qc}}{\sqrt{\Omega^2 +J^2}}) \label{coState}
\ .\end{equation}
 The total subspace of interest is
 \begin{equation}
     P_0 \rho_0 + P_1 \rho_1\ ,
 \end{equation}
 for this pair, and defining
 \begin{equation}
     \Pi_i =|\tilde{0}_i\rangle \langle \tilde{0}_i| + |\tilde{1}_i\rangle \langle \tilde{1}_i| =P_{0,i} \rho_{0,i}+ P_{1,i} \rho_{1,i}\ ,
 \end{equation}
 the subspace of interest of the whole system becomes
 \begin{equation}
     P = \prod_i \Pi_i = \prod_i\sum_{m_i}|\tilde{m}_i\rangle \langle \tilde{m}_i|
 \ .\end{equation}
Here we defined the basis within the subspace by $\prod_i |\tilde{m}_i\rangle$. If we treat it as the computational basis corresponding to bitstrings $m_1\dots m_n$, we can denote the corresponding Pauli matrices acting within the subspace as $\tilde{X}, ~\tilde{Z}$. For simplicity of notation, augment them by zeros outside the subspace so that all the matrices act in the original space. The qubit $Z$ matrices coincide with $\tilde{Z}$ within the subspace $P Z = P \tilde{Z}$, so we'll use them interchangeably. The qubit $X$ is not as trivial, it contains out-of-subspace transitions, and there is a nontrivial overlap coefficient that appears in front of $\tilde{X}$ when $X$ is expressed via it, as we will see below. Note that the gap between this subspace and the rest is $\Delta = \sqrt{\Omega^2 +J^2}$. The quantities required for our lemma can be computed using Eq. (\ref{coState}):
\begin{align}
    \chi_{i,j} = \text{tr}\rho_1 Z_{qc} = \frac{J}{\sqrt{J^2 +\Omega^2}}\ ,\quad 
    F  =\sqrt{\text{tr} \rho_0 \rho_1} = \frac{\Omega}{\sqrt{J^2 +\Omega^2}} \ ,\quad 
    i_{i,j} = \|P Z_{qc}\| = 1 
\ .\end{align}

Setting the gap $\Delta = \sqrt{\Omega^2 +J^2} =1$, the Hamiltonian defined above takes the form 
\begin{equation}
    H_m = \sqrt{1-J^2}X, \quad I_m = JZ 
\ .\end{equation}
The coupling operator $I_{i,j} = Z_{qc,i}$ independent of $j$.
The list of the values required by our theorem is as follows:
\begin{equation}
    \chi_{i,j} = J,~ F =\sqrt{1-J^2},  ~ i_{i,j} = 1\label{qcProp} 
\ .\end{equation}

The errors $\delta_J \geq \|P \delta J_i Z_{q,i} Z_{qc,i}\|=|\delta J_i|$ and $\delta_H \geq \|P \delta \Omega_i X_{q,i} Z_{qc,i}\| =| \delta \Omega_i| $ are all just control errors of the 1 and 2 qubit terms, that we consider being $\delta$ just as the errors in $f,h,t$.
Plugging it all in the theorem, we get:
\begin{equation}
    \delta  \leq \text{max}_J\frac{(s\epsilon)^2 \Delta_V (3 +\sqrt{1-J^2} +sJ^2)^{-1}}{12\cdot7n(1 + \sqrt{1-J^2}^{-1} + s J^{-2} )^2} \label{deltaQ}
\ .\end{equation}
From this, we obtain the bound on $\delta$ for all values of $s\geq 1.5, ~ \Delta_V \leq \Delta=1, ~J\in[0,1],~ \epsilon \leq 7/16$:
\begin{equation}
    \delta  \leq \frac{\epsilon^2 }{12\cdot7n(1 + \sqrt{1-J^2}^{-1} + 1.5 J^{-2} )} \leq \frac{1}{3 \cdot 2^{9} n}
\ .\end{equation}
The reduction factor we use:
\begin{equation}
    \alpha_o = \frac{s\epsilon \Delta_V}{3\cdot 7n(1+\sqrt{1-J^2}^{-1} + s J^{-2})^2}\ ,
\end{equation}
is bounded as:
\begin{equation}
    \alpha_o = \frac{J^2}{3\cdot 2^{4}n(1+\sqrt{1-J^2}^{-1} + s J^{-2})} 
\ .\end{equation}
For $v$, we can just use $\|V\|$:
\begin{align}
    \|V\| \leq n(\delta_J +\delta_H) +\alpha_o (1+F^{-1} + s\frac{1}{|\chi_{i,j},\chi_{j,i}|}))  
    \leq \frac{1}{3 \cdot 2^{8} } +\frac{1}{3\cdot 2^4  } \leq 0.023
\ .\end{align}
Thus $\Delta_V =0.95 \Delta =0.95$ can be used.

Investigating Eq. (\ref{deltaQ}) numerically, we get $10^{-7}$ required control precision from this simplified formula for  $n=40,~ s=3,~ \epsilon =0.1$. Using more careful embedding and bounds discussed in the Appendix \ref{q40} improves this result by 2 orders of magnitude.

Several checks need to be made before one can use the solution above. First, note that $f_{ij} \sim \alpha \ll 1/n$, therefore each individual interaction between mediators can be realized just as a mutual inductance between the corresponding portions of their inductors (for flux qubit mediators). If we needed $f_{ij}\sim 1$, it would be unrealistic to fit so many interactions onto a single qubit. Finally, $\alpha F^{-1} \leq 1$ which is simple here since $F$ is $n$-independent.

\subsection{results for LC circuit}

We now move on to the Harmonic oscillator (LC-circuit) mediator, which would be a warmup for the transmission line. Let's first define the Hamiltonian: 
\begin{equation}
    H_m = a^\dag a , \quad I_m = J(a + a^\dag)
\ .\end{equation}
This $I_m$ results in a shift $ \tilde{a}  =a +Z J$, where  $\tilde{a}$ defines the new ground state. The coupling operator $I_{i,j} = a_i + a_i^\dag$ independent of $j$. The quantities needed for the lemma can be computed to be:
\begin{equation}
   \Delta =1,\quad \chi = 2 J, \quad i_m =1 
\ .\end{equation}
The computation of $F$ is nontrivial: in the coordinate basis corresponding to $x= (a + a^\dag)/\sqrt{2}$, the Hamiltonian becomes:
\begin{equation}
    H_m+ZI_m = \frac{x^2+p^2-1}{2} +\sqrt{2}JZx = 
    \frac{\tilde{x}^2 + p^2-1}{2} -J^2\ ,
\end{equation}
where $\tilde{x} = x+\sqrt{2}JZ$. The wavefunctions of the groundstates for the two values of $Z$ are:
\begin{equation}
    \psi_0(x,Z) = \frac{1}{\pi^{1/4}} e^{-\frac{(x + \sqrt{2}JZ)^2}{2}}
\ .\end{equation}
Their overlap is:
\begin{equation}
    F = \int_{-\infty}^\infty  \psi_0(x,1) \psi_0(x,-1) dx =e^{-2J^2} 
\ .\end{equation}

The errors in $J$ are $\delta_I$ directly, while the meaning of $\delta_H$ is more complex. We have various errors in the form $x,x^2,p,p^2, \{x,p\}$. For a single mediator, the error operator is: 
\begin{align*}
    \delta H_m =& \delta_x x + \delta_{x2} x^2 + \delta_p p + \delta_{p2} p^2 +\delta_{px}\{p,x\} = \\ =& \delta_x (\tilde{x} -\sqrt{2}JZ) + \delta_{x2} (\tilde{x}^2 - 2 \sqrt{2}\tilde{x}JZ + 2J^2) + \delta_p p + \delta_{p2} p^2 +\delta_{px} ( \{p,\tilde{x}\} - 2\sqrt{2}JZp)\ ,\\ \|P\delta H_m\| \leq&  \frac{(|\delta_x| +|\delta_{x2}|2\sqrt{2}J) +(|\delta_p| +|\delta_{px}|2\sqrt{2}J)  }{\sqrt{2}} +   \sqrt{2}|\delta_{px}|+ \frac{\sqrt{3}(|\delta_{x2}|   + |\delta_{p2}|)}{2} + |\delta_{x2}|2J^2 
 +|\delta_x|\sqrt{2}J\leq \\
    \leq& \frac{3|\delta_x|+ |\delta_p|}{\sqrt{2}}  +(2+\sqrt{2})|\delta_{px}| +\frac{ (8+\sqrt{3}) |\delta_{x2}|+\sqrt{3}|\delta_{p2}|}{2} \leq 
    \delta_H \ ,
\end{align*}
since $J\leq 1$. The theorem will work for any bound $\delta_H\geq \|P\delta H_m\|$, but for the purposes of constraining $\Delta_V$ we define one such bound $\delta_H$ as:
\begin{equation}
    \delta_H =\text{max}\left(2.2|\delta_x| +0.8|\delta_p| +  3.5|\delta_{px}|  + 4.9|\delta_{x2}|   + 0.9|\delta_{p2}|\right)  
\ .\end{equation}
We now use the above definitions to compute the contribution to the value $v =\|V\|_c$ used in $\Delta_V$,
where $\|\cdot\|_c$ is defined in Eq. \ref{vBound}. Specifically we will compute a bound $C$ on $\|\sum_i \delta H_{m,i}\|_c$, which will require us to prove:
\begin{equation}
   - C (I +H_0)\leq \sum_i \delta H_{m,i} \leq \|\sum_i C (I +H_0)
\ .\end{equation}
If each $\delta H_{m,i}$ (we omit the index $i$ when describing individual mediator with the Hamiltonian $H_m +Z I_m +E_0$ where we choose the energy shift $E_0 = J^2$ s.t. the ground state energy is exactly zero)  obeys
\begin{equation}
  -a I -b (H_m +Z I_m +E_0) \leq  \delta H_m \leq a I +b (H_m +Z I_m +E_0)\ ,
\end{equation}
then the norm of the sum obeys:
\begin{equation}
   - naI -bH_0\leq \sum_i \delta H_{m,i} \leq naI +bH_0\ ,
\end{equation}
since $H_0 = \sum_i H_{m,i} +Z_i I_{m,i} +E_0$. The identity term will almost always dominate in the bound $C=$max$(na,b)$. Still, we compute both.
First note that:
\begin{align*}
    |x| \leq \frac{1+x^2}{2} \leq 1 + H_m, \quad |p| \leq 1 + H_m\ ,\quad 
    \pm\{x,p\} \leq x^2 +p^2\leq 1+ 2H_m \ ,\quad 
     x^2 \leq 1 +2H_m,\quad  p^2 \leq 1+2H_m
\ .\end{align*} 
The absolute value on operators possesses the properties $A\leq |A|$ and $-A\leq |A|$. By transitivity if $|A| \leq B$ then $A\leq B$ and $-A\leq B$.
The full Hamiltonian of one mediator is not just $H_m =(x^2 +p^2-1)/2$, but $H_{m,i} +Z I_{m,i} +E_0 =(p^2 +(x+\sqrt{2}JZ)^2/2 = (p^2 +\tilde{x}^2-1)/2$. 
A more careful calculation follows:
\begin{align*}
    \pm x =  \pm \tilde{x} \mp\sqrt{2}JZ \leq  \sqrt{2}J + \frac{1 + \tilde{x}^2}{2} \leq  1 +\sqrt{2}J + H_m +ZI_m +E_0 \ ,\quad 
   \pm Z \tilde{x} \leq \frac{1+\tilde{x}^2}{2} \leq 1 +H_m +ZI_m +E_0 
\ .\end{align*}

The other two terms containing $x$ are bounded as follows: 
\begin{align}
    \pm\{x,p\} = \mp 2\sqrt{2} J Z p  \pm \{\tilde{x},p\} \leq  \sqrt{2}J + (\sqrt{2} J +1)(\tilde{x}^2 +p^2) =2\sqrt{2}J+1 + 2(\sqrt{2} J +1)(H_m +ZI_m +E_0)\ ,\\
    x^2 = (\tilde{x} -\sqrt{2}
    J Z)^2 = \tilde{x}^2 -2\sqrt{2}J Z\tilde{x} +2J^2  \leq 2J^2 +2\sqrt{2}J+1+(2+2 \sqrt{2}J)(H_m +ZI_m +E_0)
\ .\end{align}
This allows us to determine $a$ and $b$:
\begin{align*}
    a\leq& \left( 1 +\sqrt{2}J\right) |\delta_x|  +  |\delta_p|  +|\delta_p|^2 +\left(2J^2 + 2\sqrt{2}J +1\right)|\delta_{x2}|  +(2\sqrt{2}J+1)|\delta_{px}| \leq  \\\leq& ~ 2.5|\delta_x| + |\delta_p| + |\delta_{p2}| + 5.9|\delta_{x2}|  + 3.9|\delta_{px}|\leq1.25 \delta_H\ ,\\
    b\leq& |\delta_x| +|\delta_p| +(\sqrt{2}J+2)|\delta_{xp}| + (2 + 2\sqrt{2}J)|\delta_{x2}| +2|\delta_{p2}| \leq
     |\delta_x| +|\delta_p| +3.5|\delta_{xp}| + 4.9|\delta_{x2}| +2|\delta_{p2}|\leq 2.3\delta_H
\ .\end{align*}

Using this we find:
\begin{align}
    \|\sum_i \delta H_{m,i}\|_c  \leq\text{max}(na,b) \leq  \text{max}(1.25 \delta_H n,2.3\delta_H) \leq 1.25 \delta_H n
\ .\end{align}
If instead of $x\leq \frac{1+x^2}{2}$ we had used $x\leq \frac{\lambda+ \lambda^{-1} x^2}{2}$, then optimizing $\lambda$ would have balanced the powers of $n$ between two arguments of the maximum at least for some of the terms, but we will find that even this weak bound will be enough for our purposes.
We see that even though we did not specify $\|\delta H_m\|_c$ in the definition of $\delta_H$, a multiple of it bounds it. The same occurs for $\|\delta I_m\|_c \leq \sqrt{2}(1+\sqrt{2}J) \delta_I n \leq 3.5 \delta_I n$.
The quantities used in the theorem are:
\begin{equation}
   i_{i,j} = \chi_{i,j} + i_m  = 1+2J, \quad \frac{i_{i,j}}{\chi_{i,j}} = \frac{1}{2J} +1  
\ .\end{equation}

Plugging it in we get the following:

\begin{align}
    \delta_H +\delta_I +  \delta (1+e^{-2J^2} +s(2J)^2) \leq  \frac{(s\epsilon)^2 \Delta_V}{12\cdot7n(1  +e^{2J^2} +s  (\frac{1}{2J} +1 )^2)^2} \label{notGood}
\ .\end{align}
Since $\Delta_V \leq \Delta$, and the full range of parameters is  $s\geq 1.5,~ \epsilon \leq 7/16, ~ J\leq 1$, we establish:
\begin{equation}
    \delta_H +\delta_I \leq \frac{c}{2^6\cdot 3^5 n}
\ .\end{equation}
The theorem uses
\begin{equation}
    \alpha_o = \frac{s\epsilon \Delta_V}{3\cdot7n(1+e^{2J^2} + s (\frac{1}{2J} +1 )^2)^2}
\ .\end{equation}
Which can be bounded as
\begin{equation}
    \alpha_o \leq \frac{1}{3n2^4(\frac{1}{2J} +1 )^2(1+e^{2J^2} + s (\frac{1}{2J} +1 )^2)}
\ .\end{equation}
Now we can compute $v$: 
\begin{align}
    v \leq n\left(1.25\delta_H + 3.5\delta_I +2 \alpha_o +\frac{2s\alpha_o}{2\text{min}{\chi_{i,j}\chi_{j,i}}}\right) 
\ .\end{align}
The inequality on $v$ is obtained using the bound on $\alpha_o$ and $\delta_H +\delta_I$:
\begin{align*}
    v \leq \frac{3.5\cdot7}{2^6\cdot 3^5} +\frac{ \left(1  +\frac{s}{4J^2}\right)}{3\cdot 2^4(\frac{1}{2J} +1 )^2(1+e^{2J^2} + s (\frac{1}{2J} +1 )^2)}
\ .\end{align*}

The last term is bounded as:
\begin{equation}
    \frac{ 4\left(\frac{J^2}{s}  +\frac{1}{4}\right) }{3 \cdot 2^4 (\frac{1}{\sqrt{2J}} +\sqrt{2J} )^4} \leq \frac{\frac{2}{3} + \frac{1}{4}}{3 \cdot 2^6}\leq 0.0048
\ .\end{equation}
The upper bound on $v$ is then 
\begin{equation}
    v \leq 0.0016  +0.0048 \leq 0.0064
\ .\end{equation}
In other words, taking $\Delta_V =0.9936\Delta$ is justified in the entire region of solutions.

We see that even though qubit is a nonlinear circuit element, and the harmonic oscillator is a linear circuit element, there's only a minute difference between them in this gadget.


\subsection{The main result for the transmission line}
\label{tlmain}

Now let's consider the transmission line. Specifically, we use the open boundary conditions transmission line of $n$ sites with the qubit attached at the start. This circuit is analyzed in App. \ref{tlSec}, resulting in the following Hamiltonian:


\begin{align}
    H_{m,i} =\sum_l^n p_{ci,l}^2 + x_{ci,1}^2+ x_{ci,n}^2 +\sum_{l=2}^{n}(x_{ci,l} -x_{ci,l-1})^2 \ ,\quad 
     I_{m,i} =J  x_{ci,1} 
\ .\end{align}
The coefficient $J\leq1$. The other $n-1$ chains have a mutual inductance at inductors $2\dots n$ in order. Specifically, for the chain $i$, the coupling to chain $j$ is at a location $r_i(j)= j+1$ for $j<i$ and $r_i(j) =j$ for $j>i$. The coupling operator is:
\begin{equation}
    I_{i,j} = x_{ci,r_i(j)} - x_{ci,r_i(j)-1}
\ .\end{equation}
Unfortunately, there is no way to couple locally to $x_{ci,r_i(j)}$ only.
For the computation of the parameters needed for the theorem, see Appendix \ref{trLin}. The results are:
\begin{align}
    \Delta =2 ~ \text{sin} \frac{\pi}{2(n+1)}~, \quad \chi = \frac{J}{n+1}\ ,\quad 
    i_{i,j} \leq \sqrt{2}\ ,\quad 
        F^{-1} = \text{exp} \frac{J^2}{4(n+1)}\sum_{k=1}^{n} \frac{\cos^2 \frac{k\pi}{2(n+1)} }{\sin\frac{k\pi}{2(n+1)}}
\ .\end{align}
The expression for $F$ is bounded by:
\begin{equation}
    F^{-1} \leq e^{J^2(0.5 + \text{ln}n)/4}
\ .\end{equation}
Since $J\leq 1$, this is upper bounded by $e^{1/8} n^{1/4}$. 

The meaning of errors $\delta_H, \delta_I$ is as follows. If $\delta_{zx}$ is the error in the specification of $J$, then  $\|\delta I_m\| = |\delta_{zx}| \|P x_{ci,1}\| \leq |\delta_{zx}| (J +\sqrt{(\text{ln}n+1)/2})$ from Eq. (\ref{used1},\ref{used2}) and we define $\delta_I = \text{max}|\delta_{zx}| (J +\sqrt{(\text{ln}n+1)/2})$ (In the main text, we use $J=1$ and a simpler expression $1 +\sqrt{\text{ln}n} \geq 1 +\sqrt{(\text{ln}n+1)/2}$ for our range $n\geq 4$). The corresponding contribution to $v$ from a single mediator is:
\begin{equation}
     \pm x_{ci,1} \leq \frac{1 +J}{2} -E_0  + \frac{1}{2}(H_{m,i} +Z_iI_{m,i} +E_0)\ ,
\end{equation}
where $E_0$ is given in App. \ref{erShift}. The leading contribution from the sum is then:
\begin{align}
   \| \sum_i x_{ci,1}\|_c \leq (\frac{1 + J}{2} -E_0)n \leq (1.5 + n )n\ ,\quad  \| \sum_i \delta I_{m,i}\|_c \leq (1.5+n)n\text{max}|\delta_j| = \frac{(1.5+n)n \delta_I}{J +\sqrt{(\text{ln}n+1)/2}}  
\ .\end{align}
As for the errors $\delta H_{m,i}$, while it is straightforward to include terms like $\sum_{l} \delta_{p,il}p_{ci,l}$ and the quadratic terms as we did for the LC circuit mediator, we will consider only the terms $\delta H_{m,l}=\sum_{l}^{n-1} \delta_{x,il}(x_{ci,l} - x_{ci,l+1})$ for simplicity:
\begin{align}
    \|P\delta H_{m,i}\| \leq \sum_{l}^{n-1} |\delta_{x,il}|\|P((x_{ci,l} - x_{ci,l+1})\| \leq \sum_{l}^{n-1} |\delta_{x,il}| i_{i,l} \leq \sqrt{2}\sum_{l}^{n-1} |\delta_{x,il}|  
\ .\end{align}
We define: $\delta_H = \sqrt{2}(n-1)\text{max}|\delta_{x,il}|$
Note that any definition of control errors the bound $\delta_H$ will contain a factor of $n$: $\delta_H  = n \text{max}\|PO_{ci,l}\|$ where $O_{ci,l}$ are some local error operators (we have used $O_{ci,l} =I_{i,l} $). 
This means that $\delta_{H} = \sqrt{2} (n-1) \delta_{H,\text{loc}}$ where $ \delta_{H,\text{loc}}$ is the constant local error, the maximum of the terms $|\delta_{x,il}|$. That would be the most optimistic estimate for our purposes. Now we use:
\begin{equation}
    \pm I_{i,j} \leq \frac{1}{2} - E_0+ \frac{1}{2}(H_{m,i} +Z_iI_{m,i} +E_0)\  ,
\end{equation}
to bound the contribution to $v$ as:
\begin{equation}
    \|\sum_i \delta H_{m,i}\|_c \leq \left(\frac{1}{2} - E_0\right)\frac{n\delta_H}{\sqrt{2}} \leq \frac{n(1+n)\delta_H}{\sqrt{2}}
\ .\end{equation}
We now use the theorem:
\begin{align}
    \delta_H +\delta_I +  \delta (1 + F +J^2s(n+1)^{-2}) \leq   \frac{\Delta_V(s\epsilon)^2}{12\cdot 7n(1  +F^{-1} +2s   J^{-2}(n+1)^2)^2}
\ .\end{align}
Before computing $\Delta_V$, we note that the $J^{-2}(n+1)^2$ term is the leading term in the $n\to \infty$ limit, while $F^{-1}\leq  n^{1/4}$ is subleading. Due to this, we take possibly suboptimal $J=1$ for our final result that was reported in Sec. \ref{sec:main}. 
We will now bound $\delta_H + \delta_I$ in the whole range:
\begin{align}
    \delta_H +\delta_I  \leq   \frac{7\Delta s^2}{2^{10}\cdot 3n(1  +F^{-1}(J) +2s    J^{-2}(n+1)^2)^2} \leq 
    \frac{7\pi}{2^{12}\cdot 3n(n+1)^5J^{-4}}\leq \frac{7\pi}{2^{12}\cdot 3n(n+1)^5}
\ .\end{align}
The value of $\alpha_o$ is:
\begin{equation}
    \alpha_o =\frac{\Delta_Vs\epsilon}{3\cdot 7n(1  +F^{-1}(J) +2s   J^{-2}(n+1)^2)^2}
\ .\end{equation}
It is bounded as:
\begin{align}
    \alpha_o \leq \frac{\Delta }{3\cdot 2^{5}n    J^{-2}(n+1)^2(1  +F^{-1}(J) +3   J^{-2}(n+1)^2)}  \leq \frac{\pi J^2 }{9\cdot 2^{5}n    (n+1)^5}
\ .\end{align}
We note that
\begin{align*}
    \|\sum_{ij}\frac{J_{ij}^*}{\chi_{i,j}\chi_{j,i}} I_{i,j} I_{j,i}\|_c =2^{-1}(n+1)^2J^{-2}\|\sum_{ij}J_{ij}^*( I_{i,j}^2 + I_{j,i}^2)\|_c \leq  2^{-1}(n+1)^2J^{-2}(-E_0 n) \leq  2^{-1}(n+1)^2J^{-2}(0.5 +n)n
\ .\end{align*}
The expression for $v$ is:
\begin{align*}
    v \leq n \left( \frac{(1.5+n) \delta_I}{J +\sqrt{(\text{ln}n+1)/2}}  + \frac{(n+1)\delta_H}{\sqrt{2}}+ \alpha_o \left(2 +\frac{(n+1)^2J^{-2} (0.5 +n)n}{2n}\right)  \right)  \leq \\ \leq (n+1)n(\delta_I +\delta_H) +  \frac{\pi (2n + (n+1)^2(0.5 +n)n/2 )}{9\cdot 2^{5}n    (n+1)^5} \leq   \frac{7\pi}{2^{12}\cdot 3(n+1)^4} +  \frac{\pi (2 + 0.5(n+1)^2(0.5 +n) )}{9\cdot 2^{5}    (n+1)^5} \leq \\ \leq  
    \left(\frac{7\pi}{2^{12}\cdot 3(n+1)^3} +  \frac{\pi (2 + 0.5(n+1)^2(0.5 +n) )}{9\cdot 2^{5}    (n+1)^4}\right)\sin \frac{\pi}{2(n+1)}  \leq 0.00052 \Delta\ ,
\end{align*}
for $n\geq 4$. Since $\Delta \leq 2\sin (\pi/10)$, we obtain:
\begin{equation}
    \Delta - \Delta_V = v(1+\Delta) \leq 0.0009\Delta 
\ .\end{equation}
In other words, it is sufficient to take $\Delta_V = 0.9991\Delta$.
We also note that the correction $\Delta -\Delta_V$ has $ n^{-1} \Delta$ scaling. The resulting scaling of errors is $n^{-6}$ for most errors and $n^{-7}$ for the errors in the implementation of the transmission line $ \delta_{H,\text{loc}}$.  The full form of the constraint on $\delta$'s is:
\begin{align}
   \sqrt{2} (n-1) \delta_{H,\text{loc}} +\delta_I +  \delta (1+F +\frac{s}{n^2}) \leq 2 ~ \text{sin} \frac{\pi}{2(n+1)} \frac{0.9991(s\epsilon)^2}{12\cdot 7n(1  +F^{-1} +2s   (n+1)^2)^2}
\ .\end{align}
In Sec. \ref{sec:main} we present a stronger inequality from which the above follows:
\begin{align}
   \sqrt{2} n \delta_{H,\text{loc}} +\delta_1(1+\sqrt{\text{ln} n}) +  3\delta \leq \frac{0.011 \epsilon^2}{n(n+1)^5}
\ ,\end{align}
where $\delta_1$ is the maximum of the error terms $|\delta_{zx,i}|$. We can also use the freedom in $\Delta_V$ as one of the ways to simplify $\alpha$. Indeed, let $\Delta_V^* \leq \Delta_V$ be the new value defined as follows:
\begin{equation}
   \Delta_V^* =\frac{\pi\Delta_V}{(n+1)\Delta(\frac{J^2(1  +F^{-1}(J))}{2s(n+1)^2} +1   )^2}, \quad  \alpha_o =\frac{\Delta_V^*\epsilon J^4 }{12\cdot 7ns(n+1)^4}
\ .\end{equation}
We can then use $\Delta_V^*$ in the expression on $\delta'$s and strengthen the factor to a constant (for $n\geq 4, s\geq 1.5, J\in [0,1]$, we use
    $(\frac{J^2(1  +F^{-1}(J))}{2s(n+1)^2} +1   )^{-2}\geq 0.934$):
  \begin{align}
     \sqrt{2} n \delta_{H,\text{loc}} +\delta_1(1+\sqrt{\text{ln} n})  +  3\delta \leq \frac{0.01 \epsilon^2}{n(n+1)^5}
\ .\end{align}  
Which is the main reported result together with the simplified expression for $\alpha_o$ for $J=1$ and $\Delta_V^* = 0.934*0.9991\pi /(n+1)= 0.933 \pi /(n+1) $:
\begin{equation}
    \alpha_o =\frac{0.035\epsilon  }{ns(n+1)^5}
\ .\end{equation}

\section{Transmission line circuit description}
\label{tlSec}

  \begin{figure}
\centering
\includegraphics[width=0.6\columnwidth]{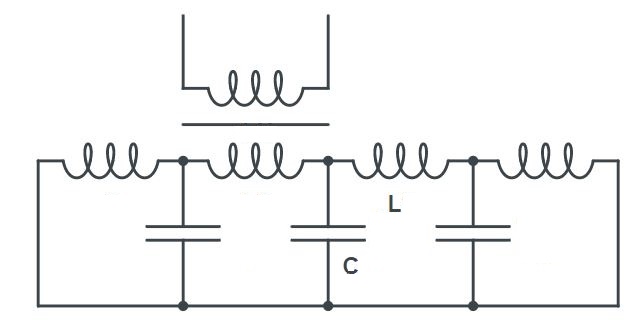}
\label{TMsketch}
\caption{A transmission line where every local inductance $L$ has a small mutual inductance $M$ with another circuit}
\end{figure}

Here we explain how a Hamiltonian of $x_i$ and $p_i$ is derived from a circuit description of the collection of transmission lines, and why coupling is to the difference $x_{i} - x_{i-1}$

The simplest model of a transmission line we use is illustrated in Fig. \ref{TMsketch}. It is made of two parallel wires, and for each portion of the length of those wires with an inductance $L$ there's a capacitance $C$ that connects the wires. The boundary conditions are that the ends of the transmission line are connected and there's one extra inductance so the circuit is reflection-symmetric.
We first investigate the mutual inductance $M$ between two inductors with inductance $L$. For currents $\mathcal{I}_u, \mathcal{I}_d$ going through them, the flux energy is given by:
\begin{equation}
    E_\Phi = \frac{L}{2} (\mathcal{I}_d^2 + \mathcal{I}_u^2)  + M \mathcal{I}_d \mathcal{I}_u
\ .\end{equation}
The flux through each of the elements is:
\begin{equation}
    \Phi_d = L \mathcal{I}_d + M \mathcal{I}_u, \quad \Phi_u = L\mathcal{I}_u + M \mathcal{I}_d 
\ .\end{equation}
Expressing the energy in terms of flux:
\begin{equation}
    E_\Phi  = \frac{L (\Phi_u^2  + \Phi_d^2)  -2M\Phi_u \Phi_d}{2(L^2 -M^2)}
\ .\end{equation}
Now let's consider the full gadget, with $j$'th transmission line having inductance $L_{j,i}$ at position $i=1..n$, and the coupling position $r_j(i)$ defined in Eq. (\ref{rDef}). The flux energy is:
\begin{equation}
E_{\Phi,\text{tot}} = \sum_{i=1}^n \sum_{j=1}^{n+1}\frac{L_{i,j}}{2} \mathcal{I}_{i,j}^2   + \sum_{i>j>1}^n M_{ij} \mathcal{I}_{i,r_i(j)} \mathcal{I}_{j,r_j(i)} + M_q \sum_{i=1}^n \mathcal{I}_{i,1} \mathcal{I}_{qi}
\ .\end{equation}
All the inductors except for the one at $j=n+1$ are coupled to something in our construction. The one at $j=0$ is coupled to a qubit  
and all the middle ones are coupled to different transmission lines by tunable couplers $M_{ij}$. 
In hardware, a tunable coupler is a complicated circuit itself, but we can consider the mutual inductance coupling to its inductance constant, and the tunability coming from the properties of its design and applied controls. We do not need to go into details of its implementation for our construction, we only observe that as long as the required $M$ is not bigger than $L$, it is feasible to realize. At the very least, one can always directly fabricate the required values without tunability allowed. Switching from currents to fluxes yields:
\begin{equation}
E_{\Phi,\text{tot}} = \sum_{i=1}^n \left(\frac{1}{2L_{i,1}} \Phi_{i,1}^2+\frac{1}{2L_{i,n+1}} \Phi_{i,n+1}^2 +\sum_{j=2}^{n}\frac{L_{i,j}}{2(L_{i,j}^2 -M_{i,j}^2)} \Phi_{i,j}^2  \right) + \sum_{i>j>1}^n \frac{-M_{i,j}}{(L_{i,j}^2 -M_{i,j}^2)} \Phi_{i,r_i(j)} \Phi_{j,r_j(i)} - \frac{M_q}{L_{i,1}} \sum_{i=1}^n \Phi_{i,1} \mathcal{I}_{qi}
\ .\end{equation}
To obtain the starting point of our construction in the main text, we need to enforce (introducing dimensionless units):
\begin{equation}
    E_{\Phi,\text{TL }j} = \sum_{i=1}^{n+1} \Phi_{j,i}^2 +\sum_{i>j>1}f_{ij} \Phi_{j,r_j(i)} \Phi_{i,r_i(j)}  + \sum_{i=1}^nj\Phi_{i,1} Z_i
\ .\end{equation}
We choose each $L_{i,j}, M_{i,j}$ as solutions of the equations:
\begin{equation}
    L_{i,1} = L_{i,n+1}=1, \quad M_q = J |\mathcal{I}_{qi}|,\quad \frac{L_{i,j}}{2(L_{i,j}^2 - M_{i,j}^2)} =1 |_{j=2\dots n} \quad  \frac{-M_{i,j}}{L_{i,j}^2 - M_{i,j}^2} =f_{ij} |_{j=2\dots n}
\ .\end{equation}
As there is a closed flux loop, we enforce $\sum_{k=1}^{n+1} \Phi_{i,k} =0$, which allows to express $\Phi_{i,n+1} = - \sum_{k=1}^{n} \Phi_{i,k} $. We next note that the voltages across the capacitors are:
\begin{equation}
    C q_{j,i} = \sum_{k=1}^i \dot{\Phi}_{j,k}
\ .\end{equation}
It is these sums that make it impossible to couple to the variable conjugate to $q$ locally. The best we can do is to couple to a difference as follows. Introducing new variables $x_{cj,l}$ such that $\Phi_{j,l} = x_{cj,l} - x_{cj,l-1}$ for $i>1$, and $\Phi_{j,1} = x_{cj,1}$, we express the sums as:
\begin{equation}
\sum_{k=1}^l \Phi_{j,k} = x_{cj,l}, \quad \Phi_{j,n+1} = -x_{cj,n}
\ .\end{equation}
We define the interaction operators $I_{i,j}  = \Phi_{i,r_i(j)} = x_{ci,r_i(j)} - x_{ci,r_i(j)-1}$. The total energy is given by:
\begin{equation}
    E_{\text{tot}} = \sum_{j=1}^n\left(\sum_{l=1}^n \frac{\dot{x}_{j,l}^2}{2C} +x_{cj,1}^2 + x_{cj,n}^2 +\sum_{l=2}^{n} (x_{cj,l}-x_{cj,l-1})^2 \right)+\sum_{i>j>1}f_{ij} I_{i,j} I_{j,i}  + \sum_{i=1}^nJx_{ci,1} Z_i
\ .\end{equation}
Finally, we find the conjugate variables $p_{i,l}$ for $x_{ci,l}$ to be $p_{i,l} = \dot{x}_{i,l}/C$. Setting $C=2$ results in the form of the Hamiltonian that was the starting point of our construction. The steps we take here appear slightly different from the common approach to circuit quantization \cite{devoret1995quantum}, but are equivalent to it.

\section{Transmission line calculation}
\label{trLin}
The Hamiltonian derived in the previous section can be split into mediators (individual transmission lines) coupled to their respective qubits, and perturbation. The terms in the unperturbed Hamiltonian are:
\begin{align}
    H_{m,i} = \sum_l^n p_{ci,l}^2 +  2x_{ci,l}^2 -2\sum_l^{n-1}x_{ci,l}x_{ci,l+1} \\
     I_m =J  x_{ci,1} 
\ .\end{align}
The perturbation includes the other $n-1$ couplings are at sites $2\dots n$ in order. Specifically, for the chain $i$, the coupling to chain $j$ is at a location $r_i(j)= j+1$ for $j<i$ and $r_i(j) =j$ for $j>i$. The coupling operator is:
\begin{equation}
    I_{i,j} = x_{ci,r_i(j)} - x_{ci,r_i(j)-1}
\ .\end{equation}
To find quantities required the theorem in Sec. \ref{general},
we will start with finding the spectrum of $H_m$. We can use the eigenvectors $v_k$ of the $n\times n$ tridiagonal matrix $T_{ij} =\delta_{i,j+1} + \delta_{i,j-1}$ (open boundary conditions) and their corresponding eigenvalues $e_k$, indexed by $k =1 \dots n$:
\begin{equation}
    v_{k,m} =\sqrt{\frac{2}{n+1}}\sin \frac{km \pi}{n+1}, \quad m =1\dots n, \quad e_k = 2 \cos \frac{k\pi}{ n+1}
\ .\end{equation}
It is normalized:
\begin{equation}
    \sum_k v_{k,j}^2 =1
\ .\end{equation}

With this we can introduce the change of variables (omitting the subscript $ci$): 
\begin{equation}
p_{ci,l} =\sum_k v_{k,l} p_k,\quad x_{ci,l} =\sum_k v_{k,l} x_k
\ .\end{equation}
It preserves the commutation relation $[p_{ci,l},x_{ci,l'}] =-i\delta_{ll'}$. The resulting Hamiltonian is:
\begin{equation}
     H_{m,i} = \sum_k^n p_k^2 + (2 - 2 \cos \frac{k\pi}{ n+1})x_k^2 +JZ_i x_k \sqrt{\frac{2}{n+1}}\sin \frac{k \pi}{n+1} \label{pkxk}
\ .\end{equation}
Or in terms of creation-annihilation operators:
\begin{align}
    p_k = i\sqrt{\omega_k/2} (a_k - a_{k}^\dag), \quad x_k = \frac{1}{\sqrt{2\omega_k}} (a_k + a_{k}^\dag)\\
    \omega_k = \sqrt{ 2(1-\cos \frac{k\pi}{ n+1})}
\ .\end{align}
Define the combination of the two transformations: 
\begin{equation}
    x_n =\sum_k g_k^n(a_k + a_k^\dag),\quad g^n_k =    \frac{1}{\sqrt{\omega_k}} \sqrt{\frac{1}{n+1}}\sin \frac{kn \pi}{n+1}
\ .\end{equation}
The linear term becomes
\begin{align}
    H_{m,i} =\sum_k \omega_k (a_k^\dag a_k +0.5) + JZ (g_k^1a_k + g_k^{1*}a_k^\dag)
\ .\end{align}

The shift in creation-annihilation operators diagonalizes the Hamiltonian
\begin{equation}
    H_{m,i}= \sum_k \omega_k (a_k + \frac{Jg_k^{1*}}{\omega_k})^\dag(a_k + \frac{Jg_k^{1*}}{\omega_k}) + \frac{\omega_k}{2} - \frac{J^2 |g_k^1|^2}{\omega_k}\ ,
\end{equation}
which suggests the magnitude of the shift in $a$:
\begin{equation}
   a_k = \tilde{a}_k - \frac{Jg_k^{1*}}{\omega_k}
\ .\end{equation}
The groundstate is annihilated by the new operators:
\begin{equation}
    \tilde{a}_k |g_{b_i}\rangle|_{b_i=1} = 0
\ .\end{equation}
 In terms of $\tilde{a}$, the relevant operators are:
\begin{align}
    H_{m,i} +E_0= \sum_k \omega_k\tilde{a}_k^\dag \tilde{a}_k, \quad \omega_k = 2\sin \frac{k\pi}{ 2(n+1)}, \quad E_0 = \sum_k -\frac{\omega_k}{2} + \frac{J^2 |g_k^1|^2}{\omega_k}\\
    I_{i,j}  =x_{ci,r_i(j)} - x_{ci,r_i(j)-1}, \quad x_{ci,r}  =\sum_{k=1}^{n} g_k^{r}a_k + g_k^{r*}a_k^\dag -\frac{J}{\omega_k}(g_k^{r}g_k^{1*} + g_k^{r*}g_k^1)
\ .\end{align}

In the following sections, we will compute the energy shift $E_0$ as well as the following:
\begin{align} 
 \chi_{i,j}=\langle g_{b_i}| I_{i,j}|g_{b_i}\rangle|_{b_i =1}\ ,\quad 
\|PI_{i,j}\| \leq i_{i,j}  \ ,\quad 
      F = \langle g_{b_i=1}|g_{b_i=-1}\rangle
\ .\end{align}
\subsection{Energy shift}
\label{erShift}
We need to know the exact energy shift such that the ground state is at zero energy. The sums involved in $E_0$ are:
\begin{align}
     -E_0 = \sum_k\frac{\omega_k}{2} - J^2\frac{ |g_k^1|^2}{\omega_k} = \sum_k \sin \frac{k\pi}{ 2(n+1)} - J^2\frac{ \sin^2 \frac{k\pi}{n+1} }{4(n+1)\sin^2 \frac{k\pi}{ 2(n+1)}} =\\= \frac{1}{2}\left( \frac{ \cos  \frac{ \pi}{4(n+1)}}{\sin \frac{ \pi}{4(n+1)}} -1 \right)- \frac{J^2}{n+1} \sum_k \cos^2\frac{k\pi}{ 2(n+1)} = \frac{1}{2}\left( \frac{ \cos  \frac{ \pi}{4(n+1)}}{\sin \frac{ \pi}{4(n+1)}} -1 \right) - \frac{J^2n}{2(n+1)} 
\ .\end{align}
This expression is $\sim n$ due to the first term. It is bounded as:
\begin{equation}
    -E_0 \leq n + \frac{1 -J^2(1+n^{-1})^{-1}}{2}
\ .\end{equation}
\subsection{Susceptibility}
For the calculation of $\chi_{ij}$ we'll need to find expectation values of $x_{ci,r}$ at various positions $r$:
\begin{align*}
    \langle g_{b_i}| x_{ci,r} |g_{b_i}\rangle|_{b_i=1} = -\sum_k  \langle g_{b_i}|\frac{J}{\omega_k}(g_k^{r}g_k^{1*} + g_k^{r*}g_k^1) |g_{b_i}\rangle|_{b_i =1}=
    -\sum_k  \frac{J(g_k^rg_k^{1*} + g_k^{r*}g_k^{1}) }{\omega_k} = J \zeta_r
\ .\end{align*}
In the limit of $J \to 0$, the expression $\zeta_r$ is the linear response susceptibility. It can be shown that the Kubo formula gives exactly this expression, in other words, the linear response here coincides with the exact response. In our notation, $\chi_{i,j}$ includes $J$ as a factor, and is the difference:
\begin{equation}
   \chi_{i,j} = J (\zeta_{r_i(j)} -\zeta_{r_i(j)-1}) 
\ .\end{equation}
More generally, define
\begin{equation}
    \zeta_{r,s} = -\sum_k  \frac{(g_k^rg_k^{s*} + g_k^{r*}g_k^{s}) }{\omega_k} \ ,
\end{equation}
such that $\zeta_r = \zeta_{r,1}$.
Plugging in the expressions for $g$ and $\omega$:
\begin{align*}
    \zeta_{r,s}=  -\frac{2}{n+1}\sum_{k=1}^{n}  \frac{\sin \frac{kr \pi}{n+1}\sin \frac{ks \pi}{n+1}}{ 2 (1 - \cos \frac{\pi k}{n+1})} 
\ .\end{align*}

The susceptibility can be computed using the change of variables $z =$exp$(k\pi/2(n+1))$ and the formula for geometric series: $1 -z^{a+1} = (1-z)\sum_{i=0}^a z^i $. The result is:
\begin{equation}
    \zeta_{r,s} = -\frac{1}{n+1} ((n+1)\text{min}(r,s) -rs )
\ .\end{equation}

For $\chi_{i,j}$ we get, independent of $j$:
\begin{equation}
    \chi_{i,j} = J(\zeta_{1,r_i(j)} - \zeta_{1,r_i(j)-1} ) = J/(n+1)
\ .\end{equation}
This concludes the calculation of $\chi$. Note that the response of the variable $x_{ci,1}$ corresponding to the flux in the inductor connected to the qubit is
\begin{equation}
     \langle g_{b_i}| x_{ci,1} |g_{b_i}\rangle|_{b_i=1}  =J\zeta_1  = -\frac{Jn}{n+1} \label{used1}
\ ,\end{equation}
and the response in the last inductor is
\begin{equation}
    - \langle g_{b_i}| x_{ci,n} |g_{b_i}\rangle|_{b_i=1}  =-J\zeta_n  = \frac{J}{n+1}
\ .\end{equation}
Flux cancellation is satisfied as expected of the definition of $x$: $ \langle g_{b_i}| x_{ci,1} |g_{b_i}\rangle|_{b_i=1} - \langle g_{b_i}| x_{ci,n} |g_{b_i}\rangle|_{b_i=1} + \sum_{j=2}^n\chi_{i,j} =0$
The fluxes in the unperturbed inductors spread evenly to cancel the perturbation-induced flux in the first inductor.

\subsection{Projected coupling}
We will choose $ i_{i,j} = |\chi_{i,j}| +i_m$  that upper bounds $|\chi_{i,j}| + \|PI_{i,j}Q\|$. Below we will show that the bound on any individual $\|P x_{ci,r}Q\|$ is:
\begin{equation}
  \|P x_{ci,r}Q\| \leq \sqrt{\sum_k {g_{k}^{r2}}} \leq \sqrt{ \frac{1}{\pi} (1+ \text{ln} n) + \left( \frac{\pi}{2} -1\right)\frac{n}{2\pi^2(n+1)}  }
\ .\end{equation}
Recall that:
\begin{align}
  \omega_k = 2\sin \frac{k\pi}{ 2(n+1)} \ ,\quad 
   g^n_k =    \frac{1}{\sqrt{\omega_k}} \sqrt{\frac{1}{n+1}}\sin \frac{kn \pi}{n+1}
\ .\end{align}
Plugging that in:
\begin{align}
  \sqrt{\sum_k |g_k^{r}|^2}
=
    \sqrt{\frac{1}{n+1}\sum_k \frac{\sin^2 kr\pi/(n+1)}{\omega_k}} \ ,
\end{align}
and:
\begin{equation}
    \sum_k g_k^{r 2} = \frac{1}{2(n+1)}\sum_k \frac{\sin^2 kr\pi/(n+1)}{\sin k\pi/2(n+1)}
\ .\end{equation}
A simple way to upper bound this is to use $\sin^2x \leq 1$ in the numerator and $\sin x \leq 2x/\pi$ in the denominator. This costs a total factor of $\pi$ in the leading term compared to the true scaling.
\begin{equation}
     \sum_k g_k^{r 2} \leq \frac{1}{2}\sum_{k=1}^n \frac{1}{k} \leq \frac{1}{2}
     (1+\text{ln}n) \label{used2}
\ .\end{equation}
A tighter bound with the correct leading term can be obtained by using $1/\sin x \leq (1+(\frac{\pi}{2}-1)(\frac{2x}{\pi})^2)/x$ for the denominator:
\begin{align}
     \sum_k g_k^{r 2} \leq \sum_{k=1}^n \frac{1}{\pi k} (1 +(\frac{\pi}{2}-1)(\frac{2}{\pi})^2) (\frac{\pi k}{2(n+1)})^2 ) \leq \frac{1}{\pi}
     (1+\text{ln}n) +(\frac{\pi}{2}-1)(\frac{2}{\pi})^2\frac{ n}{8(n+1)}
\ .\end{align}

We will not use the above bound directly, as the difference can be bounded better:
\begin{align}
    \|P (x_{ci,r}- x_{ci,r-1})Q\| \leq \sqrt{\sum_k (g_{k}^{r} - g_{k}^{r-1})^2 } \leq 
    \sqrt{\frac{1}{n+1}\sum_k\frac{(\sin \frac{kr \pi}{n+1} -\sin \frac{k(r-1) \pi}{n+1})^2}{2\sin \frac{k\pi}{2(n+1)}}}
\ .\end{align}
In the numerator, we get $4\cos^2 \frac{k(r-1/2) \pi}{n+1} \sin^2 \frac{k \pi}{2(n+1)} $, so the $\sin$ cancels with the denominator. The range of $r,n$ is $n\geq 2,~ 2\leq r\leq n$. We compute $n=2$ separately and use $n\geq 3$ to derive the bound for all the other values. We will use the following:
\begin{equation}
   \sum_{k=1}^{n} \sin\frac{k\pi}{2(n+1)} 
 =\frac{1}{2}\left( \frac{ \cos  \frac{ \pi}{4(n+1)}}{\sin \frac{ \pi}{4(n+1)}} -1 \right)
\ .\end{equation}
Plugging that in:
\begin{align}
    \|P (x_{ci,r}- x_{ci,r-1})Q\| \leq \sqrt{\frac{2}{n+1}\sum_k \cos^2 \frac{k(r-1/2) \pi}{n+1} \sin \frac{k \pi}{2(n+1)}}\leq \sqrt{\frac{2}{n+1}\sum_k  \sin \frac{k \pi}{2(n+1)}} \\=\sqrt{\frac{1}{n+1}\left( \frac{ \cos  \frac{ \pi}{4(n+1)}}{\sin \frac{ \pi}{4(n+1)}} -1 \right)}   \leq \sqrt{\frac{1}{n+1} \frac{1}{\eta \frac{ \pi}{4(n+1)}}} \leq \sqrt{\frac{4}{\eta \pi}} \leq 1.14\ ,
\end{align}
where and $\eta =\frac{\sin \pi/16}{\pi/16}\approx 0.9936$ is a factor we can use for $n\geq 3$. The bound $i_{i,j}$ is:
\begin{equation}
   \frac{J}{n+1} +  \|P (x_{ci,r}- x_{ci,r-1})Q\| \leq \sqrt{2} =i_{i,j}
\ ,\end{equation}
which can be shown for all $n,r$ in the allowed range: for $n\geq 3$, $J< 1$ we can use $1/4 + 1.14 \leq \sqrt{2}$, while for $n=2$ we just compute $\sqrt{\frac{2}{n+1}\sum_k  \sin \frac{k \pi}{2(n+1)}}  = \sqrt{\frac{1}{3}(\frac12 +\frac{\sqrt{3}}{2})} <0.7  $ and $0.7+\frac13 \leq \sqrt{2}$. which concludes the calculation of $i_{i,j}$.

\subsection{Overlap}
Using the Hamiltonian (Eq. (\ref{pkxk})) of individual transmission line written in terms of $p_k,x_k$, as well as $\omega_k, g_k^1$:
\begin{equation}
     H_{m,i} = \sum_k^n p_k^2 + \omega_k^2 x_k^2 +JZ_i x_k \sqrt{2\omega_k} g_k^1\ ,
\end{equation}
the overlap is defined as:
\begin{equation}
    F =\langle g_{b_i=1}|g_{b_i=-1}\rangle  = \prod F_k\ ,
\end{equation}
where $|g_{b_i}\rangle$ is the ground state of transmission line $H_{m,i}$ with $Z_i =b_i$ treated as a number. $F_k$ is defined as the overlap of $Z_i=\pm1$ ground states of each individual term in the sum over $k$:
\begin{align}
     H_{k\pm} =  p_k^2 +  \omega_k^2x_k^2 \pm J\sqrt{2\omega_k}  g_k^1 x_k = p_k^2 + \omega_k^2\left(x_k \pm \frac{J g_k^1}{\omega_k^{3/2}\sqrt{2}}\right)^2 + \text{const}
\ .\end{align}
The ground state wavefunctions are:
\begin{equation}
    \psi_{k\pm} (x) =\left( \frac{\omega_k}{\pi}\right)^{\frac{1}{4}} \text{exp} (- \omega_k (x\pm x_{0k})^2/2) = \frac{\text{exp} (- \omega_k (x\pm x_{0k})^2/2)}{\left( \int_{-\infty}^\infty dx \text{exp} (- \omega_k x^2)\right)^{\frac{1}{2}}}\ ,
\end{equation}
where $x_{0k} = \frac{J g_k^1}{\sqrt{2}\omega_k^{3/2}}$. The overlap is then:
\begin{equation}
    F_k = \frac{\int e^{-\omega_k((x-x_{0k})^2 +(x+x_{0k})^2)/2}dx}{\int e^{- \omega_k x^2}dx} = e^{-\omega_k x_{0k}^2} =\text{exp}  -\frac{J^2 |g_k^{1}|^2}{2\omega_k^{2}}\ ,
\end{equation}
and the total overlap:
\begin{equation}
    F =\text{exp}  -\frac{J^2}{2}\sum_k\frac{ |g_k^1|^{2}}{\omega_k^{2}}
\ .\end{equation}
The sum that appears in the overlap:
\begin{equation}
    \sum_k \frac{|g_k^{1 }|^2}{\omega_k^2} =\frac{1}{8(n+1)}\sum_{k=1}^{n} \frac{\sin^2 \frac{k\pi}{n+1} }{\sin^3\frac{k\pi}{2(n+1)}} = \frac{1}{2(n+1)}\sum_{k=1}^{n} \frac{\cos^2 \frac{k\pi}{2(n+1)} }{\sin\frac{k\pi}{2(n+1)}} =   \frac{1}{2(n+1)}\sum_{k=1}^{n} \frac{1 }{\sin\frac{k\pi}{2(n+1)}}  - \sin\frac{k\pi}{2(n+1)}
\ .\end{equation}
We have already used the second sum in the previous section:
\begin{equation}
   \frac{2(n+1)\cos\frac{\pi}{16} }{\pi} - 1/2\leq\sum_{k=1}^{n} \sin\frac{k\pi}{2(n+1)} 
 =\frac{1}{2}\left( \frac{ \cos  \frac{ \pi}{4(n+1)}}{\sin \frac{ \pi}{4(n+1)}} -1 \right) \leq \frac{2(n+1)}{\eta \pi}\ ,
\end{equation}
again for $n\geq 3$ and $\eta =\frac{\sin \pi/16}{\pi/16}\approx 0.9936$. $n=2$ case can be checked independently. The remaining sum can be bounded as follows:
\begin{equation}
  (n+1) \frac{2}{\pi}\text{ln} n\leq \sum_{k=1}^{n} \frac{1 }{\frac{k\pi}{2(n+1)}}  \leq \sum_{k=1}^{n} \frac{1 }{\sin\frac{k\pi}{2(n+1)}} \leq   \sum_{k=1}^{n} \frac{1 }{\frac{2}{\pi}\frac{k\pi}{2(n+1)}} \leq (n+1)(1+\text{ln}n)
\ .\end{equation}
The bound on $\sum_k \frac{|g_k^{1 }|^2}{\omega_k^2}$ is:
\begin{equation}
    \frac{1}{\pi}\text{ln} n - \frac{1}{\eta \pi}\leq \sum_k \frac{|g_k^{1 }|^2}{\omega_k^2} \leq \frac{1}{2}(1+\text{ln}n) -  \frac{\cos\frac{\pi}{16} }{\pi} + \frac{1}{4(n+1)}
\ .\end{equation}
For $n=2$ we find $\sum_k \frac{|g_k^{1 }|^2}{\omega_k^2} \approx0.298$ and the above is satisfied as $-0.1 \leq 0.298 \leq 0.62$. We can slightly weaken the bounds to simplify the expressions:
\begin{equation}
    \frac{1}{\pi}\text{ln} n - \frac{1}{3}\leq \sum_k \frac{|g_k^{1 }|^2}{\omega_k^2} \leq \frac{1}{2}\text{ln}n  +\frac{1}{4}\ ,
\end{equation}
where we confirmed the inequality directly for $n=2,3$, and it is weaker than the previous version for $n\geq 4$.

For the theorem, we will need the inverse overlap $F^{-1}$. Its exact value is:
\begin{equation}
    F^{-1} = \text{exp} \frac{J^2}{4(n+1)}\sum_{k=1}^{n} \frac{1 }{\sin\frac{k\pi}{2(n+1)}}  - \sin\frac{k\pi}{2(n+1)}\ ,
\end{equation}
and the bounds are:
\begin{equation}
  n^{\frac{J^2}{2\pi}} e^{- \frac{J^2}{6}}\leq F^{-1}\leq n^{ \frac{J^2}{4}}e^{\frac{J^2}{8}}
\ .\end{equation}
Since $J\leq 1$, the power of $n$ is at most $\frac{1}{4}$. This concludes the calculation of the overlap.

\section{40 qubit example, full calculation}
\label{q40}
  \begin{figure}
\centering
\includegraphics[width=0.6\columnwidth]{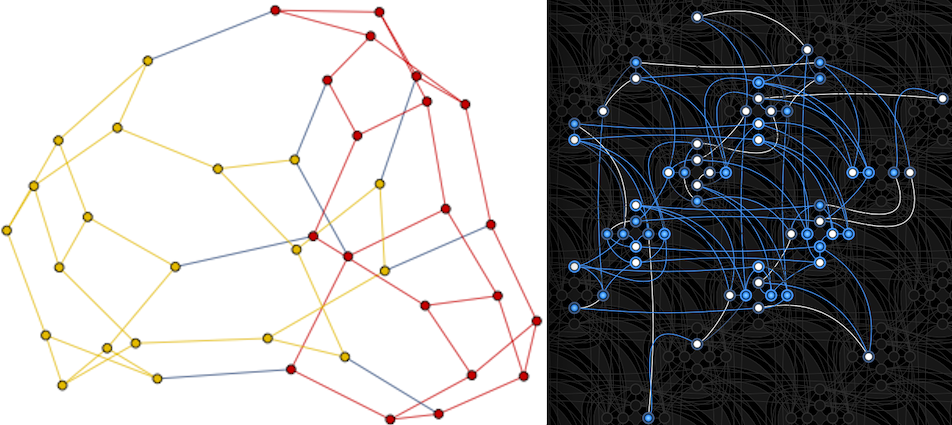}
\label{40g}
\caption{Left: the specific instance of $n=40,~ s=4$ random graph that we study. The colors represent a partition into 20 qubit subsystems that has the least interactions between subsystems. Right: An embedding of this problem into D-Wave Pegasus architecture. White links correspond to the interactions between qubits and their mediators. Blue links encode the problem interactions.}
\end{figure}

We would like to repeat the calculation one more time for a specific instance of $n=40$, degree $2s=4$ that requires a minimum of $16$ interactions to be cut to split the graph into two $20$ qubit partitions. The problem graph and the corresponding gadget with single qubit couplers as mediators are illustrated in Fig. \ref{40g}. Since the allowed Hamiltonian error is $\epsilon n s$, neglecting the $16$ interactions of strength $\leq 1$ between subsystems would correspond to $\epsilon =0.1$ of error. We take this value as the target precision for our gadget and check what quantum hardware capabilities are needed for the operation at that target precision. We assume that a Pegasus graph is used \cite{boothby2020next}. An embedding (see Fig. \ref{40g}, right) exists where $18$ pairs of hardware qubits are used to represent $18$ logical qubits, and the remaining $22$ are represented by a single hardware qubit. One of each of those pairs will be the coupler in our construction. We would also need to count the required qubit-coupler and coupler-coupler interactions (the qubit coupler is different in that it has $f_{ij} \sim 1/\chi$ instead of $1/\chi^2$). We choose the coupler out of the pair to minimize the number of those interactions and maximize the direct ones. In fact, it is possible to choose the couplers so that no coupler-coupler interactions are needed. The minimal number of qubit-coupler interactions is $31$. We also note that the degree of the Pegasus graph is $15$, and each interaction has some control error, even if we are not using them. We will assume that the effect of the unused interactions is incorporated into the local field error. 

We note that classically one can neglect the 16 interactions and simulate each 20 qubit subsystem via an exact diagonalization. 
There is, however, no out-of-the-box algorithm for $\epsilon <0.1$. Of course, the classical state of the art is much better than that, and it is not our goal to review it here. We still consider beating the out-of-the-box classical algorithm an important milestone towards being competitive with the state of the art in the future. 

The Hamiltonian of the $n=40,~ 2s=4$ gadget for this graph is as follows. Some qubits of the gadget are coupled to a qubit coupler.
The bare Hamiltonian is just the qubit couplers and their couplings to the qubit:
  \begin{equation}
     H_0 = \sum_{i\in vq} JZ_{q,i} Z_{qc,i} + \sqrt{1-J^2} X_{qc,i}, \quad \Delta =1
 \ .\end{equation}
Here $vq$ is the set of indices of qubits with a coupler.
The qubit couplers are extended objects that are coupled to one or a few other qubits directly. Let's denote the set of these couplings $v$:
 \begin{equation}
    V_c= \sum_{i,j\in v}f_{ij}Z_{qc,i}Z_{q,j}
 \ .\end{equation}
Each qubit of the simulator has a local field, and some of the qubits are coupled directly. Let's denote the set of direct couplings $vd$:
 \begin{equation}
    V_q = \sum_i h_i^cZ_{q,i} +t_i^c X_{q,i} + \sum_{i,j\in vd}J_{ij}^cZ_{q,i} Z_{q,j}
 \ .\end{equation}
 Note that $vq$ is the set of the unique first elements of $v$. 
 The perturbation also includes the noise of strength $\delta$ in the bare Hamiltonian. Note that the terms that are part of the perturbation are also considered noisy with the same strength $\delta$, but we assume that their noisy values don't exceed the range of their ideal values. The total perturbation is:
 \begin{equation}
     V =V_q + V_c + \sum_{i\in vq}  \delta_{Zi}Z_{q,i} Z_{qc,i} + \delta_{Xi}  X_{qc,i}, \quad |\delta_{Zi}|,~ |\delta_{Xi}| \leq \delta
 \ .\end{equation}
Compare this with the general definition of precision used previously:
\begin{align}
    \|P\delta H_m\| \leq \delta_H , ~ \|P \delta I_m\| \leq \delta_I, \quad \delta H_m = \delta_X \cdot X_{qc}, ~ \delta_I  =\delta_Z \cdot Z_{qc}
\ .\end{align}
Note that for any unitary $U$ and any projector $P$ the norm $\|PU\|=1$. In particular, $\|PX_{q}\|=\|PZ_q\| = \|PZ_{qc}Z_q\| =\|PX_{qc}\|=1.$ Thus $\delta_H, \delta_J=\delta$ for this system.
The properties of the qubit coupler have been found before in Eq. (\ref{qcProp}):
\begin{equation}
    \chi_{i,j}  = J,~ F =\sqrt{1-J^2}, ~ i_{i,j}=1
\ .\end{equation}
The required $f_{ij} = \alpha J_{i,j}/ \chi_{i,j}, ~ J_{ij}^c = \alpha J_{ij}, ~ h_{i}^c =\alpha h_i, ~ t_i^c = \alpha F^{-1} t_i$ for qubits with a coupler or $\alpha t_i$ for qubits without a coupler.
The proof of the general theorem allows for a minor modification: instead of every qubit coupling to $s$ others, we now have $|vq|$ qubits out of the total $n$ coupling to $s=1$ others through the mediator, and $|vd|$ direct couplings are included in the perturbation.

The bound on the error $r =H_{\text{targ}}-H_{\text{eff}}$ is given by:
\begin{align}
    \|r\| \leq & \delta \left(\sum_i  \| PZ_iP\| + \|PX_iP\| + \sum_{i\in vq} \|PZ_{q,i} Z_{qc,i}P\| +    \|PX_{qc,i}P\| +\right.\\ &\left. + \sum_{i,j\in vd}\|P Z_{q,i} Z_{q,j}P\|+\sum_{ij\in vq} \|PZ_{qc,i} Z_{q,j}P\| \right) +   2c\frac{\|PV\|^2}{\Delta -(1+\Delta) \|V\|}\label{genErr2}
\ .\end{align}
In this section, we use the value $c=3.5$ unless otherwise specified. It would be convenient to renormalize $\frac{c}{\Delta - (1+\Delta)\|V\|} \leq c^*$. We will try $c^* = 1.01 c$, and then check if the resulting $\|V\| = 2\delta |vq| + (n+ |vq|F^{-1} +|vd|+|v|J^{-1} +sn -|v|)\alpha$ satisfies the inequality.

Using
\begin{equation*}
    \|PX_{q\in vq} P\| =\sqrt{1-J^2} =F, \quad\|PZ_{qc} P\| =J, \quad\|PX_{qc} P\| =\sqrt{1-J^2}, \quad \|PZ_{q} P\| =\|PX_{q\in vd} P\|=1, \quad \|PZ_{qc}Z_{qc} P\| =J
\ .\end{equation*}
The first term is just:
\begin{equation}
    (J+F)\delta |vq| +(n +|vq|F + |vd|+|v|J+(ns -|v|))\delta
\ .\end{equation}
The $\|PV\|$ contains:
\begin{align}
    \|PV\| \leq \delta \sum_{i\in vq}(\|PZ_{q,i} Z_{qc,i}\| +  \|PX_{qc,i}\|) +  \sum_i |h_i|  +|t_i|  +\sum_{i,j\in vd}\|PJ_{i,j}Z_{q,i} Z_{q,j}\|+\sum_{ij\in vq} \|P f_{ij} Z_{qc,i} Z_{q,j}\|  \\
    \leq 2\delta|vq|  + \alpha (n  +|vq|F^{-1} + |vd| +\frac{|v|}{J} + (sn-|v|))
\ .\end{align}

Plugging the expressions above into Eq. (\ref{genErr2}), we get a quadratic inequality on $\alpha$:
\begin{align}
   (J+F)\delta |vq| +(n +|vq|F + |vd|+|v|J+(ns -|v|))\delta + 2c^*(2\delta|vq|  + \alpha (n  +|vq|F^{-1} + |vd| +\frac{|v|}{J} + (sn-|v|)))^2\leq \alpha \epsilon ns
\ .\end{align}
We arrive at the same form as in the proof:
\begin{equation}
    nG_1(\alpha + G_2)^2 +G_4 -\alpha s \epsilon \leq 0 \ ,
\end{equation}
where now:
\begin{align}
    G_1 =2 c^* (n  +|vq|F^{-1} + |vd| +\frac{|v|}{J} + (sn-|v|))^2/n^2 \\
    G_2 = \frac{2\delta |vq|}{(n  +|vq|F^{-1} + |vd| +\frac{|v|}{J} + (sn-|v|))} \\
   G_4 =  (J+F)\delta \frac{|vq|}{n} +(1 +\frac{|vq|F + |vd|+|v|J+(ns -|v|)}{n})\delta 
\ .\end{align}
 For the $n=40,~ 2s=4$ problem we study here the values $|vq| =18, ~ |vd| =n- |vq| =22, ~ |v| =31$.

 We use the full expression from the proof of the theorem:
\begin{equation}
    s \epsilon = \text{min} 2(nG_1 G_2 +\sqrt{nG_1 G_4 + n^2G_1^2 G_2^2})
\ .\end{equation}
For $\epsilon =0.1$, this is satisfied for $J=0.76,~ \delta =4.1\cdot10^{-7}$. We also check that for this value of $\delta$ the chosen $c^* =1.01 c$ satisfies $\frac{c}{\Delta - (1+\Delta)\|V\|} =1.006 c  \leq c^*$, and that $x =2\|PV\|/(\Delta -(1+\Delta)\|V\|)=0.0064< 1/16$.

 Now we will show that a bigger precision is also allowed, using the version of the theorem that contains $\|PVQ\|(\|V-  QVQ\|)$ instead of $2\|PV\|$. For that, we need to calculate $\|V- QVQ\|$ and $\|PVQ\|$. For each system of a qubit and its qubit coupler, $P$ and $Q$ are given by
\begin{equation*}
   P = P_0 \rho_0 + P_1 \rho_1,\quad Q =  P_0 q_0 + P_1 q_1, \quad \rho_{0,1} = \frac{1}{2}(1 -\sqrt{1-J^2} X_{qc} \pm JZ_{qc}), \quad  q_{0,1} = 1 - \rho_{0,1} = \frac{1}{2}(1 +\sqrt{1-J^2} X_{qc} \mp JZ_{qc})
 \ .\end{equation*}
 The terms in $V$ are $X_q,Z_q, X_{qc}, Z_{qc}Z_q, Z_qZ_q$, and the norms of the required projections (obtained in Mathematica \cite{urlCode}) are:
 \begin{align}
     &\|PX_{q\in vq} Q\| = \|\rho_0 q_1\| = J, \quad \|X_q - QX_qQ\| =a_J =\sqrt{\frac{1}{2} (1+ J^2 + \sqrt{1+2J^2 -3J^4})} \\
    & \|PZ_q Q\| = 0, \quad \|Z_q - QZ_qQ\| =1 \\
     &\|PX_{qc} Q\| =  J, \quad \|X_{qc} - QX_{qc}Q\| = b_J =\frac{1}{2}(\sqrt{1-J^2} + \sqrt{1+3J^2}) \\
      q,qc\in v: \quad &\|PZ_{qc} Z_q Q\| =  (1-J^2)^{\frac{1}{4}}, \quad \|Z_{qc}Z_q - QZ_{qc}Z_qQ\| =c_J = \frac{1}{2}(J + \sqrt{4-3J^2})
 \ .\end{align}
With these, the norms $\|PVQ\|$ and $\|V- QVQ\|$ take the form:
\begin{align}
    \|PVQ\|  \leq \delta |vq| ((1-J^2)^{\frac{1}{4}} + J) + \alpha |vq| \frac{J}{\sqrt{1-J^2}}    \\ \|V - QVQ\| \leq \delta |vq| (c_J + b_J)  +\alpha(n +|vq|\frac{a_J}{\sqrt{1-J^2}} + |vd| + \frac{|v|}{J} + (sn-|v|))
\ .\end{align}
The inequality on $\alpha$ is:
\begin{align*}
   (J+F)\delta |vq| +(n +|vq|F + |vd|+|v|J+(ns -|v|))\delta +\\ c^*(\delta |vq| ((1-J^2)^{\frac{1}{4}} + J) + \alpha |vq| \frac{J}{\sqrt{1-J^2}})(\delta |vq| (c_J + b_J)  +\alpha(n +|vq|\frac{a_J}{\sqrt{1-J^2}} + |vd| + \frac{|v|}{J} + (sn-|v|)))\leq \alpha \epsilon ns\ ,
\end{align*}
where we now used the exact adjustment of the gap $c^* =c/(1-2\|V\|)$. We will collect it into the form:
 \begin{equation}
 G_1(\alpha +G_2)(\alpha+G_3) +G_4 -\alpha sn\epsilon \leq 0\ ,
\end{equation}
where
\begin{align}
    G_1 = c^* |vq| \frac{J}{\sqrt{1-J^2}} (n +|vq|\frac{a_J}{\sqrt{1-J^2}} + |vd| + \frac{|v|}{J} + (sn-|v|)) \\
    G_2 = \delta |vq| ((1-J^2)^{\frac{1}{4}} + J) \frac{\sqrt{1-J^2} }{ |vq|J} \\
G_3 =\frac{\delta |vq| (c_J + b_J)  }{n +|vq|\frac{a_J}{\sqrt{1-J^2}} + |vd| + \frac{|v|}{J} + (sn-|v|)}
    \\
   G_4 =  (J+F)\delta |vq| +(n +|vq|F + |vd|+|v|J+(ns -|v|))\delta 
\ .\end{align}
The quadratic polynomial in $\alpha$ has coefficients:
\begin{equation}
    G_1 \alpha^2 + ( G_1 (G_2+G_3) - sn\epsilon)\alpha + G_4 +G_1G_2 G_3 \leq 0
\ .\end{equation}
Searching for the zero of:
\begin{equation}
   ( G_1 (G_2+G_3) - sn\epsilon)^2 -4 G_1G_4 -4G_1^2G_2 G_3 = 0\ ,
\end{equation}
unfortunately, leads to $x = \|V- QvQ\|/(\Delta -(1+\Delta)\|V\|)>1/16$, so instead we set $x=1/16$ and seek the numbers $J,\delta, \alpha$ that satisfy both $x=1/16$ and $ G_1 \alpha^2 + ( G_1 (G_2+G_3) - sn\epsilon)\alpha + G_4 +G_1G_2 G_3=0$. The largest $\delta$ we were able to find numerically is $0.9\cdot10^{-5}$, with $\alpha =0.000256,~ J =0.37$.

Let's try to repeat the calculation using the finite-dimensional Lemma.

The inequality on $\alpha$ now contains $\|V\| = 2\delta |vq| + \alpha(n+ |vq|F^{-1} +|vd|+|v|J^{-1} +sn -|v|)$ as follows:
\begin{align*}
   (J+F)\delta |vq| +(n +|vq|F + |vd|+|v|J+(ns -|v|))\delta +\\ c(\delta |vq| ((1-J^2)^{\frac{1}{4}} + J) + \alpha |vq| \frac{J}{\sqrt{1-J^2}})(2\delta |vq| + \alpha(n+ |vq|F^{-1} +|vd|+|v|J^{-1} +sn -|v|))\leq \alpha \epsilon ns
\ .\end{align*}
 We will collect it into the form: 
 \begin{equation}
 G_1(\alpha +G_2)(\alpha+G_3) +G_4 -\alpha sn\epsilon \leq 0\ ,
\end{equation}
where
\begin{align}
    G_1 = c |vq| \frac{J}{\sqrt{1-J^2}}(n+ |vq|F^{-1} +|vd|+|v|J^{-1} +sn -|v|) \\
    G_2 = \delta |vq| ((1-J^2)^{\frac{1}{4}} + J) \frac{\sqrt{1-J^2} }{ |vq|J} \\
G_3 =\frac{2\delta |vq|  }{n+ |vq|F^{-1} +|vd|+|v|J^{-1} +sn -|v|}
    \\
   G_4 =  (J+F)\delta |vq| +(n +|vq|F + |vd|+|v|J+(ns -|v|))\delta 
\ .\end{align}
The solution is again on the boundary given by
\begin{equation}
   \|V\| =\frac{1}{16}, \quad G_1 \alpha^2 + ( G_1 (G_2+G_3) - sn\epsilon)\alpha + G_4 +G_1G_2 G_3 =0
\ .\end{equation}
We numerically find that the control precision $\delta =1.1\cdot 10^{-5}$ is allowed at $J=0.42$. The corresponding $\alpha =0.000302$.


 This noise level is not unrealistic when compared to the reported control noise \cite{boothby2021architectural} that is between $10^{-2}$ and $10^{-3}$ depending on the specific point in the parameter space. Recall that our rigorous bounds are not tight and a more noisy simulator can still output accurate answers, just without theoretical guarantees.

\end{document}